\documentclass[aps,prl,singlecolumn,10pt,superscriptaddress,floatfix]{revtex4-2}
\usepackage[utf8]{inputenc}
\usepackage{graphicx}             
\usepackage{amsmath,amssymb}      
\usepackage[dvipsnames]{xcolor}
\usepackage{newunicodechar}
\DeclareUnicodeCharacter{202F}{~}
\usepackage{bibunits} 
\usepackage[justification=raggedright,compatibility=false]{caption}    

\usepackage{times} 

\usepackage{fancyhdr}
\usepackage{ragged2e}
\usepackage{float}

\newcommand{\defined}{\mathrel{\vcenter{\offinterlineskip
  \hbox{\scriptsize ---}\vskip0.15ex
  \hbox{\scriptsize ---}\vskip0.15ex
  \hbox{\scriptsize ---}}}}

\usepackage[colorlinks=true, linkcolor=blue, citecolor=blue, urlcolor=blue]{hyperref}

\begin{document}
\title{Ultrafast Dynamics of Spin–Orbit Entangled Excitons Coupled\\ to Magnetic Ordering in van der Waals Antiferromagnet \texorpdfstring{NiPS$_3$}{NiPS3}}

\author{Sidhanta Sahu}
\affiliation{Department of Physical Sciences, Indian Institute of Science Education and Research, Kolkata, West Bengal, India, 741246}

\author{Anupama Chauhan}
\affiliation{Department of Physical Sciences, Indian Institute of Science Education and Research, Kolkata, West Bengal, India, 741246}

\author{Poulami Ghosh}
\affiliation{Department of Physical Sciences, Indian Institute of Science Education and Research, Kolkata, West Bengal, India, 741246}

\author{Sayan Routh}
\affiliation{Department of Condensed Matter and Materials Physics, S. N. Bose National Centre for Basic Sciences, Kolkata, West Bengal, India, 700 106}
\author{Ruturaj Puranik}
\affiliation{Department of Condensed Matter Physics and Materials Science, Tata Institute of Fundamental research,  Mumbai, Maharashtra, India, 400005}

\author{Setti Thirupathaiah}
\affiliation{Department of Condensed Matter and Materials Physics, S. N. Bose National Centre for Basic Sciences, Kolkata, West Bengal, India, 700 106}

\author{Siddhartha Lal}
\affiliation{Department of Physical Sciences, Indian Institute of Science Education and Research, Kolkata, West Bengal, India, 741246}

 \author{Shriganesh Prabhu S}
\affiliation{Department of Condensed Matter Physics and Materials Science, Tata Institute of Fundamental research,  Mumbai, Maharashtra, India, 400005}

\author{Chiranjib Mitra}
\affiliation{Department of Physical Sciences, Indian Institute of Science Education and Research, Kolkata, West Bengal, India, 741246}

\author{N. Kamaraju}
\email{nkamaraju@iiserkol.ac.in}
\affiliation{Department of Physical Sciences, Indian Institute of Science Education and Research, Kolkata, West Bengal, India, 741246}

\date{\today}

\begin{abstract}
Spin-orbit entangled excitons (SOEE) in two-dimensional (2D) antiferromagnets provide direct access to explore unconventional many body interactions in correlated electron systems. In this work, we carry out a detailed investigation using non-degenerate isotropic and anisotropic pump–probe reflection spectroscopy to probe the ultrafast dynamics of SOEE and their coupling to spin fluctuations in NiPS$_3$. Transient reflectivity data reveals acoustic phonon oscillations at $\sim$27\,GHz, along with two distinct relaxation timescales: fast (1\textbf{--}9\,ps) and  slower components (1\textbf{--}4\,ns) associated with SOEE coherence and spin reordering, respectively. Both timescales exhibit pronounced temperature dependence  near the exciton dissociation ($T_{\mathrm{ED}}
=120\,K$) and Néel ($T_{\mathrm{N}}
=155\,K$) temperatures. The SOEE coherence shortens from $\sim 8\text{\textbf{--}}9\,\mathrm{ps}$ at $T < T_{\mathrm{ED}}$ to $\sim$ 3\,ps at $T > T_{\mathrm{ED}}$ with a finite tail persisting beyond $T_{\mathrm{N}}$. The spin reordering time grows near 120\,K, and shows critical slowing down around $T_{\mathrm{N}}$. Pump fluence studies further corroborate their spin origin. Our findings uncover the direct interplay between the excitonic and spin degrees of freedom across ultrafast and longer timescales, offering new opportunities to probe and engineer emergent many-body interactions in 2D antiferromagnets.


\end{abstract}

\maketitle
\begin{bibunit}[apsrev4-2]
Unconventional excitons emerge in correlated electron systems, where they are entangled with spin and orbital degrees of freedom~\cite{11,13}. The discovery of magnetic 2D van der Waals (vdW) materials has opened new avenues for exploring such novel SOEE~\cite{11,12,13,15,21,31,SOEE1,SOEE2,SOEE3,SOEE4,SOEE5}. For example, NiPS\textsubscript{3}, a quasi-2D antiferromagnetic (AFM) charge-transfer insulator with an indirect band gap of $\sim$1.8 eV~\cite{5,33,21,7}, has recently attracted intense interest as a platform to study a wide range of correlated phenomena. This includes strong spin--charge coupling~\cite{5}, photoinduced magnetic anisotropy~\cite{6}, exciton-driven antiferromagnetic metallicity~\cite{7}, and magnetically brightened electron-phonon bound states~\cite{8}. The negative photoinduced conductivity at THz frequencies~\cite{9}, sharp excitonic features in photoluminescence~\cite{11,12,15,21,31,SOEE1,SOEE2,SOEE3} and magnon-like dispersion observed in resonant inelastic X-ray scattering (RIXS)~\cite{12}, highlight the strong coupling between excitons and the spin background in NiPS\textsubscript{3}.

The nature of these excitons, however, remains under debate: they have been variously attributed to Hund excitons~\cite{12}, charge-transfer excitons~\cite{17}, and coherent many-body excitons~\cite{11}. Regardless of their precise origin, these excitons can be tuned via optical excitation~\cite{7,9,allington2025distinct,38} and magnetic fields~\cite{15,26,31}, enabling direct pathway to explore spin–exciton coupling. In NiPS\textsubscript{3}, the AFM spin ordering effectively enhance the Coulomb attraction between electrons and holes in forming excitons~\cite{11}, analogous to phonon-driven Cooper pairing in superconductors~\cite{43}. More recently, Hamad et. al.~\cite{16}, theoretically demonstrated that these excitons can even be dressed by the AFM background, and propagate in the material as “singlet polarons”. Despite these advancements, direct experimental evidence for dynamic coupling — the nonequilibrium interplay between these novel excitonic coherence and the AFM order — remains elusive, especially in the time-resolved spectroscopic studies.

Here, we track the dynamics of both SOEE and spin reordering by photoexciting NiPS\textsubscript{3}  above the charge-transfer gap, across a wide temperature range of 5–296 K, using time-resolved reflectivity in the near-infrared (NIR) regime. We observe a clear signature of critical slowing down~\cite{29,30} of spin fluctuations near N\'eel temperature ($T_{N}=155\,\text{K}$), accompanied by the opening and collapse of a spin-wave gap. A phase coherent exciton state is, on the other hand, observed below a exciton dissociation temperature $T_{ED}=120\,\text{K}$. Evidence for a dynamical coupling between the excitonic and magnetic degrees of freedom is obtained from a loss of coherence of the SOEE state due to the critical slowing down of spin fluctuations.
The observed dynamical coupling between SOEE and the spin ordering in NiPS\textsubscript{3} provides a new paradigm to explore nonequilibrium critical behaviour in 2D quantum magnets and paves the way for ultrafast control of correlated low energy excitation dynamics in van der Waals antiferromagnets.

\begin{figure}[!htbp]
    \centering
    \includegraphics[width= 1\linewidth]{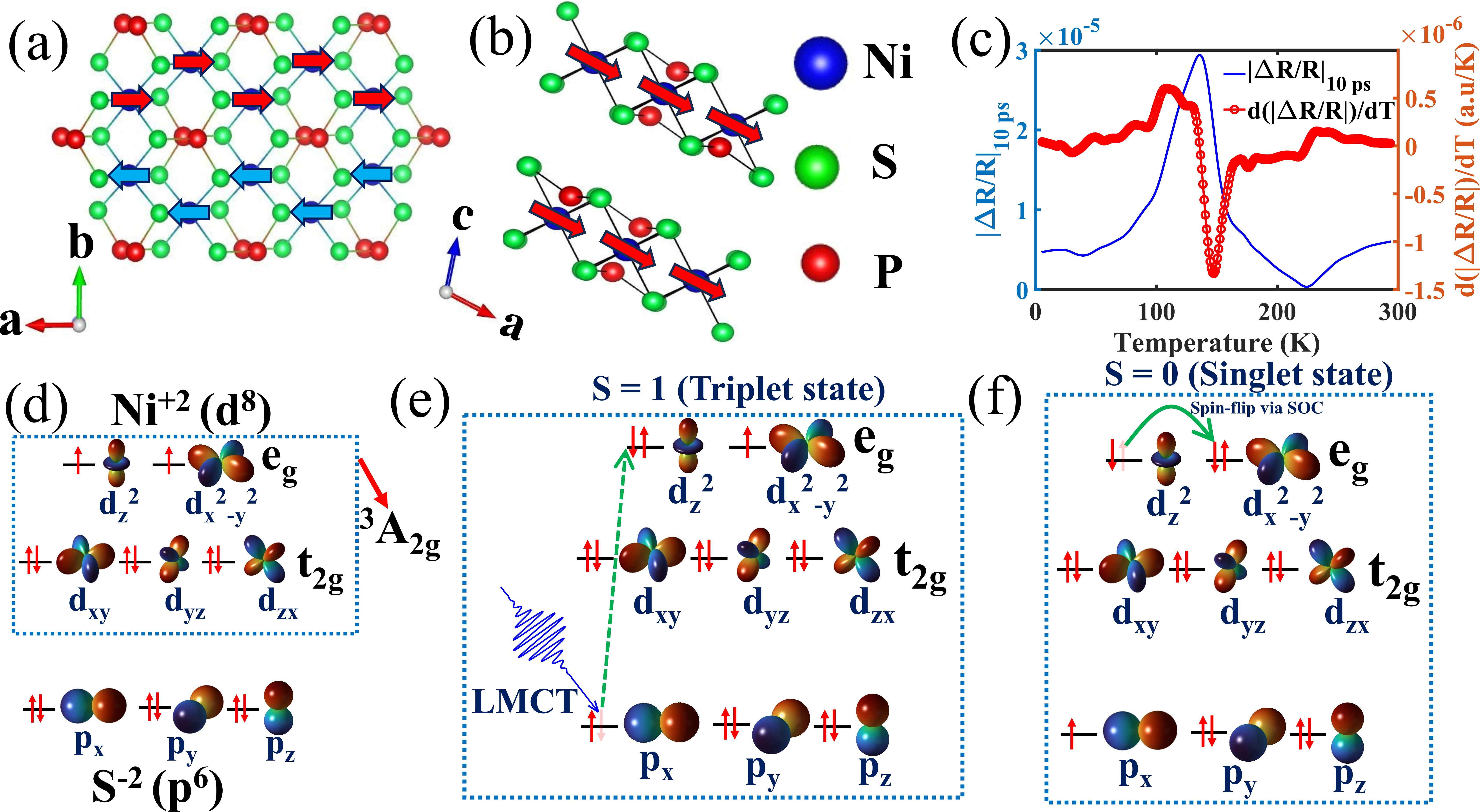} 
    \caption{\justifying(a) Crystal and magnetic structure of NiPS\textsubscript{3}: (a) in the \textit{ab} plane where Ni atoms (blue spheres) form a honeycomb lattice with zigzag AFM ordering along the \textit{a}-axis (red and cyan arrows); (b) in the \textit{ac} plane along the \textit{c}-axis, showing weak interlayer ferromagnetic coupling.  (c) The peak transient reflectivity at the time delay of 10 ps vs temperature (blue line) capturing the AFM-PM transition along with its temperature derivative (red circles). (d) Electronic ground state of NiPS\textsubscript{3} (\protect{$^3$A\protect{$_{2g}$}), where Ni\protect{$^{2+}$} 3d-orbitals are split into lower-energy t\protect{$_{2g}$} and higher-energy e\protect{$_g$} levels by the octahedral crystal field from surrounding S\protect{$^{2-}$} ligands. (e) Upon 3.1\,eV photoexcitation, a LMCT occurs, promoting an electron from the S 3p-orbitals to the Ni e\protect{$_g$} manifold. (f) SOC enables a transition within the e\protect{$_g$} levels, where the excited electron flips and goes to d\protect{$_{x^2 - y^2}$} from d\protect{$_{z^2}$}, forming a bound SOEE state.}
    }
     \label{fig:1a} 
   \label{fig:1b}
    \label{fig:1c}
     \label{fig:1d}
      \label{fig:1e}
       \label{fig:1f}
\end{figure}

High-quality single crystals of NiPS\textsubscript{3} of dimensions 5\,\textit{mm}\,$\times$ 5\,\textit{mm}\,$\times$ 0.5\,\textit{mm}
 were grown by a chemical vapor transport method (See Supplemental Material(SM) Sec. S1 for details).
NiPS$_3$ is a quasi-two-dimensional antiferromagnet in which Ni$^{2+}$ ions form a honeycomb lattice of edge-sharing NiS$_6$ octahedra~\cite{5,13}. Below $T_N$, spins on Ni sites align ferromagnetically along the crystallographic \textit{a}-axis and couple antiferromagnetically between zigzag chains in the \textit{a}-\textit{b} plane [Fig.~\hyperref[fig:1a]{\ref*{fig:1a}(a)}]. The interlayer exchange along the \textit{c}-axis is weak and favours ferromagnetic alignment [Fig.~\hyperref[fig:1b]{\ref*{fig:1b}(b)}]~\cite{5,11,12,13}. Prior literature~\cite{5,11,17} identifies NiPS$_3$ as a charge-transfer material, where excitons mainly originate from ligand (Sulphur)-to-metal (Nickle) charge transfer (LMCT), specifically from S 3p to Ni 3d orbitals, due to strong orbital hybridization~\cite{33}. These excitons are strongly influenced by the underlying AFM order~\cite{11,12,13,15,38}, making them a sensitive probe of spin and charge dynamics.

We performed non-degenerate isotropic and anisotropic pump-probe reflectivity measurements (See SM Sec. S2 and S5 for details of the experimental procedure) on NiPS$_3$ single crystals. 
An ultrashort ($\sim$ 90\,fs~\cite{foot1}
) optical pump pulse centred at 3.14\,eV, initiates the LMCT process, in which an electron is excited from the ligand 3p-orbitals into the nickel 3d orbitals (d$_{z^2}$, and d$_{x^2-y^2}$) within the e$_g$ manifold, as illustrated in Fig.~\hyperref[fig:1e]{\ref*{fig:1e}(e)}. Following photoexcitation, spin-orbit coupling (SOC) facilitates~\cite{16} an intra e$_g$ orbital transition (between d$_{z^2}$ and d$_{x^2-y^2}$) of the excited electron[Fig.~\hyperref[fig:1f]{\ref*{fig:1f}(f)}]. The resulting configuration, with one hole spin confined to the Ni 3d orbital and the other spread over the 3p orbitals of the surrounding S ligands, evolves into a SOEE through strong electronic correlations~\cite{7,13}. The transient reflectivity ($\Delta R/R$) is recorded using a probe pulse at 1.57\,eV, which is sensitive to the SOEE dynamics~\cite{7}. All measurements were carried out 
over a temperature range of 5--294\,K. The pump (230~$\mu$J/cm$^2$) and probe (9~$\mu$J/cm$^2$) fluences were kept constant across all temperatures to ensure the linear response regime.


\begin{figure}[!htbp]
  \centering
  \includegraphics[width= 1\linewidth]{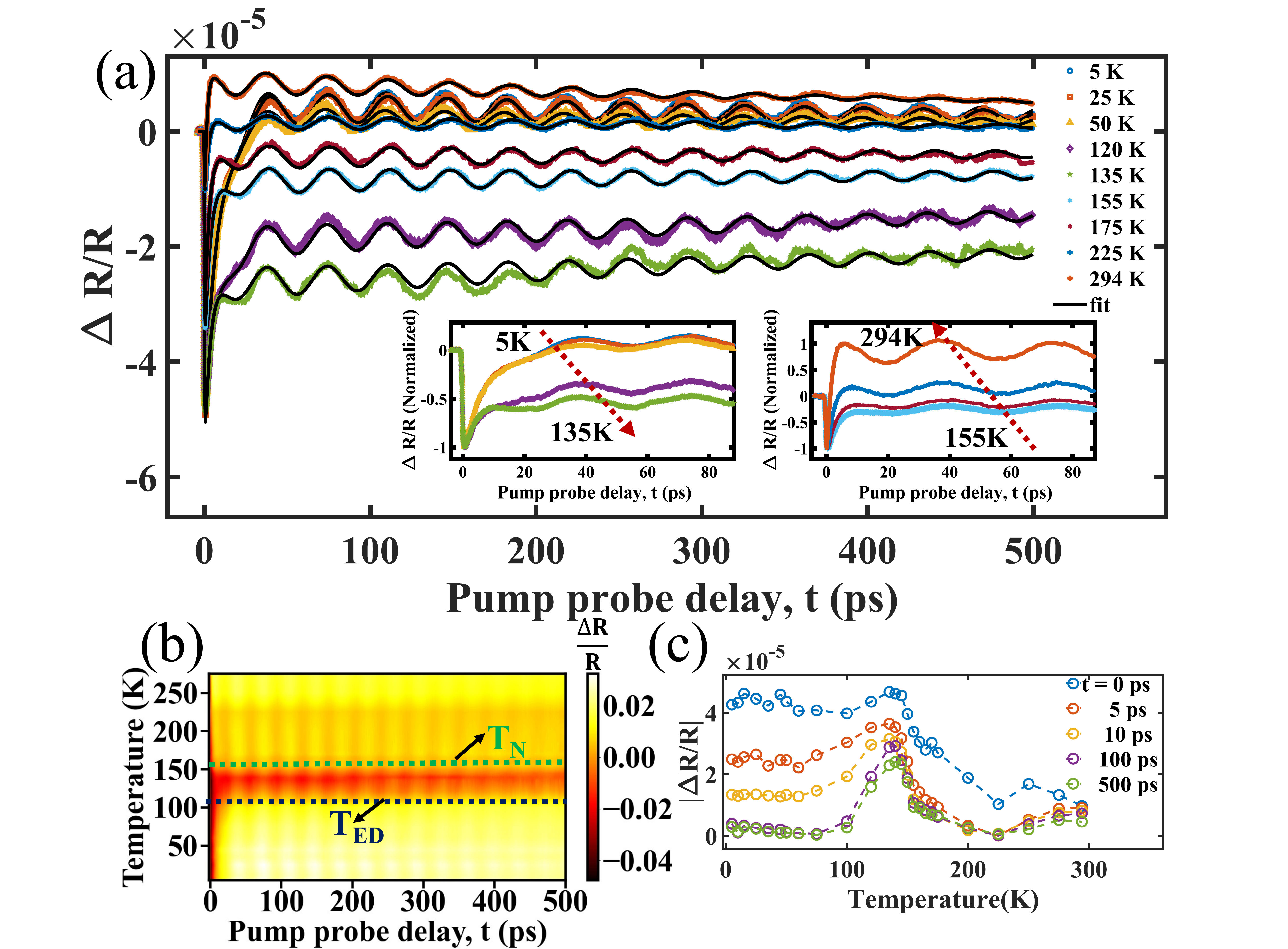}
  \caption{\justifying (a) Temperature-dependent photoinduced differential reflectivity ($\Delta R/R$) traces of NiPS$_3$ (coloured lines) along with fits to Eq.~\hyperref[eq:1]{(\ref*{eq:1})} (black solid lines). The insets show the same data, normalized to their respective negative peak amplitudes, allowing direct comparison of relaxation dynamics at $5~\text{K} < T < 135~\text{K}$ and $155~\text{K} < T < 294~\text{K}$. (b) 2D false-colour map showing the evolution of transient reflectivity as a function of pump-probe delay (0--500\,ps) and temperature. The horizontal green and black dotted lines indicate $T_N$ and $T_{ED}$ respectively. (c) $\Delta R/R$ as a function of temperature for five discrete pump-probe delays to highlight the temperature-dependent evolution of the instantaneous transient reflectivity signal.}
  \label{fig:2a} 
    \label{fig:2b} 
      \label{fig:2c} 
\end{figure}

Figure~\hyperref[fig:2a]{\ref*{fig:2a}(a)} illustrates the temperature-dependent temporal evolution of the photoinduced differential reflectivity ($\Delta R/R$) in NiPS$_3$, capturing the relaxation dynamics following optical pump pulse excitation. At 5\,K, the transient reflectivity exhibits a sharp negative peak with a rise time of $\sim$293\,fs, indicating ultrafast SOEE formation~\cite{17,18}. The subsequent dynamics reveal a bi-exponential decay superimposed with an oscillation at 27\,GHz, which can be assigned to the longitudinal coherent acoustic phonon (LCAP) based on earlier Brillouin--Mandelstam spectroscopy study~\cite{19}. By tracking these dynamics from 5\,K to 294\,K [see inset of Fig.~\hyperref[fig:2a]{\ref*{fig:2a}(a)}], we identify three relaxation regimes. At the lower temperatures (5--75\,K), the decay profiles remain nearly unchanged, indicating a temperature-insensitive relaxation behaviour in this range. As the sample warms into the intermediate regime (75--150\,K), the overall recovery dynamics slow down. Upon further heating above $T_N$, the recovery becomes faster.

\begin{figure}[!htbp]
  \centering
  \includegraphics[width= 1\linewidth]{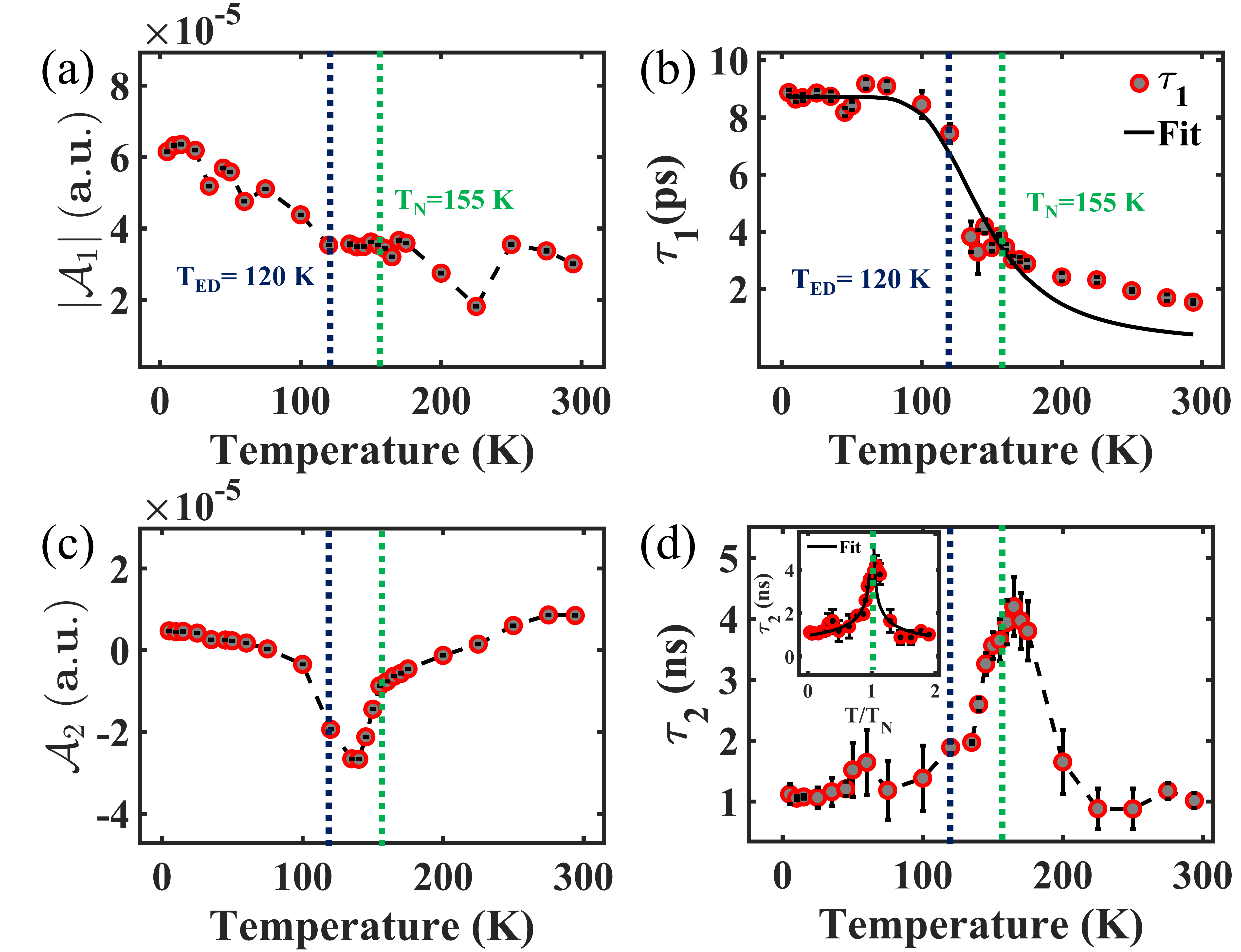}
   \caption{\justifying (a) Temperature dependence of the amplitude associated with the coherence of SOEE. (b) The coherence time ($\tau_1$) of SOEE, Solid black line represents the fit (see text). (c) and (d) show the temperature dependence of the amplitude and lifetime of the slower relaxation component ($\tau_2$), attributed to spin-ordering dynamics. The inset in (d) presents a power-law fit (black solid line) to $\tau_{2}$, revealing critical slowing down as the system approaches the N\'eel temperature. The blue and green vertical dotted line indicates the $T_{ED} \sim 120$\,K and $T_N \sim 155$\,K respectively. The black dashed line connecting the symbols is drawn as a guide to the eye.
}

  \label{fig:3a} 
    \label{fig:3b} 
      \label{fig:3c} 
        \label{fig:3d}
         \label{fig:3} 
\end{figure}

To elucidate the impact of magnetic ordering on the transient reflectivity signal, we construct a 2D false colour plot of $\Delta R/R$ as a function of pump--probe delay (0--500\,ps) and temperature [Fig.~\hyperref[fig:2b]{\ref*{fig:2b}(b)}]. A bright, rectangular band appears in the map between $T_{ED} \sim 120$ K and $T_N \sim 155$ K, highlighting the regime where signal is most responsive to underlying interactions. We further analyse the temperature dependence by extracting $\Delta R(t)/R$ at five discrete pump--probe delays (t= 0\,ps, 5\,ps, 10\,ps, 100\,ps, and 500\,ps).
[Fig.~\hyperref[fig:2c]{\ref*{fig:2c}(c)}]. In each trace, ${\Delta R}/{R}$  grows as the sample warms toward $T_N$, peaks near $T_N$, and then falls off at higher temperatures, producing a clear hump-shaped temperature dependence. This behaviour is attributed to strong spin-charge and spin-phonon coupling in NiPS$_3$~\cite{5,37}, which enables the magnetic phase transition to modulate the dielectric function and, subsequently, the time-resolved reflectivity/transmission signal~\cite{20,23,tian2016ultrafast}. The effect can be understood as the narrowing of the spin-wave gap near $T_N$, which allows high-energy phonons generated by carrier relaxation to excite spin states across the gap~\cite{20}.
To understand the underlying physical processes after photoexcitation, the $\Delta R/R$ data were fitted with a model detailed below:
\begin{equation}
\resizebox{\columnwidth}{!}{$
\begin{aligned}
\frac{\Delta R}{R}(t) 
  &= \Bigg(\sum_{i=1}^{2} \mathcal{A}_i e^{-t/\tau_i}\,(1 - e^{-t/\tau_r})
   + \mathcal{A}_{\mathrm{ph}} e^{-t/\tau_{ph}}\cos(2\pi f t + \phi)\Bigg)
     \otimes g(t)
\end{aligned}
$}
\label{eq:1}
\end{equation}
Here, the first term signifies the biexponential relaxation, where ($\mathcal{A}_{1}, \tau_1$) and ($\mathcal{A}_{2}, \tau_2$) denote the amplitudes and decay constants of two distinct channels with a rise time of $\tau_r$ (see SM Sec. S3). The second term is the LCAP contribution, expressed as a damped harmonic oscillator with amplitude ($\mathcal{A}_{ph}$), damping time ($\tau_{ph}$), frequency ($f$), and initial phase ($\phi$). The instrument response function $g(t)$ is modeled as a Gaussian profile $g(t) = \tfrac{1}{\sigma \sqrt{2\pi}} e^{-\frac{t^{2}}{2\sigma^{2}}}$, where $\sigma$ is determined experimentally from the pump--probe cross-correlation. The extracted parameters corresponding to the bi-exponential decay are presented in Fig.~\hyperref[fig:3]{\ref*{fig:3}}, while the LCAP contribution is discussed in SM Sec. S4. After $\Delta R/R$ reaches its negative maximum, the relaxation proceeds via two channels: a faster decay ($\tau_1$), attributed to SOEE coherence, and a slower decay ($\tau_2$), associated with spin reordering, as discussed below.

The faster component, $\tau_1$, decreases from $\sim$9\,ps at low temperatures (T$<$75\,K) to $\sim$4\,ps near 120\,K, followed by a more gradual decrease till room temperature [Fig.~\hyperref[fig:3b]{\ref*{fig:3b}(b)}]. We attribute $\tau_1$ to the relaxation of SOEE, which remain coherent at low temperatures and progressively lose coherence as spin fluctuations begin to develop near 120\, K\,[Fig.~\hyperref[fig:3c]{\ref*{fig:3c}(c)}], well below T$_N$. Photoluminescence (PL) linewidths\cite{23,11,12,22,31,SOEE4} provide an important benchmark for this assignment. Ho et al.~\cite{23} in recent micro-thermoreflctance studies identified A$_1$ band-edge excitons~\cite{23} with a narrow linewidth of $\sim$6\,meV, corresponding to a sub-picosecond lifetime ($\sim$0.6\,ps). Even broader A$_2$ and B excitonic PL linewidths ($>$100\,meV) imply faster decay of $\sim 40$\,fs. These timescales are much shorter than our observed $\tau_1$, effectively ruling out band-edge excitons as the dominant origin of the dynamics. In contrast, Kang \textit{et al.}\cite{11} reported an ultranarrow PL linewidth of $\sim$400\,$\mu$eV at 5\,K, corresponding to a SOEE lifetime of $\sim$10\,ps, which broadens with increasing temperature due to decoherence effects, particularly beyond $\sim$ 120\,K. Comparable timescales are also found in optical pump–terahertz probe (OPTP) studies ($\sim$17\,ps)\cite{9} and time-resolved PL experiments ($\sim$10–11\,ps)\cite{21,31}. The close agreement between these values and our measured $\tau_1$ strongly supports its assignment to SOEE coherence. At elevated temperatures ($T>T_N$), however, the faster $\tau_1$ is likely to arise from a mixture of relaxation channels, including carrier–carrier relaxation\cite{e-h}, short-range SOEE sustained by residual magnetic correlations~\cite{Saha}. The fast-component amplitude $\mathcal{A}_{1}$ [Fig.~\hyperref[fig:3a]{\ref*{fig:3a}(a)}] decreases with increasing temperature, reflecting the diminishing contribution of excitonic coherence at elevated temperatures.

The trend of $\tau_1$ is well captured by a phenomenological expression adapted from the Rothwarf--Taylor (RT) model~\cite{40}, originally formulated to describe quasiparticle--phonon bottlenecks in superconductors~\cite{41}, and later extended to excitonic systems~\cite{42}. The RT framework has proven effective in modelling relaxation dynamics across an energy gap, particularly in systems exhibiting recombination bottlenecks and phonon-mediated recovery delays. Accordingly, $\tau_1$ is fitted to the following expression~\cite{40}: $\tau_1(T) = 1/\!\left[ K + L\sqrt{\Delta_E T}\, e^{-\Delta_E/(k_B T)} \right]$.  The parameters $K$ and $L$ were treated as fitting parameters, while $\Delta_E$ represents the characteristic energy scale associated with the exciton gap governing the observed dynamics (see SM Sec. S8). Fitting the fast relaxation time ($\tau_1$) to the RT model yields $\Delta_E = 66\pm12$\,meV, that is approximately half of the exciton binding energy ($\sim$132\,meV) reported in previous THz TDS and OPTP studies on NiPS$_3$~\cite{38}, suggesting a possible correlation between the observed relaxation process and excitonic dissociation.

Notably, the slower relaxation component, $\tau_2$ ($\sim$1--4\,ns), exhibits pronounced critical slowing down near $\sim\,T_N$ [Fig.~\hyperref[fig:3d]{\ref*{fig:3d}(d)}]. A power-law fit [black solid line, inset of Fig.~\hyperref[fig:3d]{\ref*{fig:3d}(d)}; see also Fig.~S8 in SM for clarity] $\tau_2(T) = [\Delta (1 - T/T_N)^m]^{-1}$~\cite{23}, yields an exponent $m = 0.44 \pm 0.18$, consistent with the universal critical exponent of the three-dimensional Heisenberg model~\cite{27,28}, thereby confirming the magnetic origin of the observed critical slowing down. It also gives an energy scale $\Delta \approx 1.07 \pm 0.03$~meV, in agreement with spin-wave gaps reported from electron spin resonance and neutron scattering on NiPS$_3$~\cite{24,25,26}, and a critical temperature $T_N = 157 \pm 4.5$~K, agrees with magnetic susceptibility ($\chi_M$) measurements (see SM, Sec.~S1.3). Meanwhile, the amplitude of the slow component is minimal and positive ($\mathcal{A}_{2} \approx 0.006 \pm 7.1 \times 10^{-6}$) at $5\,\mathrm{K}$ [Fig.~\hyperref[fig:3c]{\ref*{fig:3c}(c)}], reflecting rigid long-range spin order that suppresses energy transfer between spins. With increasing temperature, softened spin stiffness~\cite{20} promotes such transfer, generating stronger spin fluctuations that act as a dephasing channel for SOEE coherence, as reflected in the evolution of $\tau_2$ and $\mathcal{A}_{2}$ (blue dotted line).

Above $T_N$, thermal disorder suppresses spin coherence, leading to a reduction in amplitude ($\mathcal{A}_{2}$), which reflects the coupling strength between exciton and spin order parameters (see End Matter). The observed sign reversal of $\mathcal{A}_{2}$ from negative to positive near $T_N$ provides a clear signature of a second-order magnetic phase transition~\cite{30}, accompanied by a non-monotonic variation of $\mathcal{A}_{2}$ for $120\,K < T < 175\,K$, indicative of evolving exciton–spin coupling across the critical region. Further, the anisotropic temperature-dependent time-resolved reflectivity data exhibit similar dynamics, indicating that the same underlying processes are at play (See SM Sec. S5 for details). The $\tau_2$ component, which appears consistently across both isotropic and anisotropic time resolved measurements, is thus attributed to spin reordering dynamics~\cite{Lovinger}, supported by its temperature-dependent evolution. A Ginzburg–Landau theory for these observations is presented in the End Matter and SM. 

\begin{figure}[!htbp]
  \centering
  \includegraphics[width= 1\linewidth]{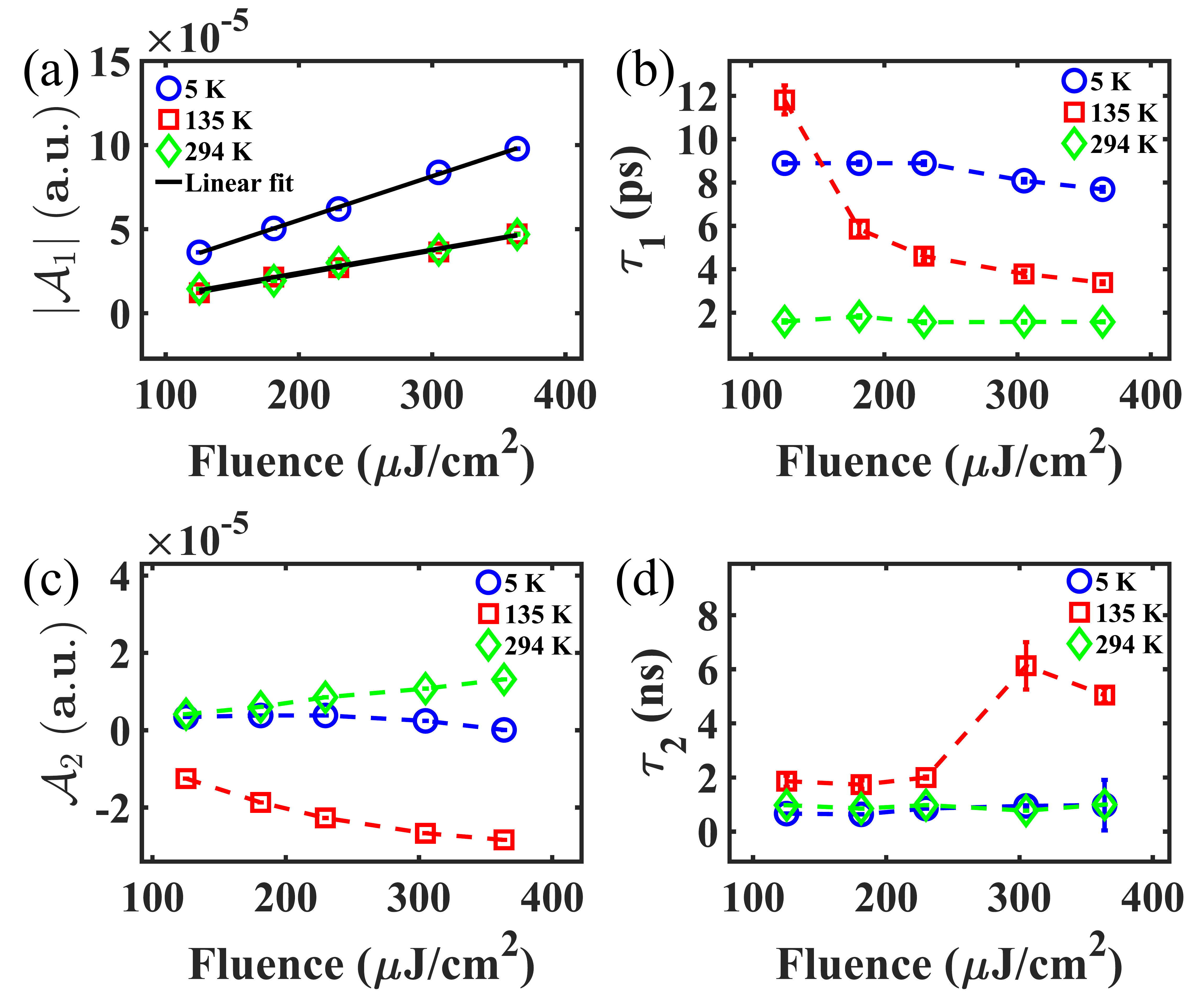}
    \caption{\justifying Parameters extracted as a function of pump excitation fluence. (a) Amplitude and (b) coherence time of the SOEE at different temperatures. (c) and (d) shows the amplitude and spin-ordering dynamics respectively.}

  \label{fig:4a} 
    \label{fig:4b} 
      \label{fig:4c} 
        \label{fig:4d} 
\end{figure}

Next, we quantify carrier density--dependent relaxation dynamics by performing fluence dependent measurements at 5\,K, 135\,K, and 294\,K with pump fluences ranging from $120\,\mu\mathrm{J/cm}^2$ to $364\,\mu\mathrm{J/cm}^2$ and a fixed probe fluence of $9\,\mu\mathrm{J/cm}^2$. 
The resulting transient reflectivity traces were fitted to Eq.~\hyperref[eq:1]{(\ref*{eq:1})}, as shown in SM Fig.~S2.4 (solid black lines), and the extracted fit parameters are compiled in Fig.~\hyperref[fig:4a]{\ref*{fig:4a}(a)--(d)}. The fit parameters of fast($\tau_1$) and slow components($\tau_2$)  display distinct trend: At 5\,K, $\tau_1$  [Fig.~\hyperref[fig:4b]{\ref*{fig:4b}
(b)}] remains nearly constant, consistent with coherent SOEE dynamics in the long-range AFM phase. Here, SOEE coherence remains robust: even at exciton densities as high as $N_{\mathrm{exciton}} \approx 1.2 \times 10^{20}\,\mathrm{cm}^{-3}$ (see SM Sec. S8), screening does not affect its coherence. In the paramagnetic regime (294\,K), $\tau_1$ is short ($<1.2$\,ps) and fluence independent consistent with intrinsic electron–hole plasma dynamics where carrier relaxation is governed by carrier–carrier and carrier–phonon scattering~\cite{e-h}. Around 135\,K, however, $\tau_1$ decreases markedly from $\sim 12$\,ps to $\sim 4$\,ps with increasing fluence. This reduction stems from enhanced carrier and exciton densities at high fluence, which strengthen Coulomb screening and destabilize SOEEs. Concurrently, many-body channels such as exciton–exciton annihilation~\cite{45} and carrier-induced photoionization~\cite{46}, together with weakened AFM correlations near 135\,K, further amplify SOEE instability and drive the marked decrease in $\tau_1$.


 In parallel, the slow decay constant $\tau_2$ remains nearly unchanged at 5\,K and 294\,K  but exhibhits a pronounced enhancement near $T_{\mathrm{N}}$ [Fig.~\hyperref[fig:4d]{\ref*{fig:4d}(d)}]. At 135\,K, the pump excitation disrupts the spin fluctuations, eroding the correlated background and dissociating SOEE, thereby prolonging relaxation. In effect, the lifetime shortening seen in $\tau_1$ is mirrored by a lifetime prolongation in $\tau_2$ — a clear fingerprint of competing many-body channels. These contrasting trends indicate that screening destabilizes SOEEs and reveal a dynamical coupling between SOEE excitations and AFM order. These results uncover a nonequilibrium regime in NiPS$_3$ where ultrafast exciton decoherence coexists with long-lived spin disorder, establishing an optical pathway to control correlated excitations.
 
The fluence dependence of amplitudes  $\mathcal{A}_{1}$ and $\mathcal{A}_{2}$ is shown in  [Fig.~\hyperref[fig:4a]{\ref*{fig:4a}(a)} and \hyperref[fig:4c]{\ref*{fig:4c}(c)}]. At 5\,K, $\mathcal{A}_{1}$ rises sharply as a function of fluence, with a slope of $0.4\ \mathrm{a.u./(mJ/cm^{2})}$, indicating enhanced SOEE population. At 135\,K, excitons are largely dissociated, $\mathcal{A}_{1}$ is nearly three times smaller than at 5~K and, accordingly, increases with fluence at a reduced slope of $0.2\ \mathrm{a.u./(mJ/cm^{2})}$. For 294\,K, both the magnitude and trend of $\mathcal{A}_{1}$ versus fluence are similar to 135\,K, suggesting the emergence of an electron–hole plasma. $\mathcal{A}_{2}$ remains near zero with a milder variation, reflecting weak spin involvement at 5\,K. At 294\,K, $\mathcal{A}_{2}$ shows a positive, linear increase with fluence. Near $T_{\mathrm{N}}$, however, $\mathcal{A}_{2}$ becomes increasingly negative with fluence, indicating growing energy transfer from dissociating SOEE to the spin system driven by enhanced spin fluctuations. While not central to the present discussion, we note that the 27\,GHz acoustic phonon mode exhibits temperature-dependent frequency shifts consistent with anharmonic decay, with negligible variation under pump fluence (see SM, Sec. S4).

The peak reflectivity ($|\Delta R/R|_{peak}$) scales linearly with pump fluence, $|\Delta R/R| = \eta F + R_0$ (SM Fig. S2.4), confirming that the measurements remain in the linear regime. Since $\Delta R/R$ reflects the total density of photoexcited carriers (free and bound), the slope $\eta$ directly tracks quasiparticle generation efficiency. Remarkably, $\eta$ remains nearly constant from 5 K to $\sim 120$ K, but drops sharply near and above $T_N$ (SM Fig. S2.5(d)). This reduction in $\eta$ signals reduced exciton stability, likely caused by spin disorder disrupting excitonic coherence (see end matter and SM, Sec. S2).

In summary, we have investigated the ultrafast dynamics of SOEE in the quasi-two-dimensional antiferromagnet NiPS$_3$ using time-resolved reflectivity. Our measurements reveal sub-picosecond SOEE formation and a pronounced decoherence near the N\'eel temperature, coinciding with the critical slowing down of spin fluctuations and the collapse of the spin-wave gap. These findings provide direct evidence of strong dynamical coupling between magnetic order and excitonic coherence, mediated by spin fluctuations. Our work positions NiPS$_3$ as a model platform to explore the interplay of spin, charge, and orbital degrees of freedom in nonequilibrium regimes, and opens pathways to optically engineer excitonic phenomena in 2D correlated magnets.

\begin{acknowledgments}
The authors thank the Ministry of Education (MoE), Government of India, for funding and IISER Kolkata for the infrastructural support to carry out the research. S.S. acknowledge University Grants Commission (UGC); A.C. acknowledges Council of Scientific and Industrial Research (CSIR); P.G. acknowledges Department of Science and Technology (DST)-Innovation in Science Pursuit for Inspired Research (INSPIRE). SL thanks the SERB, Govt. of India for funding through MATRICS grant MTR/2021/000141 and Core Research Grant CRG/2021/000852. SSP wishes to thank DAE, Government of India and TIFR for supporting the work vide grant no. RTI4003. N. K thanks the SERB, Govt. of India for funding through Core Research Grant CRG/2021/004885.
\end{acknowledgments}

\putbib[references]   
\end{bibunit}

\newpage
\newpage
\section*{End Matter}
\subsection*{Ginzburg–Landau theory}
\noindent To interpret the observed temperature dependence of the two relaxation components, we adopt a Ginzburg-Landau framework of coupled order parameters describing SOEE coherence ($\psi$) and AFM spin ordering ($\zeta$): 

\begin{equation}
F(\psi, \zeta, T) = a_{\psi}(T) |\psi|^{2} + a_{\zeta}(T) \zeta^{2} 
+ b_{\psi} |\psi|^{4} + b_{\zeta} \zeta^{4} + \lambda |\psi|^{2} \zeta^{2}
\end{equation}
where, $a_{\psi}(T) = a_{\psi0}(T - T_{\mathrm{ED}})$ governs exciton dissociation for $T\gtrsim T_{\mathrm{ED}}$ and $a_{\zeta}(T) = a_{\zeta0}(T - T_{\mathrm{N}})$ vanishes at $T_{\mathrm{N}}$, producing critical slowing down of the spin dynamics.

The coupling term \( \lambda |\psi|^2 \zeta^2 \) governs the interaction between excitonic coherence ($\psi$) and spin order ($\zeta$). 
This coupling gives rise to a feedback mechanism, such that the growth of one order parameter influences the stability of the other: a negative \( \lambda \) favors coexistence of the two orders, lowering the free energy and indicating a cooperative interaction. On the other hand, 
a positive \( \lambda \) implies competition, where the presence of one suppresses the other. Experimentally, the effective sign of \( \lambda \) is reflected in the temperature dependence of the slow-component amplitude \( A_2 \), as discussed above. The critical slowing down seen in $\tau_2$ can be understood by solving the time dependent Ginzburg Landau (TDGL) equation~\cite{28} 
\begin{equation}
\frac{\partial \zeta}{\partial t} = -\Gamma_{\zeta} \frac{\partial F}{\partial \zeta} = -\Gamma_{\zeta} (a_{\zeta}(T)\zeta+b_{\zeta}\zeta^{3}+2\lambda |\psi|^{2}\zeta)~,
\end{equation}\\
where $\Gamma_{\zeta}$ is the damping rate of spin fluctuations. To analyze the relaxation rate near $T_{N}$, one linearizes this equation by keeping only the term proportional to $\zeta$ and neglecting the higher-order contributions, $\zeta^{3}$ and $\lambda |\psi|^{2} \zeta$. 
As $T \to T_{N}$, both the spin stiffness $a_{\zeta}(T)$ and the coupling to the SOEE order parameter $\lambda |\psi|^{2}$ become vanishingly small. 
The linear relaxation rate $\partial \zeta/\partial t$ vanishes,  and $\zeta$ becomes quasi-static — this is the critical slowing down observed in $\tau_2$. By contrast, the fast component $\tau_1$ is likely associated with the relaxation of the phase of the (complex) excitonic order parameter $\psi$: the growth in the magnetic order parameter $\zeta$ as temperature is lowered below $T_{N}$ increases the timescale ($\tau_{1}$) over which phase coherence is maintained in the exciton condensate. The stabilisation of the exciton order parameter $\psi$ as $T\to T_{ED}$ then concurs with the saturation of $\tau_{1}$.

\clearpage
\begin{bibunit}[apsrev4-2]
\section*{}

\begin{center}
  \fontsize{15pt}{17pt}\selectfont \textbf{Appendix-SM}  
\end{center}

\fontsize{15pt}{17pt}\selectfont \textbf{S1. Synthesis and Characterization of NiPS$_3$ Single Crystal}
\label{sec:S1}

\vspace{0.5em}
{\fontsize{12pt}{16pt}\selectfont \hspace*{1em}\textbf{S1.1: Sample Preparation}}
\phantomsection
\label{sec:S1.1}

\vspace{0.5em}
{\fontsize{12pt}{16pt}\selectfont \hspace*{0.5em}{
NiPS\textsubscript{3} single crystals used in this study were synthesized via the chemical vapor transport (CVT) method. Stoichiometric amounts of high-purity Ni, P, and S powders (4N, Alfa Aesar), along with iodine (4N, Alfa Aesar) as the transport agent, were sealed in a quartz tube under high vacuum (\(\sim 10^{-6}\,\mathrm{mbar}\)). 
The sealed quartz tube was kept in a two-zone tube furnace for crystal growth. The hot-zone and cold-zone temperatures were maintained at 
750\,\(^\circ\mathrm{C}\) and 650\,\(^\circ\mathrm{C}\), respectively, for 8 days. 
Larger crystals with dimensions up to 5\,mm were successfully grown.}}

\vspace{0.5em}
{\fontsize{12pt}{16pt}\selectfont \hspace*{1em}\textbf{S1.2: X-ray Diffraction of single crystal and energy-dispersive x-ray analysis}}
\phantomsection
\label{sec:S1.2}

\vspace{0.5em}
{\fontsize{12pt}{16pt}\selectfont \hspace*{0.5em}{The crystal structure and orientation of NiPS$_3$ were confirmed using X-ray diffraction (XRD), and its elemental composition was confirmed using energy-dispersive X-ray analysis(EDX). The XRD pattern [Fig.~S1.2.1] shows clear (00$l$) reflections, consistent with the expected layered structure and indicating that the crystal is oriented along the $c$-axis. The peak positions match well with previously reported data~\cite{XRD1,XRD2,XRD3}, confirming high crystal quality. EDX measurements show the presence of Ni, P, and S in near-stoichiometric amounts, confirming the chemical purity of the sample [Fig. S1.2.2].

\vspace{0.5em}
The elemental composition of NiPS$_3$ single crystals as determined by EDX as shown in Fig.~S1.2.2.

Exact composition determined from EDX analysis is: NiP$_{1.03}$S$_{2.86}$

\begin{figure}[H]
  \centering
  \includegraphics[width= 0.65\linewidth]{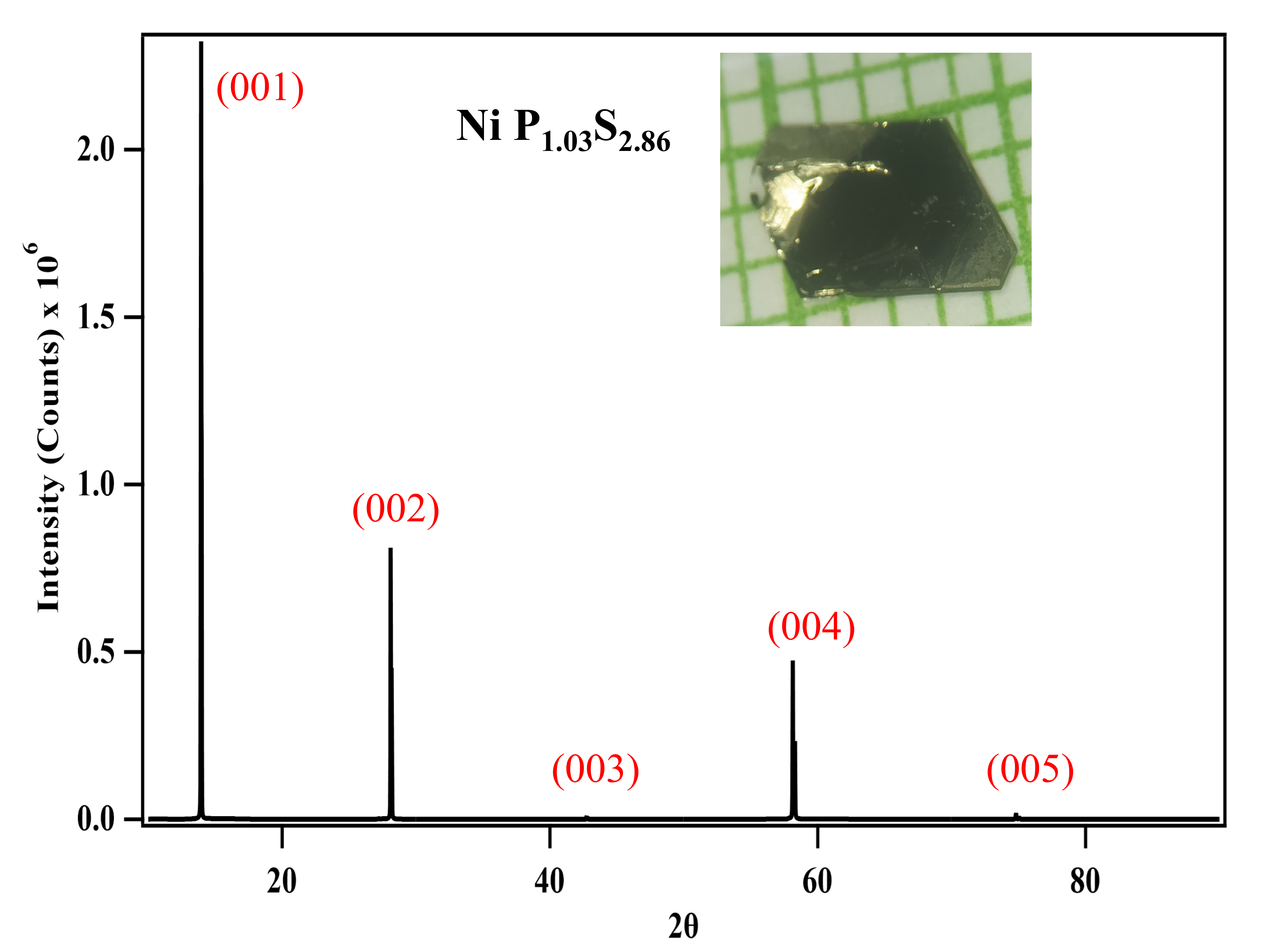}
    \caption*{\justifying FIG.~S1.2: X-ray diffraction spectrum of NiPS$_3$. Room-temperature XRD pattern of a NiPS$_3$ single crystal showing (001), (002), (003),(004), and (005) reflections, confirming $c$-axis orientation and in good agreement with literature reports~\cite{XRD1,XRD2,XRD3}.}
  \label{fig:S1.2.}
\end{figure}
}}

\begin{figure}[H]
  \centering
  \includegraphics[width= 0.7\linewidth]{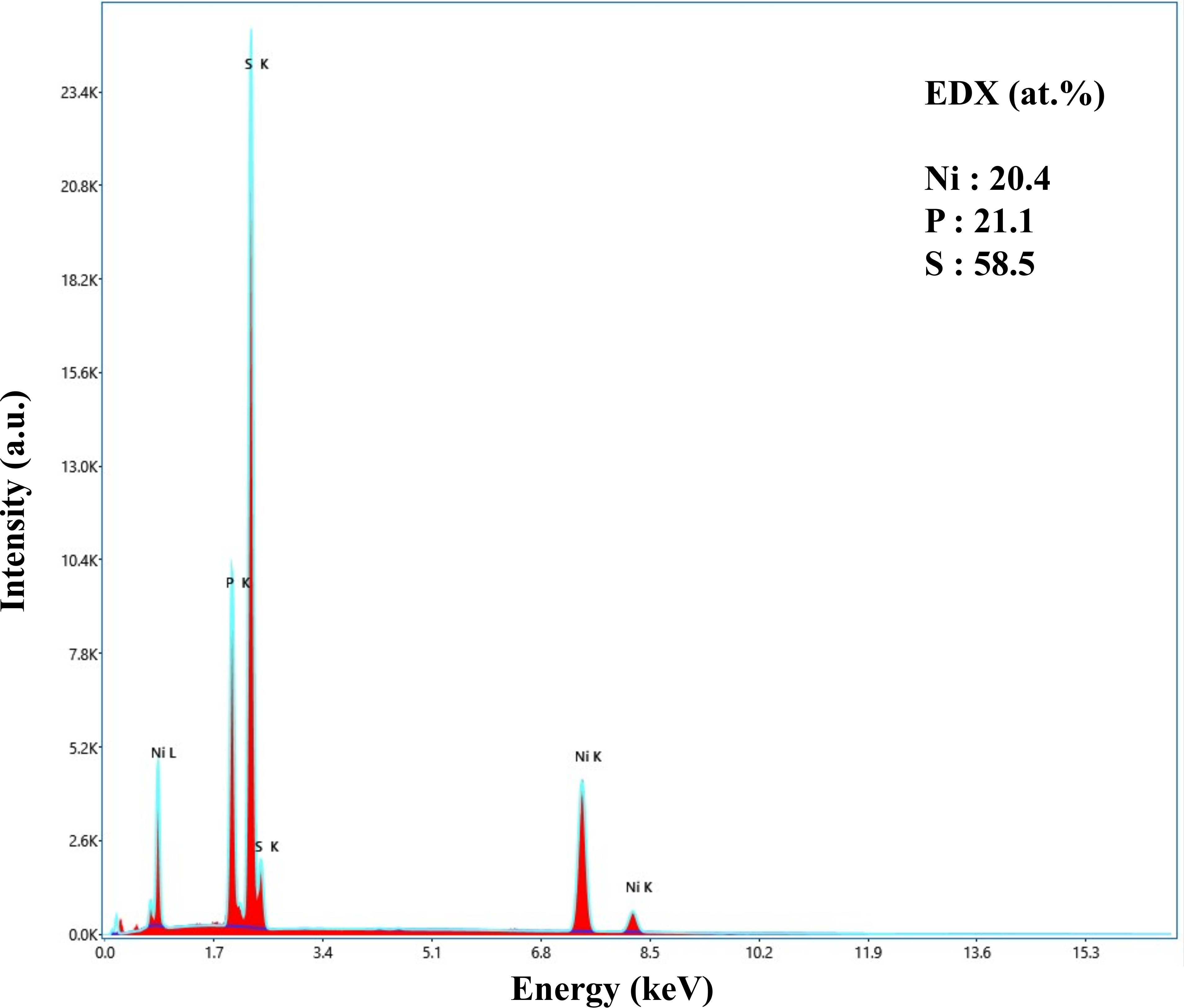}
  \caption*{FIG.~S1.2.1: EDX spectrum of NiPS$_3$ showing the characteristic peaks of Ni, P, and S.}
  \label{fig:S1.2.1}
\end{figure}

\newpage
\vspace{0.5em}
{\fontsize{12pt}{16pt}\selectfont \hspace*{1em}\textbf{S1.3: Magnetic susceptibility of NiPS$_3$}}
\phantomsection
\label{sec:S1.3}

\vspace{0.5em}
{\fontsize{12pt}{16pt}\selectfont \hspace*{0.5em}{
Magnetic susceptibility measurements as a function of temperature were performed on NiPS$_3$ single crystals using a SQUID magnetometer. During the measurement, a constant magnetic field of 5~kOe was applied along the $c$-axis of the crystal. The temperature-dependent magnetic susceptibility data are shown in Fig.~S1.3 A clear antiferromagnetic transition is observed in the $\chi(T)$ curve at the N\'eel temperature ($T_\mathrm{N}$) of 153\,K, as highlighted in the inset of Fig.~S1.3, where $\mathrm{d}\chi/\mathrm{d}T$ is plotted. Similar magnetic susceptibility measurements on NiPS$_3$ single crystals have been reported previously, with the N\'eel temperature estimated to be approximately 157\,K~\cite{MT}.

\begin{figure}[H]
  \centering
  \includegraphics[width= 0.7\linewidth]{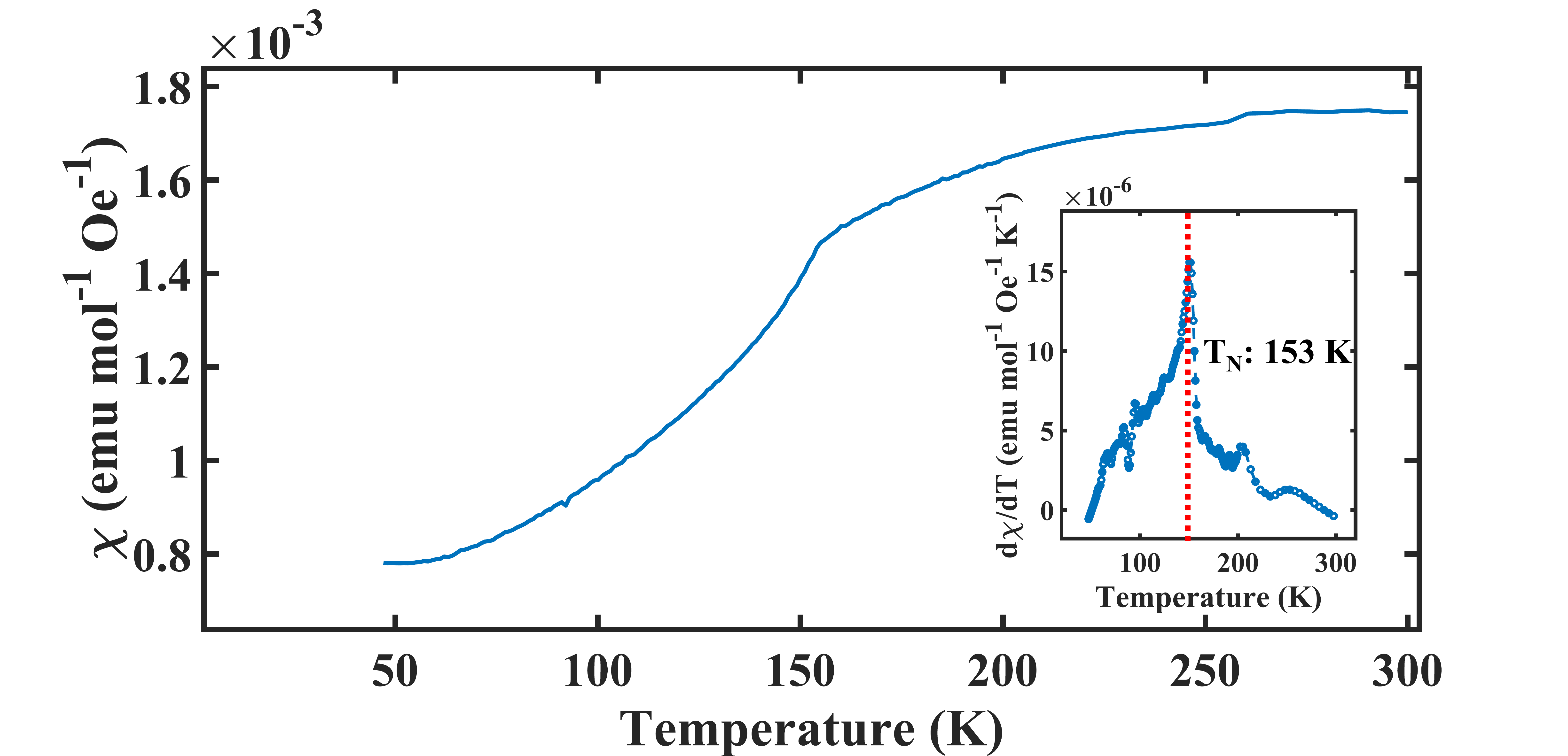}
   \caption*{\justifying FIG.~S1.3: Magnetic susceptibility of NiPS$_3$ crystal. In-plane magnetic susceptibility as a function of temperature, measured under a magnetic field of 5~kOe applied along the $c$-axis (H $\parallel$ $c$). The inset shows its derivative with respect to temperature, from which $T_{\mathrm{N}} \sim 153\,\mathrm{K}$ is identified.}
  \label{fig:S1.3}
\end{figure}
}}

\vspace{0.5em}
{\fontsize{12pt}{16pt}\selectfont \hspace*{1em}\textbf{S1.4: Raman spectroscopy of NiPS$_3$}}
\phantomsection
\label{sec:S1.4}

\vspace{0.5em}
{\fontsize{12pt}{16pt}\selectfont \hspace*{0.5em}{
Raman spectroscopy was performed on bulk NiPS$_3$ single crystals using a Horiba Jobin Yvon HR800 system with 488\,nm laser excitation. The scattered signal s were collected with a 50$\times$ objective and dispersed by a 600\,grooves/mm diffraction grating. The resulting spectrum is presented in Fig.~S1.4. Multi-Lorentzian fitting of the spectrum identified six distinct phonon modes at 132.03, 176.77, 281.42, 384.04, 559.99, and 588.84~cm$^{-1}$. These modes are assigned to the $A_g$ and $B_g$ symmetries of the $C2h/m$ point group, in agreement with previous reports~\cite{Raman}. The extracted peak positions and full width at half maximum (FWHM) values are summarized in Table~TS1.1. In addition, two weaker peaks at 220.98~cm$^{-1}$ and 435.01~cm$^{-1}$ are observed (labeled $N_{1}$ and $N_{2}$ in Fig.~S1.4), consistent with previous literature~\cite{Raman,Raman2,Raman3,Raman4}, and are attributed to intramolecular vibrations of the $(\mathrm{P}_{2}\mathrm{S}_{6})^{4-}$ bipyramidal structure~\cite{Raman2}.}

\begin{table}[H]
\centering
\renewcommand{\arraystretch}{1.5} 
\setlength{\tabcolsep}{16pt}      
\large                             

\begin{tabular}{|c|c|c|c|}
    \hline
    \textbf{Mode} & \textbf{Irreducible Representation} & \textbf{Wave number (cm$^{-1}$)} & \textbf{FWHM (cm$^{-1}$)} \\
    \hline
    M1 & $A_g$, $B_g$ & $132.03 \pm 0.11$ & $6.09 \pm 0.31$ \\
    M2 & $A_g$, $B_g$ & $176.77 \pm 0.04$ & $3.83 \pm 0.12$ \\
    M3 & $A_g$, $B_g$ & $281.42 \pm 0.07$ & $5.28 \pm 0.21$ \\
    M4 & $A_g$        & $384.04 \pm 0.04$ & $5.53 \pm 0.12$ \\
    M5 & $A_g$, $B_g$ & $559.99 \pm 0.45$ & $8.78 \pm 1.18$ \\
    M6 & $A_g$        & $588.84 \pm 0.08$ & $4.51 \pm 0.23$ \\
    \hline
\end{tabular}
\vspace{0.5em}
\captionsetup{labelformat=empty}
\caption{\textnormal{TS1.4: Raman spectroscopy of NiPS$_3$: The parameters extracted from multi-Lorentzian fitting of the Raman spectrum are summarized here.}}
\end{table}

\begin{figure}[H]
  \centering
  \includegraphics[width= 0.75\linewidth]{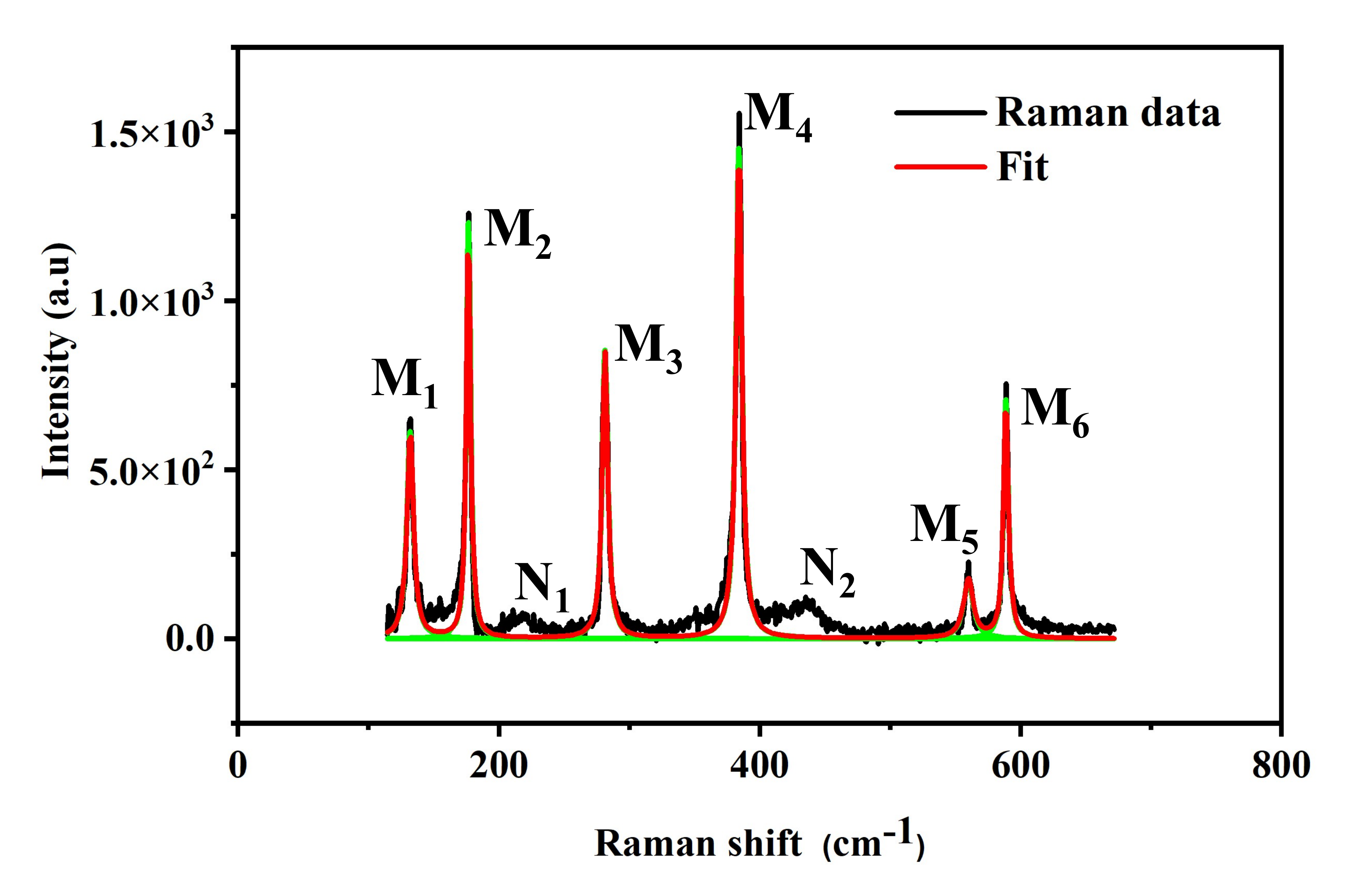}
   \caption*{\justifying FIG.~S1.4: Room-temperature Raman spectroscopy. Raman data (black lines) of bulk NiPS$_3$ along with its fit. The corresponding multi-Lorentzian fit is shown by the red line. The Raman modes are labeled M$_1$–M$_6$, corresponding to the $A_g$ and $B_g$ symmetries, while the other modes are labeled as N$_1$ and N$_2$.}
  
  \label{fig:S1.4}
\end{figure}

}

\newpage

\fontsize{15pt}{30pt}\selectfont \textbf{S2. Isotropic pump probe dynamics and data analysis}
\label{sec:S2}

\vspace{0.5em}
{\fontsize{12pt}{16pt}\selectfont \hspace*{0.5em}{
Temperature-dependent time-resolved non-degenerate transient reflectivity measurements were performed to investigate the dynamics of photoexcited spin–orbit entangled excitons (SOEE) in NiPS$_3$ single crystals. The experiments were conducted in a reflection geometry, as illustrated in Fig.~S2.1.

The laser pulses used in the experiment is from a femtosecond amplifier system (RegA 9050, Coherent Inc.) operating at 790\,nm, 250\,kHz, with 60\,fs pulses of 6.4\,\textmu J energy. The output was split at 80:20; the higher-intensity beam is frequency-doubled to 395\,nm using a type II $\beta$-barium borate (BBO) crystal for use as the pump, while the lower-intensity beam served as the probe at the fundamental wavelength 790\,nm. A motorized delay stage was incorporated into the probe beam path to introduce a controllable time delay between the pump and probe pulses at the sample point. The pump and probe beams were spatially overlapped on the sample surface, which was mounted in a closed-cycle cryostat (Oxford OptistatDry BLV) with a capability to change the sample temperature in the range of 5--294\,K. The pump-induced changes in the reflectivity of the probe beam were detected using a balanced photodetector. To enhance the signal-to-noise ratio, phase-sensitive detection was employed using a lock-in amplifier synchronized with an optical chopper that  modulated the pump beam at a frequency of 713\,Hz.

The full width at half maximum (FWHM) diameters of the pump and probe beams at the sample position were approximately  59\,$\mu$m and 54\,$\mu$m, respectively. For fluence-dependent measurements, the pump fluence was varied from 120 to $360~\mu\mathrm{J/cm^2}$, whereas in temperature-dependent studies, it was fixed at $230~\mu\mathrm{J/cm^2}$. The probe fluence remained constant at $9~\mu\mathrm{J/cm^2}$ in both cases. The fluences used in this study are such that the sample is not damaged by irradiation. The instrument response, determined from the pump--probe cross-correlation by type II BBO crystal, was approximately 120\,fs and corresponds to an actual pulse width of $\sim 90\,\mathrm{fs}$ (for a $\mathrm{sech}^2$ pulse shape).

\setlength{\textfloatsep}{2pt} 
\begin{figure}[H]
  \centering
  \includegraphics[width= 0.7\linewidth]{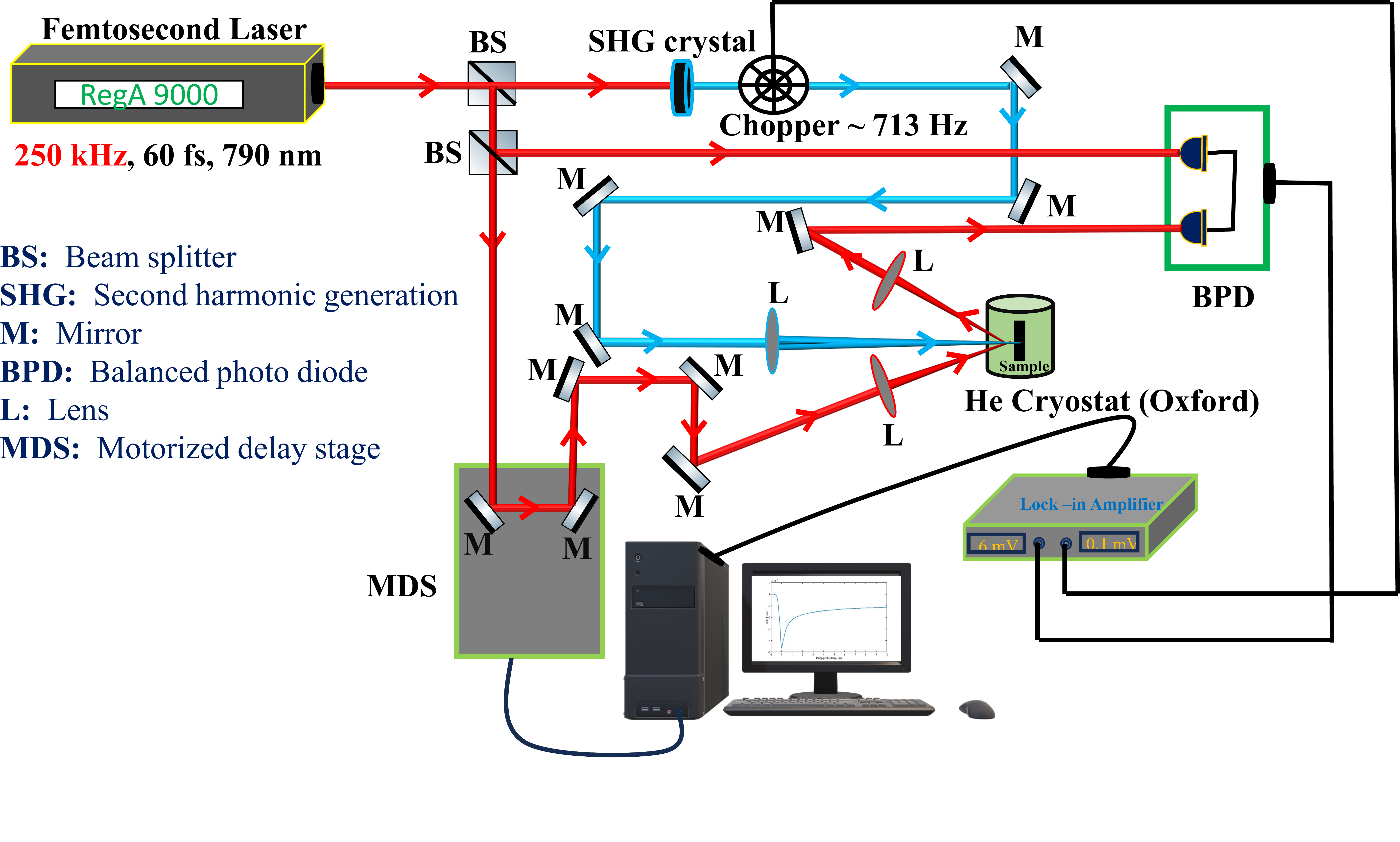}
    \caption*{\justifying FIG.~S2.1: Schematic representation of the experimental setup for non-degenerate pump-probe spectroscopy performed in reflection geometry.}
  \label{fig:S2.1}
\end{figure}
\noindent As shown in Fig.~S2.2, the two runs show excellent agreement, confirming the high reproducibility of the experiment.

\begin{figure}[H]
  \centering
  \includegraphics[width= 0.75\linewidth]{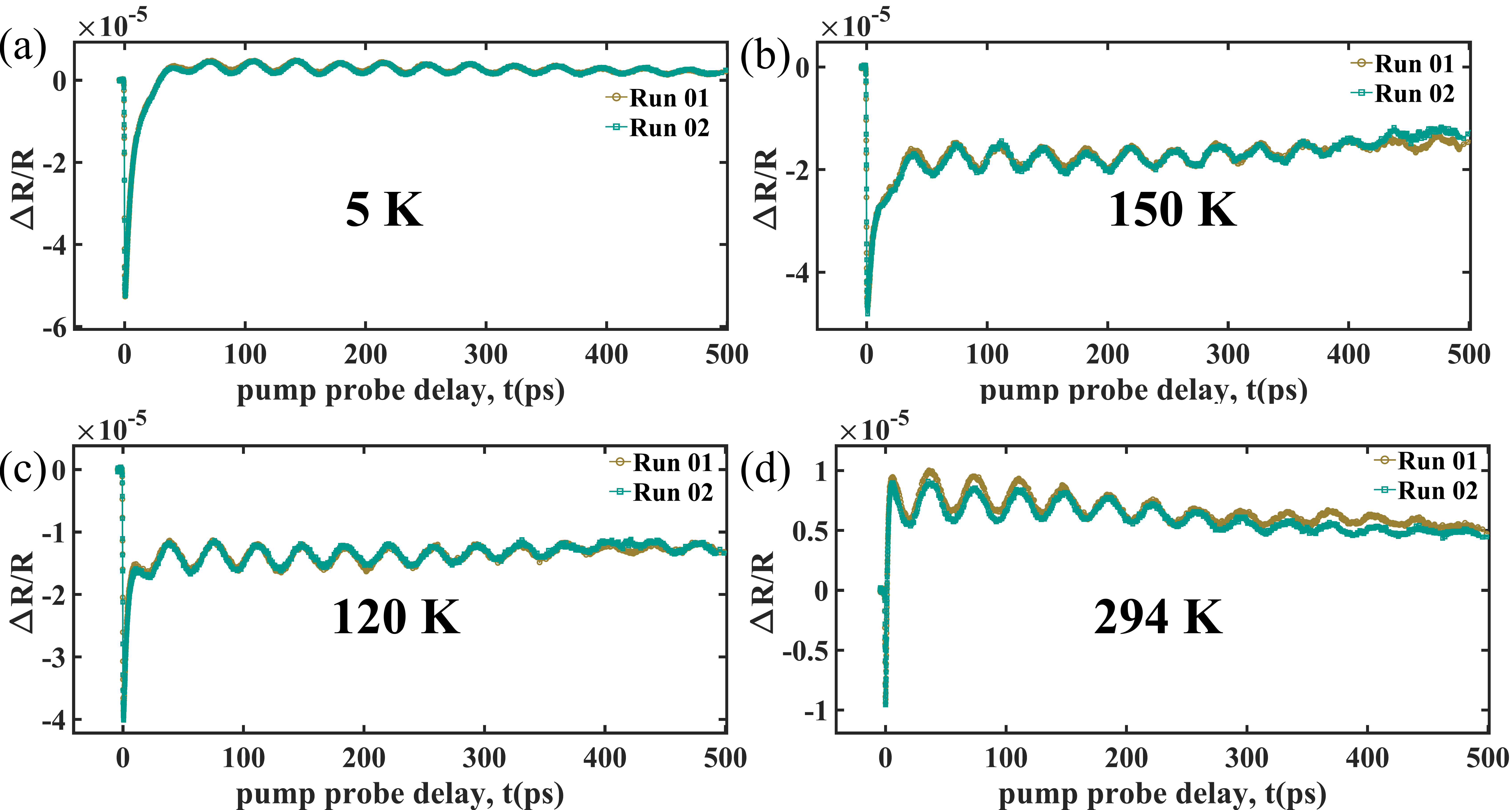}
    \caption*{\justifying FIG.~S2.2: Repeatability of pump--probe measurements at (a) 5K (b) 120\,K, (c) 150\,K,(d) 294\,K corresponding to low temperature, exciton dissociation regime, magnetic ordering transition, and room temperature, respectively.}
  \label{fig:S2.2}
\end{figure}

\begin{figure}[H]
  \centering
  \includegraphics[width= 0.7\linewidth]{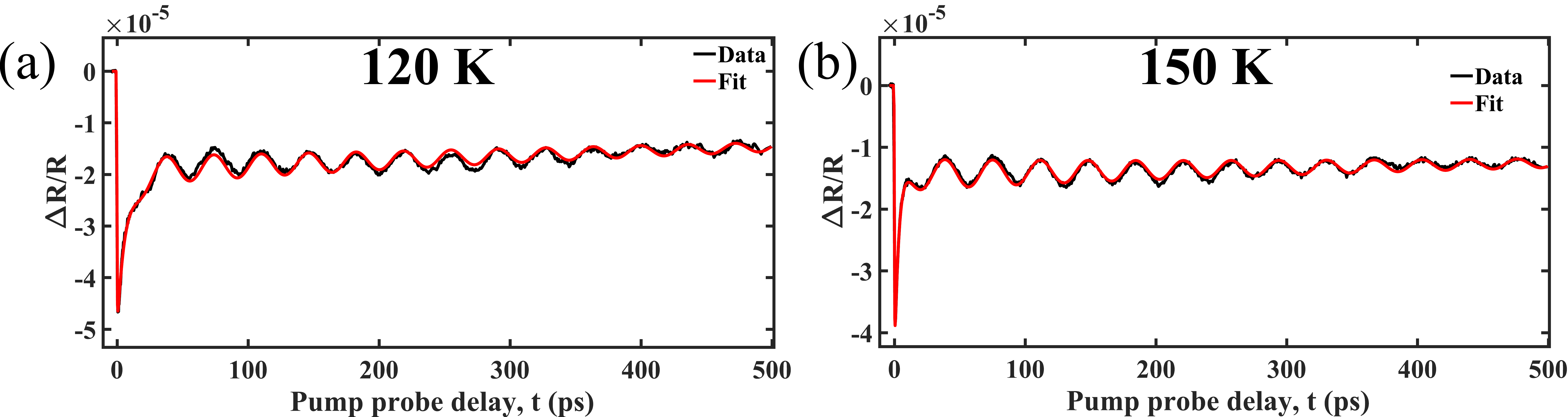}
    \caption*{\justifying FIG.~S2.3: Transient differential reflectivity data measured at (a) 120\,K and (b) 150\,K (black lines). 
The red curves show fits obtained using Eq.~(1) defined in the main text.
}
  \label{fig:S2.3}
\end{figure}

\begin{figure}[H]
  \centering
  \includegraphics[width= 0.7\linewidth]{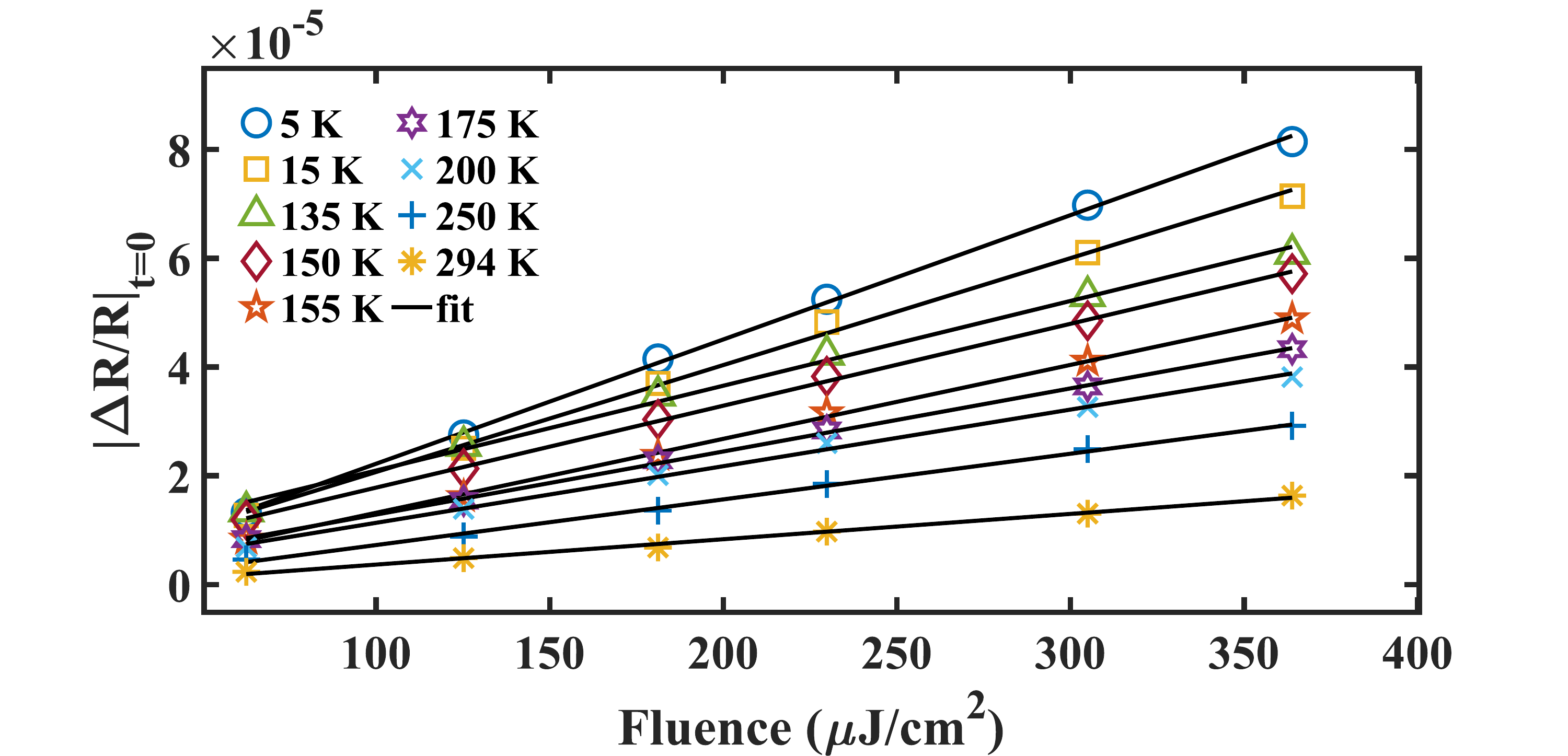}
    \caption*{\justifying FIG.~S2.4: Transient differential reflectivity peak $({\Delta R}/{R})_{t=0}$ as a function of pump fluence at different temperatures. The solid black line represents a linear fit, as discussed in the main text.}
  \label{fig:S2.4}
\end{figure}

\begin{figure}[H]
  \centering
  \includegraphics[width= 1\linewidth]{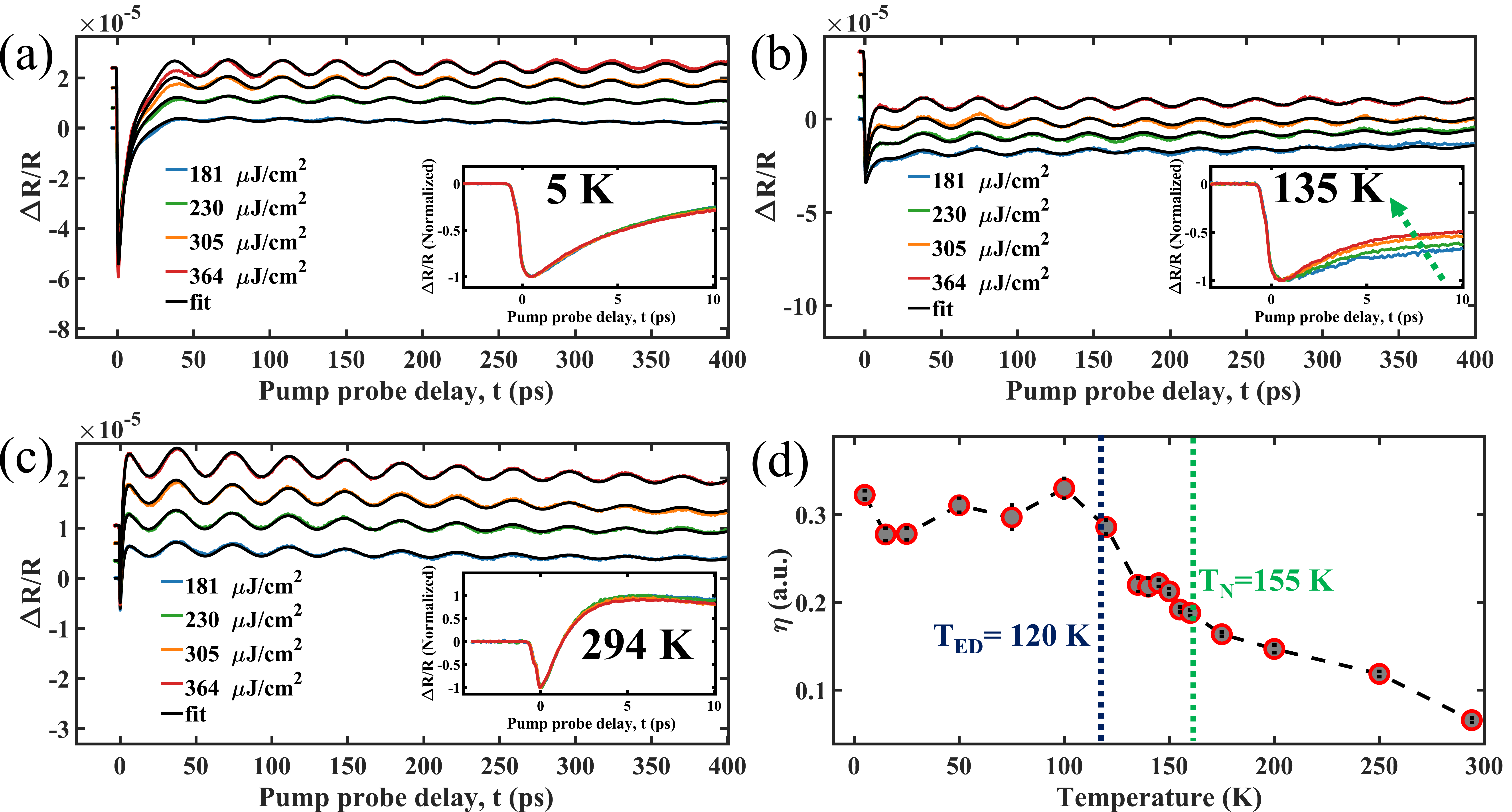}
    \caption*{\justifying FIG.~S2.5: Time-resolved $\Delta R/R$ signals measured at different pump fluences ranging from $181\,\mu\mathrm{J/cm}^2$ to $364\,\mu\mathrm{J/cm}^2$ (different coloured symbols represent different fluences) at different temperature (a) 5\,K (b) 135\,K (c) 294\,K (Black solid line: fit to Eq.~(1), see main manuscript). The data are vertically offset for clarity and the inset highlights the fluence dependence of the fast decay dynamics as the system approaches the N\'eel temperature. (d) Temperature dependence of the slope clearly exhibits significant changes around $T_{\mathrm{ED}}$ (blue dashed line) and $T_{\mathrm{N}}$ (green dashed line), respectively.
}

  \label{fig:S2.5(a)} 
    \label{fig:S2.5(b)} 
      \label{fig:S2.5(c)} 
        \label{fig:S2.5(d)} 
\end{figure}

}}
\newpage
\fontsize{15pt}{30pt}\selectfont \textbf{S3. Rise-Time Dynamics of Transient Reflectivity in NiPS$_3$}
\label{sec:S3}

{\fontsize{12pt}{16pt}\selectfont \hspace*{0.5em}{The rise time ($\tau_r$) was obtained by fitting the data with Eq.~(1) defined in the main manuscript. Fits at 5\,K, 155\,K, and 294\,K are presented in Fig.~S3(a), and the extracted $\tau_r$ as a function of temperature is plotted in Fig.~S3(b). We observe a decrease of $\tau_r$ from $293 \pm 6$ fs at 5 K to $224 \pm 18$ fs at 294 K [Fig.S3(b)]. These sub-picosecond values are substantially faster than those reported in earlier broadband transient reflectivity studies on NiPS$_3$ flakes, where SOEE-related rise times ranged from $\sim$5 ps at 1.65 eV pump to $\sim$1 ps at 2.3 eV pump\cite{17}. This variation has been attributed to excitation-energy-dependent charge-transfer processes involving S 3p and Ni 3d orbitals. In our case, excitation at 3.1 eV yields distinctly faster dynamics (224–293 fs), consistent with the general trend of shorter rise times at higher excitation photon energies~\cite{17}. The precise microscopic origin of these ultrafast dynamics, however, remains to be established.

\begin{figure}[H]
  \centering
  \includegraphics[width= 0.9\linewidth]{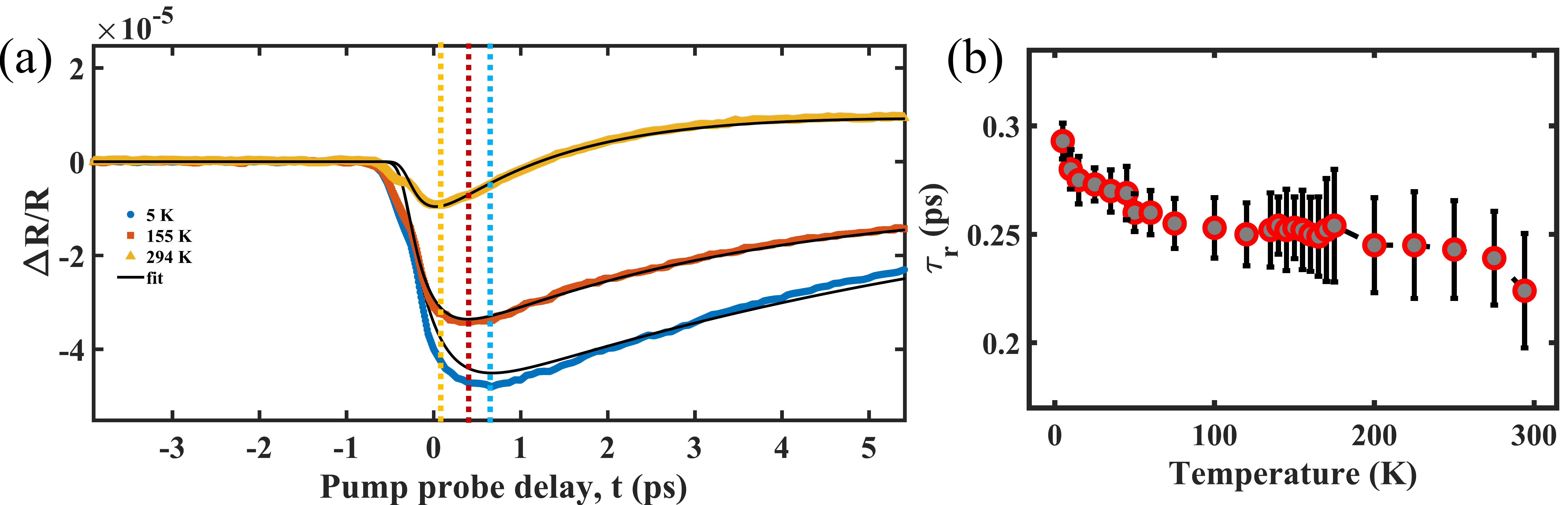}
   \caption*{\justifying FIG.~S3: Temperature dependence of the rise time. (a) Transient differential reflectivity ($\Delta R/R$) traces of NiPS$_3$ measured at 5\,K, 155\,K, and 294\,K, together with fits to the initial rising edge. The extracted rise times are $293 \pm 6$\,fs, $252 \pm 13$\,fs, and $224 \pm 18$\,fs, respectively. The vertical dashed lines (blue, red, and yellow) mark the peak positions corresponding to the maximum signal. (b) Extracted rise time plotted as a function of temperature.}
  \label{fig:S3}
\end{figure}


}}

\fontsize{15pt}{30pt}\selectfont \textbf{S4. Temperature and Fluence Dependence of Coherent Phonons in NiPS$_3$}
\label{sec:S4}

{\fontsize{12pt}{16pt}\selectfont \hspace*{0.5em}{To investigate the lattice dynamics, we analyzed the temperature and fluence dependence of the longitudinal coherent acoustic phonon (LCAP) parameters obtained from the transient reflectivity measurements. The extracted LCAP parameters were obtained by fitting the data using eq~(1), as described in the main text. Their temperature and fluence dependence are presented in Figs.~S4.1 and S4.2, respectively.

The LCAP amplitude is almost constant with temperature, and the phonon frequency softens from $\sim27.8$\,GHz at 5\,K to $\sim27.0$\,GHz at 294\,K ($\sim3\%$), indicating a temperature-dependent softening of the mode. The phonon lifetime $\tau_{\rm ph}$ decreases with increasing temperature [Fig. S4.1(c)], indicative of enhanced anharmonic interactions that accelerate phonon decay at elevated temperatures. The linewidth $\Gamma_{\rm ph}$ increases quadratically with temperature. Whereas the phonon phase remains constant with temperature[Fig. S4.1(d)]

\begin{figure}[H]
  \centering
  \includegraphics[width= 0.7\linewidth]{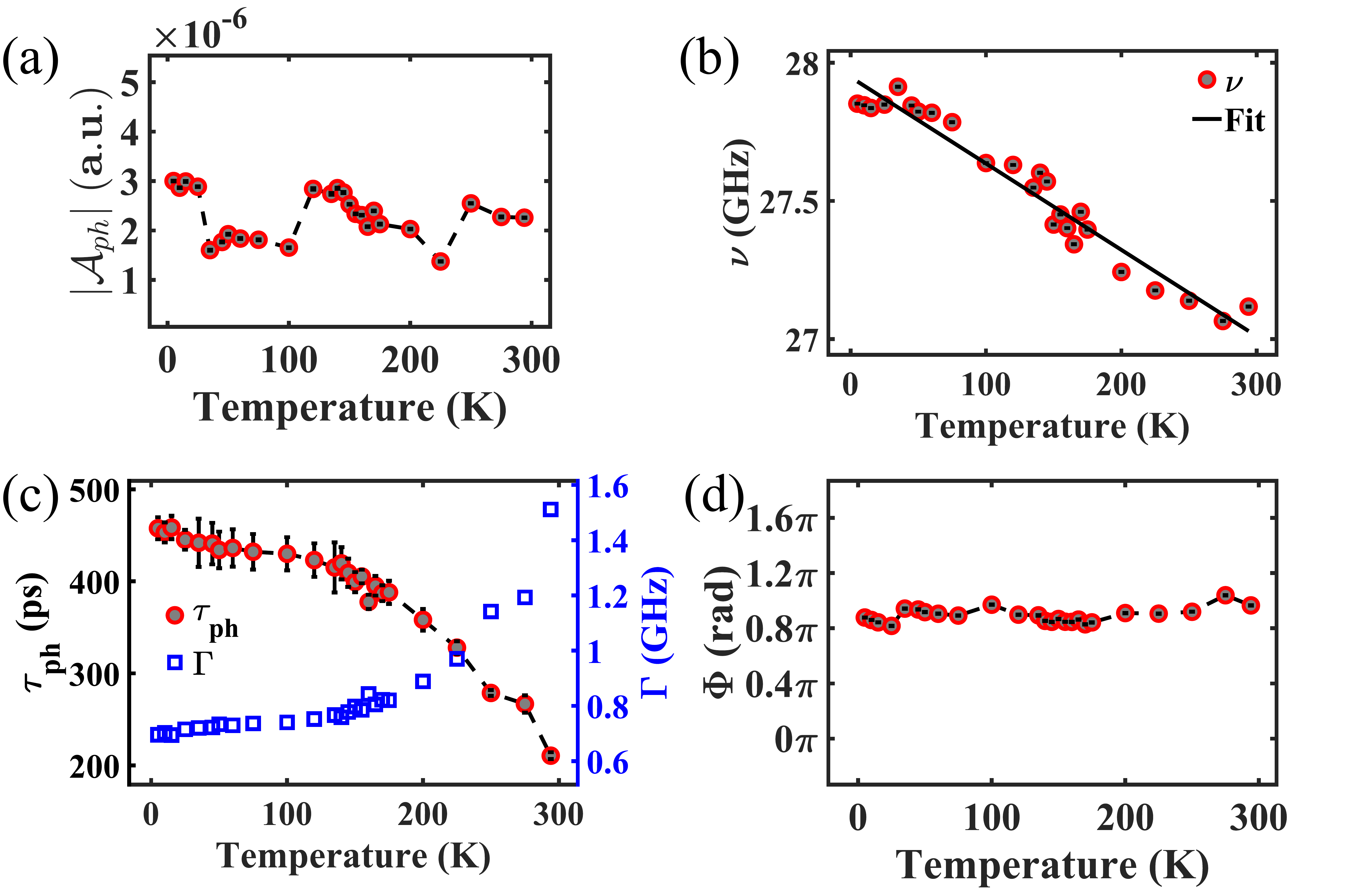}
    \caption*{\justifying FIG.~S4.1: Temperature dependence of LCAP parameters. (a) Amplitude of the oscillations, (b) oscillation frequency, (c) damping time, and (d) phase. Black dashed lines are included as guides to the eye.}
  \label{fig:S4.1}
\end{figure}

The temperature dependence of the phonon mode frequency can be quantitatively described using the model presented in Refs.~\cite{Anh1,Anh2,Anh3}, which accounts for anharmonic lattice effects. According to this model, the frequency variation with temperature is given by: 
\[
\nu_{\text{LA}}(T) = \nu_0 + \Delta\nu_{\text{anh}}(T)
\]
Here $\nu_0$ represents the intrinsic phonon frequency at absolute zero temperature, while $\Delta \nu_{\mathrm{anh}}(T)$ describes the intrinsic temperature-dependent frequency shift arising from anharmonic phonon-phonon interactions. This shift originates from the real component of the phonon self-energy and reflects the decay of an acoustic phonon into two (cubic anharmonicity) or three (quartic anharmonicity) lower-energy acoustic phonons.

Considering only the cubic anharmonic contribution, the frequency shift $\Delta\nu_{\text{anh}}(T)$ takes the form:
\[
\Delta\nu_{\text{anh}}(T) = C\left[1 + \frac{2}{e^{\frac{h\nu_{0}}{2k_B T}} - 1}\right]
\]
Here, $h$, $k_B$ are the Planck’s constant and the Boltzmann constant respectively, $T$ is the absolute temperature, and $C$ quantifies the anharmonic self-energy arising from cubic phonon-phonon interactions~\cite{Selfenergy}. The fit of this model to the experimental data is illustrated by the black line in Fig.~S4.1(b), from which the intrinsic phonon frequency $\nu_{0} = 27.95 \pm 0.02$\,GHz and the anharmonic coupling constant  $C = (-1.04 \pm 0.05) \times 10^{-3}$\,GHz were extracted.

The LCAP amplitude ($\mathcal{A}_{ph}$) shows linear dependence on carrier density, suggesting a contribution from thermoelastic stress or deformation potential mechanisms~\cite{Cd3As2}. The phonon lifetime ($\tau_{\mathrm{ph}}$) remains almost constant with carrier density at all measured temperatures (5 K, 135 K, and 294 K). Similarly, both the phonon frequency and phase show no observable dependence on carrier density throughout the temperature range studied. The generation of longitudinal coherent acoustic phonons (LCAPs) originates from the impulsive stress induced by ultrafast optical excitation, which launches a strain pulse propagating through the crystal lattice. This strain locally modulates the dielectric function, producing periodic oscillations in the probe reflection~\cite{Thomsen}.

\begin{figure}[H]
  \centering
  \includegraphics[width= 0.7\linewidth]{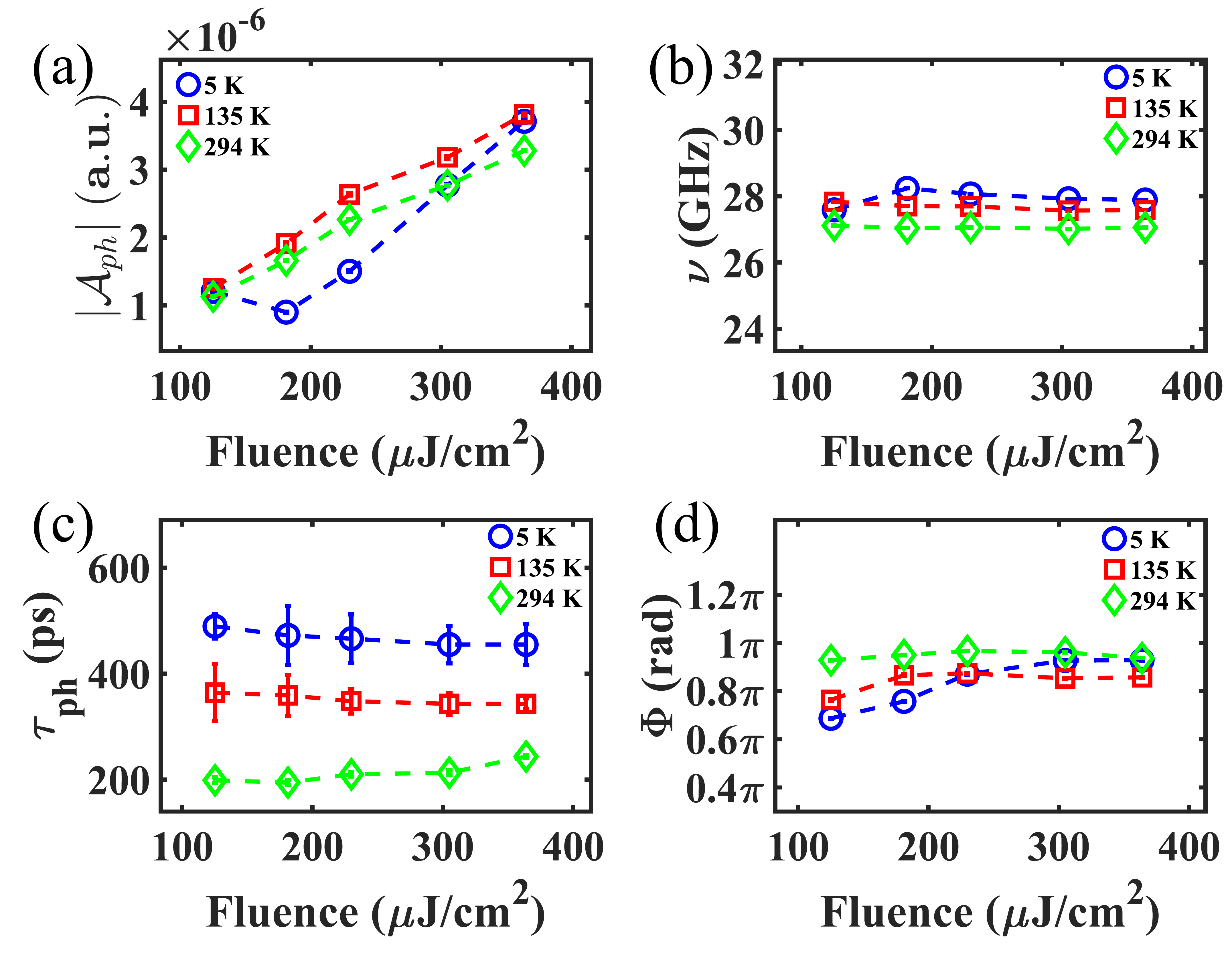}
   \caption*{\justifying FIG.~S4.2: Fluence-dependent of LCAP at 5\,K, 135\,K and 294\,K respectively. (a) amplitude, (b) frequency, (c) damping time, and (d) phase of the oscillations. Solid coloured lines are included as guides to the eye.}
  \label{fig:S4.2}
\end{figure}

The LCAP frequency remains robust against pump fluence and shows only weak temperature dependence, underscoring the stability of NiPS$_3$ for thermoelectric applications~\cite{Cd3As2}.

}}

\newpage

\fontsize{15pt}{30pt}\selectfont \textbf{S5. Anisotropic Transient Reflectivity: Experimental Setup and Data Analysis}
\label{sec:S5}

\vspace{0.5em}
{\fontsize{12pt}{16pt}\selectfont \hspace*{0.5em}{To emphasize the spin-mediated nature of the dynamics, temperature-dependent anisotropic pump–probe measurements were performed. In these measurements, the reflected probe beam was directed through a quarter-wave plate (QWP) and a Wollaston prism (WP). In these measurements,  the probe polarization is kept at 45 deg with respect to the QWP fast axis when the pump beam is blocked. This results in circular polarization and this probe beam after going through the WP, splits it into orthogonal polarization components, that are detected by a balanced photo-diode. In the absence of the pump, the output of the balanced photo-diode is zero.  This detection scheme, in the presence of pump beam, enables high-sensitivity in detecting transient polarization changes, such as rotation or ellipticity, in the reflected signal, also called as Kerr rotation signal. Apart from this polarization-resolved detection scheme, the rest of the experimental setup remained identical to that shown in Fig.~S2.1. The modified detection path is illustrated in Fig.~S5.1.

\begin{figure}[H]
  \centering
  \includegraphics[width= 0.7\linewidth]{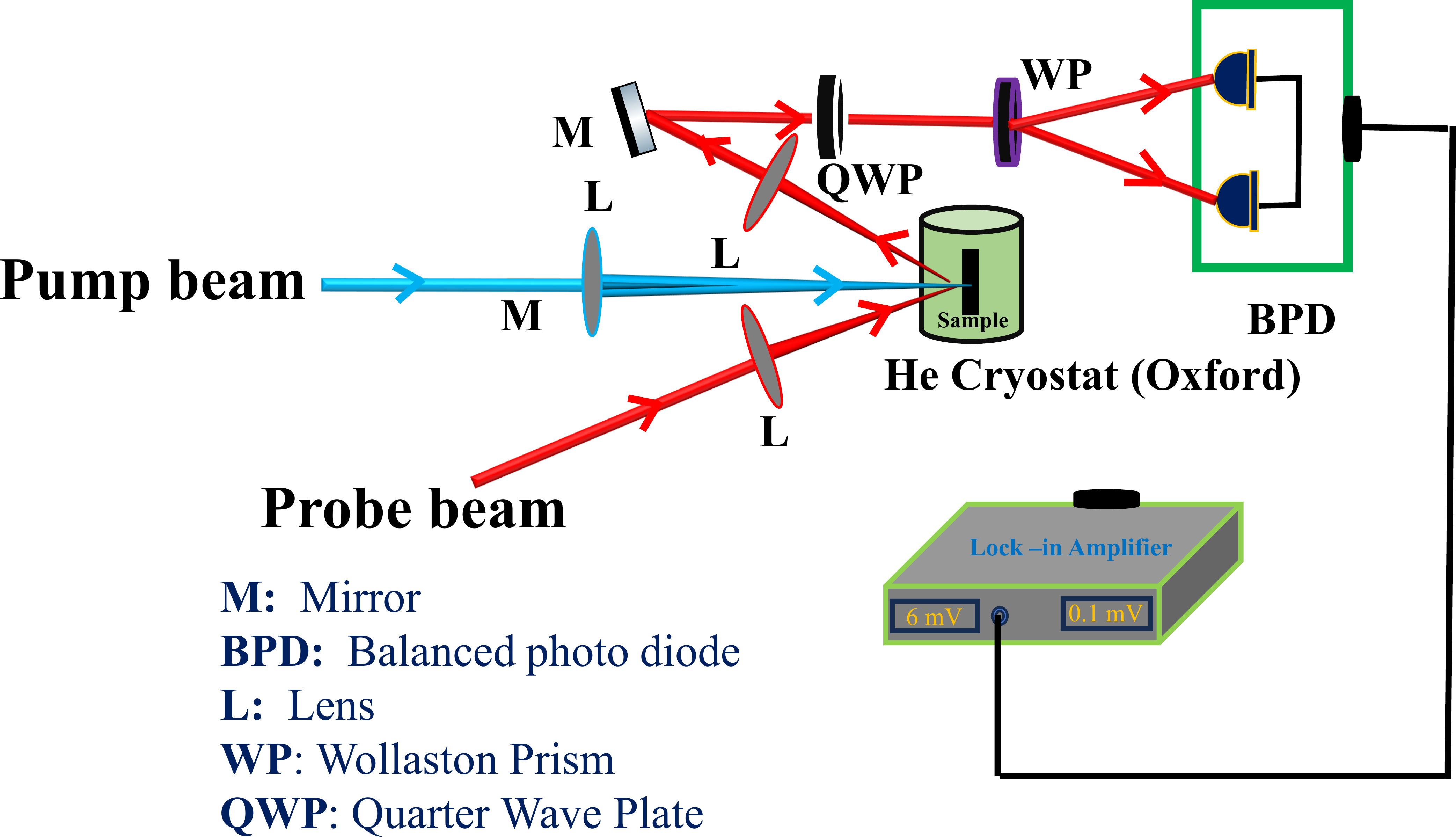}
   \caption*{\justifying FIG.~S5.1: Schematic diagram of the anisotropic pump--probe setup highlighting the detection region. All other components and configurations are identical to Fig.~S2.1}
  \label{fig:S5.1}
\end{figure}

The transient Kerr signal presented in Fig.~S5.2(a) exhibits relaxation dynamics that closely resemble those observed in the isotropic pump--probe measurements discussed in the main text. A quantitative analysis of the temporal response yields characteristic decay times, which are plotted as a function of temperature in Figs.~S5.2(b) and~S5.2(c). Remarkably, the extracted time constants follow the same temperature dependence in both measurement configurations. This consistency suggests a common underlying mechanism involving spin in the observed relaxation processes, rather than a purely electronic or lattice degrees of freedom in isotropic set up.

\begin{figure}[H]
  \centering
  \includegraphics[width= 0.7\linewidth]{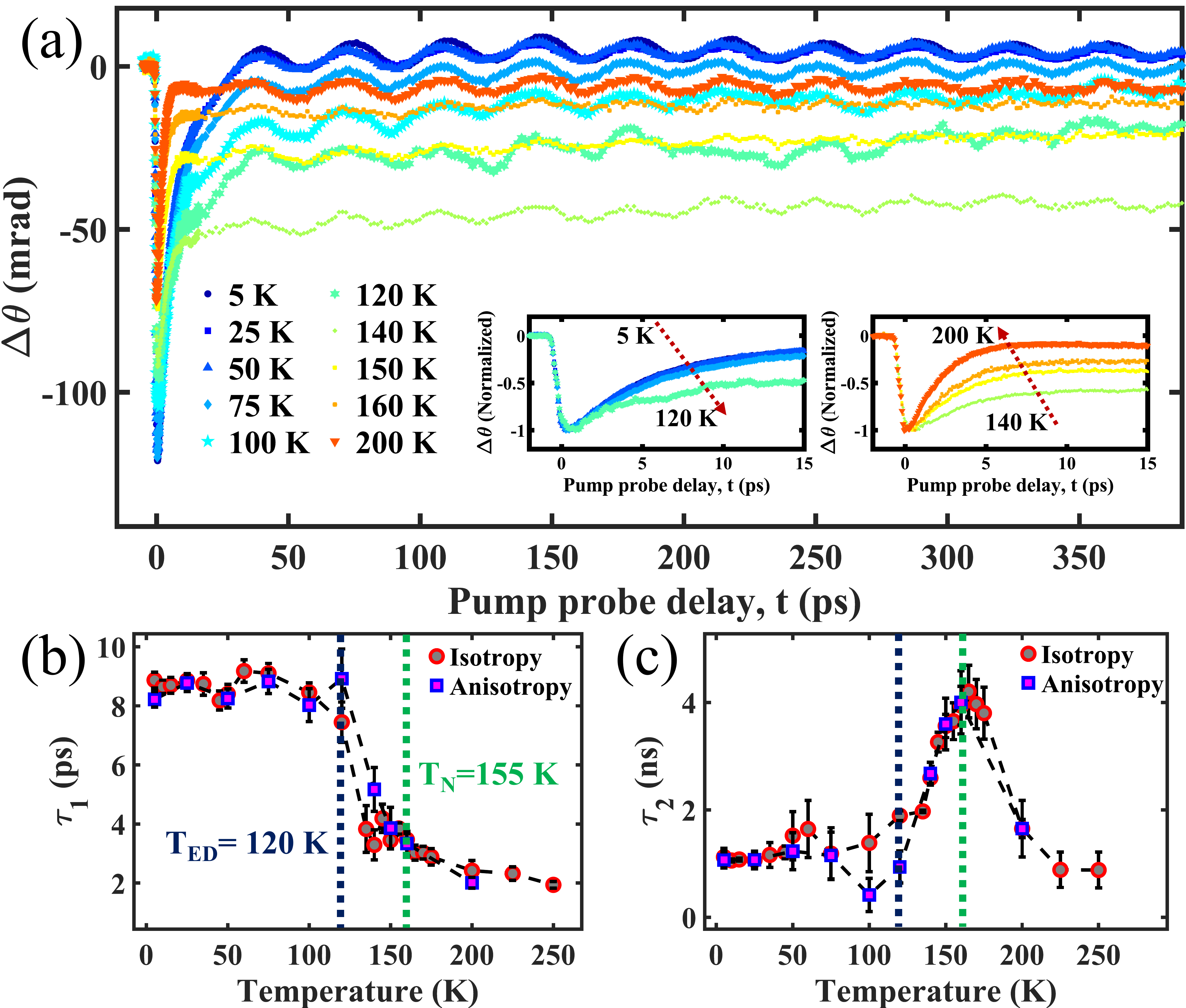}
    \caption*{\justifying FIG.~S5.2: (a) Time-resolved Kerr dynamics at different temperatures. The main panel shows the temperature-dependent Kerr rotation, with insets (Normalized with peak value) highlighting the decay dynamics for clarity at $5~\text{K} < T < 120~\text{K}$ and $140~\text{K} < T < 200~\text{K}$. (b) Comparison of the coherence time of SOEE. (c) Dynamics of AFM spin fluctuations or spin ordering. Blue and green dashed lines denote $T_{ED} \sim 120$\,K (exciton dissociation temperature) and $T_{N} \sim 155$\,K (Néel temperature); the black dashed line through the data is a guide to the eye.}
  \label{fig:S5.2}
\end{figure}

\begin{figure}[H]
  \centering
  \includegraphics[width= 0.7\linewidth]{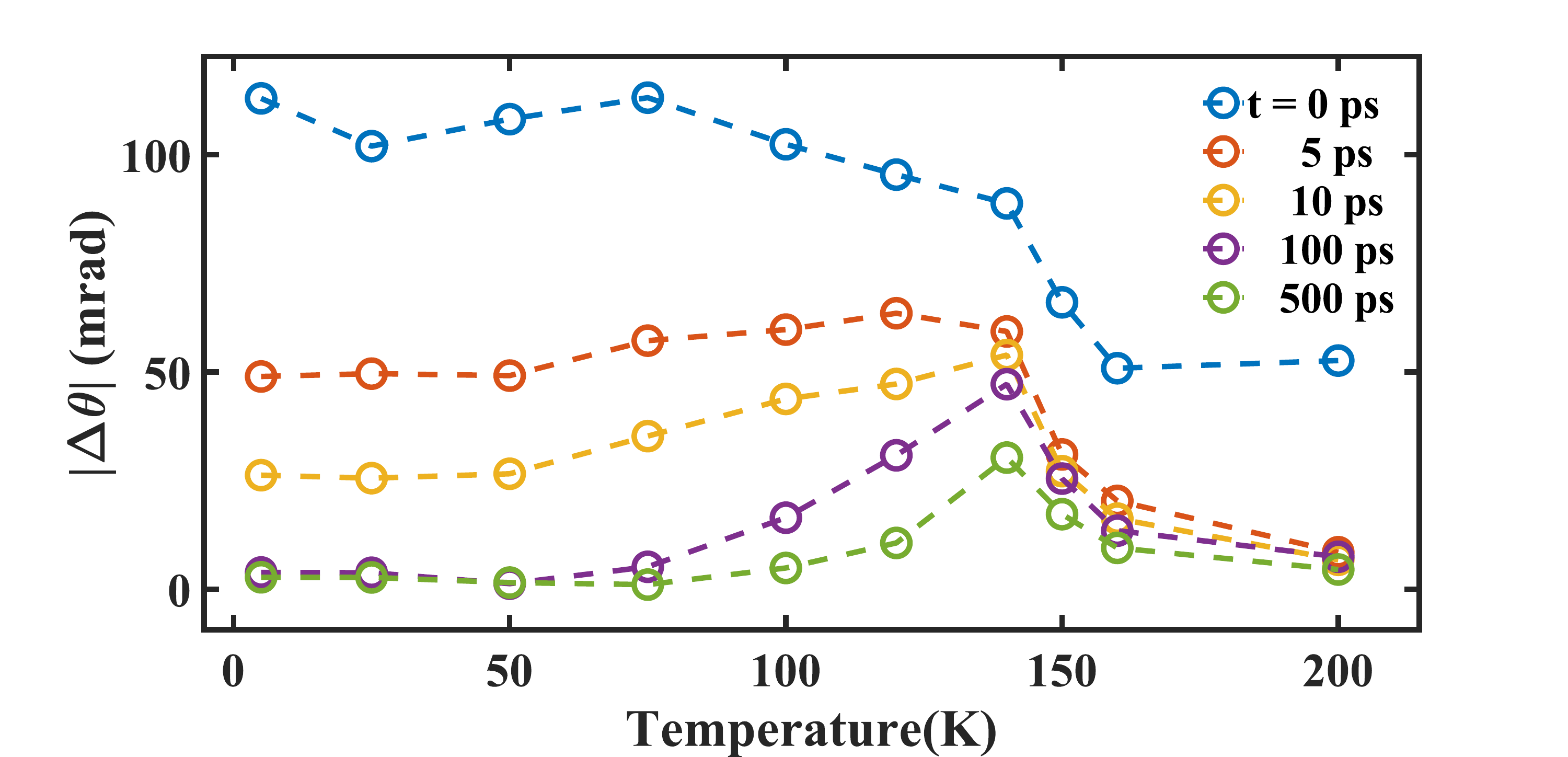}
    \caption*{\justifying FIG.~S5.3: Temperature dependence of the Kerr rotation at selected pump--probe delays ($t = 0$\,ps, 5\,ps, 10\,ps, 100\,ps, 500\,ps), resembling the isotropic $\Delta R/R$ shown in Fig.~2(c) of the main text.}
  \label{fig:S5.3}
\end{figure}

\newpage
The observed Kerr rotation in our experiments is primarily magnetic, getting weaker as the sample temperature nears N\'eel temperature ($\sim 200$~K) consistent with the antiferromagnetic--paramagnetic transition. In addition, coherent acoustic phonons modulate the Kerr signal, reflecting the coupling between lattice dynamics and magnetic degrees of freedom. The temperature dependence indicates that the magnetic contributions dominate at low temperatures, while phonon effects coexist, highlighting the interplay of magnetic and lattice dynamics in ultrafast Kerr measurements.

}}

\fontsize{15pt}{30pt}\selectfont \textbf{S6. Estimation of Photoexcited Carrier Density}
\label{sec:S6}

\vspace{0.5em}
{\fontsize{12pt}{16pt}\selectfont \hspace*{0.5em}{

The photoexcited carrier density, $N$, is estimated from the number of pump photons absorbed per unit volume of the material:

\[
N = \frac{\alpha (1 - R) F}{E_{\mathrm{pump}}}
\]
 
Here, $\alpha$ is the absorption coefficient, $R$ is the reflectivity at the pump wavelength, $F$ is the pump fluence, and $E_{\mathrm{pump}}$ is the pump photon energy.

\begin{table}[H]
\centering
\renewcommand{\arraystretch}{1.5} 
\setlength{\tabcolsep}{12pt}      
\large     
\begin{tabular}{|c|c|}
    \hline
    \textbf{Pump fluence ($\mathbf{\mu\mathrm{J}/\mathrm{cm}^2}$)}
 & \textbf{$N$ (cm$^{-3}$)} \\
    \hline
    120 & 4.1$\times$10$^{19}$ \\
    181 & 6.2$\times$10$^{19}$ \\
    230 & 7.8$\times$10$^{19}$ \\
    305 & 1.0$\times$10$^{20}$ \\
    364 & 1.2$\times$10$^{20}$ \\
    \hline
\end{tabular}

\vspace{0.5em}
\text{TS6.} Estimated photoexcited carrier densities for all pump fluences.
\end{table}
Since the binding energy of the SOEE in NiPS$_3$ is $\sim132$~meV, the estimated $N$ can be approximated as the number of excitons at $T < T_{\mathrm{ED}}$.

}}
\newpage
\vspace{1.5em}
\fontsize{15pt}{30pt}\selectfont \textbf{S7. Understanding Exciton Coherence–AFM Coupling}
\label{sec:S7}

\vspace{0.5em}
{\fontsize{12pt}{16pt}\selectfont \hspace*{0.5em}{
To understand the coupling mechanisms between exciton coherence and AFM ordering, we employ the Ginzburg--Landau formalism, expressing the free energy in terms of the two coupled order parameters as:
\[
F(\psi, \zeta, T) = a_{\psi}(T) |\psi|^{2} + a_{\zeta}(T) \zeta^{2} 
+ b_{\psi} |\psi|^{4} + b_{\zeta} \zeta^{4} + \lambda |\psi|^{2} \zeta^{2}
\]

Here $a_{\psi}(T) = a_{\psi0}(T - T_{\mathrm{ED}})$ and $a_{\zeta}(T) = a_{\zeta0}(T - T_{\mathrm{N}})$, with $a_{\psi0}$, $a_{\zeta0}$, $b_{\psi}$, and $b_{\zeta}$ as temperature-independent constants. $\psi = |\psi|e^{-i\Theta}$ and $\zeta$ denoting the complex and real order parameters for exciton coherence and AFM order, respectively. $T_{\mathrm{ED}}$ and $T_{\mathrm{N}}$ are the respective critical temperatures, and $\lambda$ quantifies the coupling strength between the two orders.

The equilibrium values of $\psi$ and $\zeta$ are obtained by minimizing the free energy $F$ with respect to both order parameters, which yields solutions of the form:
\[
\zeta = 0, \pm\sqrt{\frac{Z}{b_{\zeta}}}, \quad \text{where} \quad Z = -\left(a_{\zeta} + 2\lambda \psi^{2}\right),
\]
and
\[
\psi = 0, \pm \sqrt{\frac{-a_{\psi}}{b_{\psi}}}, \quad \pm \sqrt{\frac{-(a_{\psi} b_{\zeta} + 2\lambda a_{\psi})}{b_{\zeta} b_{\psi} + 4\lambda^{2}}}
\] 

To capture the relaxation dynamics and temporal evolution of the coupled order parameters, we use the time-dependent Ginzburg--Landau equation in the form:

\[
\frac{\partial \zeta}{\partial t} = -\Gamma_{\zeta} \frac{\partial F}{\partial \zeta}, \quad \frac{\partial |\psi|}{\partial t} = -\Gamma_{\psi} \frac{\partial F}{\partial |\psi|}
\]
Where $\Gamma_{\zeta}$ and $\Gamma_{\psi}$ are the damping rates.~\cite{spin-shear}  \\
Solving both equations at $T \to T_{N}$ and $T \to T_{ED}$, the solutions are:  
\[
\zeta(t) = \zeta_{0} e^{-a_{\zeta} \Gamma_{\zeta} t}, \quad 
|\psi(t)| = |\psi_{0}| e^{-a_{\psi} \Gamma_{\psi} t},
\]
where 
\[
\tau_{\zeta} \ \defined\ \frac{1}{a_{\zeta}\Gamma_{\zeta}}, \quad
\tau_{\psi} \ \defined\ \frac{1}{a_{\psi}\Gamma_{\psi}}
\]

The excitonic coherence can be described by a complex order parameter 
$\psi = |\psi| e^{-i\Theta}$, where $|\psi|$ represents the amplitude and $\Theta$ the phase. 
Within the Ginzburg--Landau framework, the free energy depends only on $|\psi|^2$; thus, the phase degree of freedom does not explicitly contribute to the relaxation time  $\tau_\psi$. 
The fast relaxation channel $\tau_1$ 
observed in this work can instead be associated with the decay of the phase coherence of the excitonic condensate. Physically, this means that while the amplitude 
$|\psi|$ governs the overall strength of the condensate, it is the phase dynamics that dominate the short-lived coherent response to femtosecond excitations observed by us. As the temperature decreases below the magnetic transition temperature 
$T_{N}$, the growth of the magnetic order parameter $\zeta$ acts to stabilize the excitonic condensate against phase fluctuations. This magnetic–excitonic coupling effectively prolongs the phase coherence timescale, reflected in the increase of $\tau_1$ in the temperature range $135<T<150$ K (see Fig. 3 (b) of the main manuscript). In other words, stronger magnetic correlations enforce a more rigid phase locking of $\psi$, thereby extending the timescale over which the excitonic system maintains coherence ($\approx 20-25\%$ increase in $\tau_1$). Finally, the excitonic order parameter $\psi$ is stabilised for $T<T_{ED}$. This naturally leads to a saturation of $\tau_1$: once long-range excitonic order is well established, phase relaxation channels are strongly suppressed.

}}

\fontsize{15pt}{30pt}\selectfont \textbf{S8. Fitting Details for $\tau_1$ and $\tau_2$}
\label{sec:S8}

\vspace{0.5em}
{\fontsize{12pt}{16pt}\selectfont \hspace*{0.5em}{The fitting of $\tau_{1}$ using the Rothwarf--Taylor model has been discussed in detail in the main manuscript; additional parameters obtained from this analysis are listed here;
\[
\begin{aligned}
K &= 0.114 \pm 0.003 \\
L &= 7.184 \pm 5.806 \\
\Delta_{E} &= 66 \pm 12 \,\text{meV}
\end{aligned}
\]

The divergence of $\tau_{2}$ across the transition was modeled using the power-law formula discussed in the main text. The temperature dependence of $\tau_{2}$ was fitted on both sides of $T_{N}$ [Fig. S8], and the extracted parameters are summarized in Table~S8. To further examine the scaling behavior, we analyzed the data in a log–log plot of $\tau_{2}$ versus the reduced temperature, $T_{\mathrm{red}} = |T-T_{N}|/T_{N}$ [inset of Fig.~S8]. In this form, $\tau_{2}$ exhibits a linear dependence over the range 5–294 K, evidencing critical behavior near $T_{N}$. The slope of this linear interval gives $m = 0.44 \pm 0.04$, in agreement with the universal critical exponent expected for the three-dimensional Heisenberg universality class~\cite{criticalexponent}.

\begin{figure}[H]
  \centering
  \includegraphics[width= 0.7\linewidth]{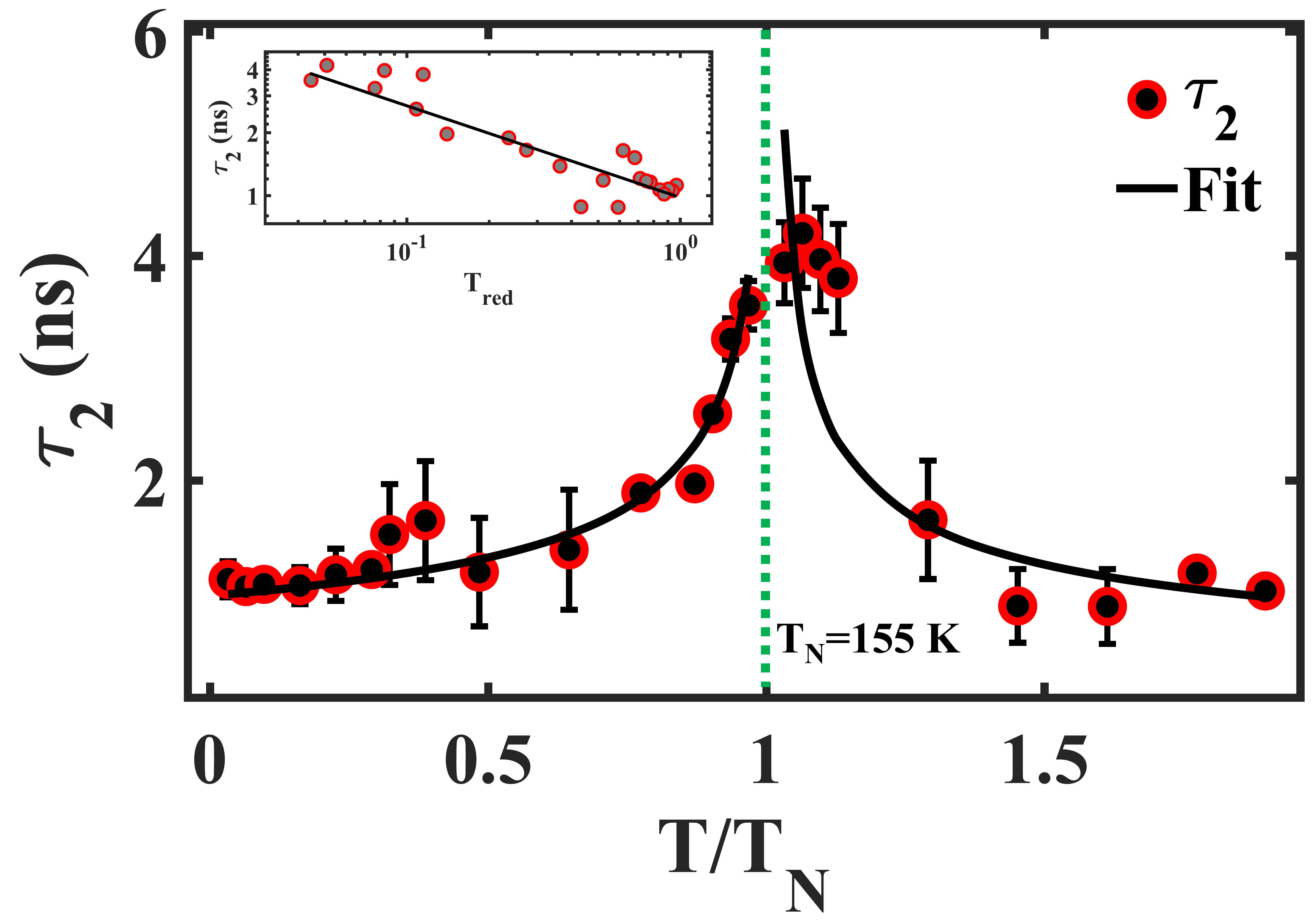}
    \caption*{\justifying FIG.~S8. Temperature dependence of $\tau_{2}$ (red circles) with a fit capturing its divergence near $T_{N}$ (green dashed line). Inset: the same data on a log--log scale, where the linear behavior confirms a power-law dependence.}
  \label{fig:S8}
  \end{figure}

  \begin{table}[H]
\centering
\renewcommand{\arraystretch}{1.5} 
\setlength{\tabcolsep}{16pt}      
\large     
\begin{tabular}{|c|c|}
    \hline
    \textbf{Below $\mathbf{T_N}$} & \textbf{Above $\mathbf{T_N}$} \\
    \hline
    $m = 0.44 \pm 0.08$ & $m = 0.43 \pm 0.28$ \\
    $\Delta = 1.03 \pm 0.08~\text{meV}$ & $\Delta = 1.10 \pm 0.50~\text{meV}$ \\
    $T_N = 157 \pm 3.65~\text{K}$ & $T_N = 157 \pm 5.30~\text{K}$ \\
    \hline
\end{tabular}

\vspace{0.5em}
\text{TS8.} Fit parameters extracted below and above $T_{N}$. 
\end{table}

The average values of the extracted parameters are  
$m = 0.44 \pm 0.18$, $\Delta = 1.07 \pm 0.29~\text{meV}$, and $T_N = 157 \pm 4.5~\text{K}$. The critical exponent (m) obtained here is consistent with the universality class of the three-dimensional Heisenberg model~\cite{criticalexponent,spin-shear}. The spin-wave gap agrees well with literature values~\cite{SWG1,SWG2,SWG3}, while the Néel temperature closely matches that obtained from our $M$--$T$ measurements (see inset of Fig.~S1.3).

}}
\newpage
\fontsize{15pt}{30pt}\selectfont \textbf{S9. Data fitting}
\label{sec:S9}

\vspace{0.5em}
{\fontsize{12pt}{16pt}\selectfont \hspace*{0.5em}{Data fitting was carried out using the Nelder--Mead simplex algorithm, implemented in \textsc{MATLAB} through the \texttt{fminsearch} function. The experimental $\Delta R/R$ data were fitted by using Eq.~(1) of the main manuscript. The fit was obtained by minimizing the residual $S = \Delta R/R - (\Delta R/R)_{\text{biexp-conv}}$. The fitted results are shown in Sec. S3 for two temperature and others not shown here for clarity. Error bars were estimated using the orthogonal distance regression (ODR) method~\cite{1,2}. In \textsc{MATLAB}, this was implemented by calculating the Jacobian matrix followed by the variance--covariance matrix to extract uncertainties in the fit parameters.

\vspace{0.5em}
{\fontsize{12pt}{16pt}\selectfont \hspace*{1em}\textbf{S9.1: Temperature-Dependent and Fluence-Dependent Fitting}}
\phantomsection
\label{sec:S9.1}

\vspace{0.5em}
{\fontsize{12pt}{16pt}\selectfont \hspace*{0.5em}{

To evaluate the influence of the slow relaxation component ($\tau_{2}$) on fit quality, we systematically varied $\tau_{2}$ while fitting time-resolved reflectivity data at selected temperatures near $T_N$: 135\,K, 155\,K, 170\,K, and 225\,K. Fits with $\tau_{2}$ values outside the optimal range exhibited noticeable deviations from the experimental traces, allowing identification of the best-fit $\tau_{2}$ that minimized residuals and reliably captured the slow exponential decay (see Sec.~S9.1.1 (a)–(d)). To further assess the sensitivity of $\tau_{2}$ to pump fluence, fluence-dependent measurements at 135~K were analyzed for two pump fluences using the same procedure (see Fig.~S9.1.2 (a)-(b)).

\begin{figure}[H]
  \centering
  \includegraphics[width= 1\linewidth]{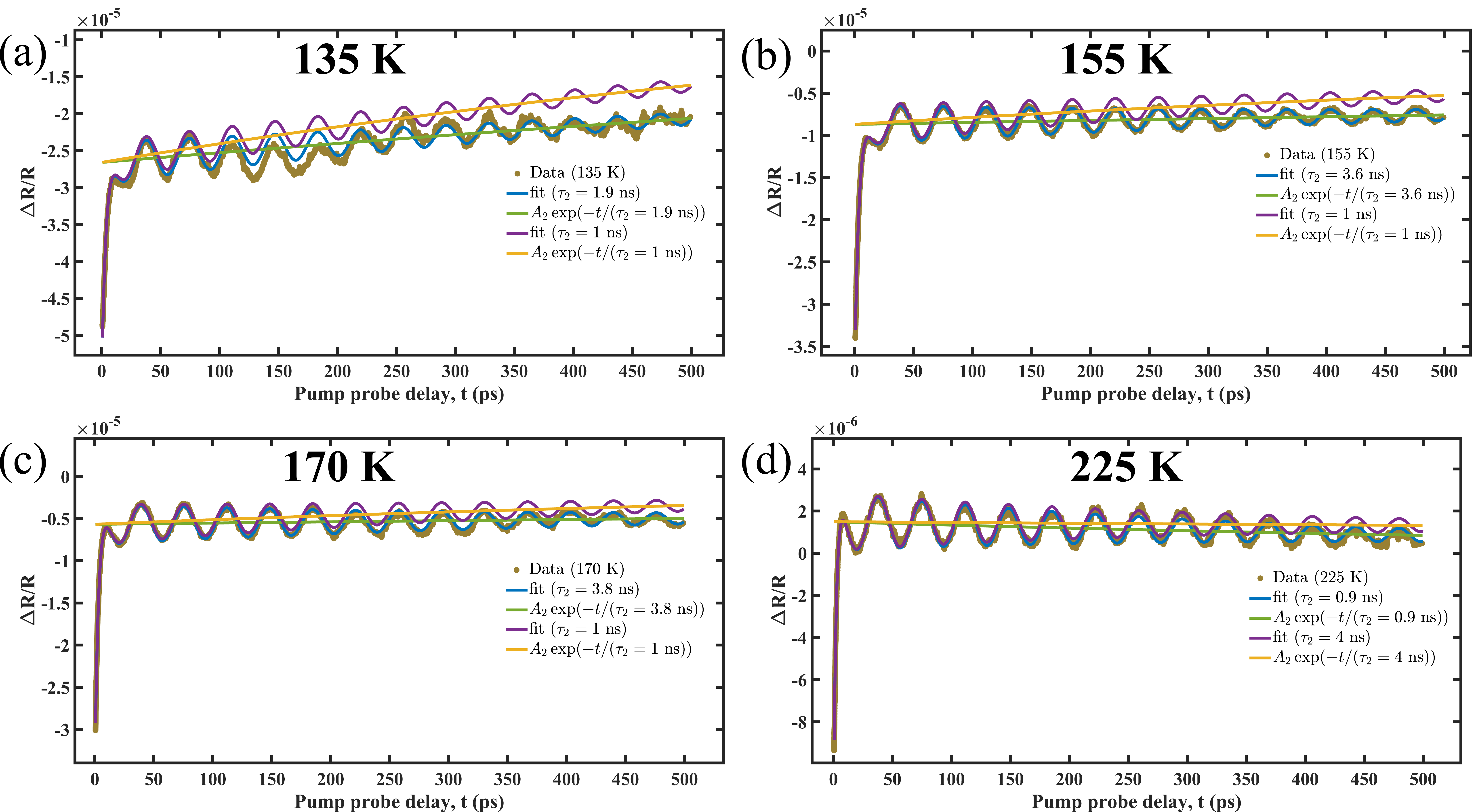}
    \caption*{\justifying FIG.~S9.1.1. Time-resolved reflectivity data of NiPS$_3$ at (a) 135\,K, (b) 155\,K, (c) 170\,K, and (d) 225\,K, together with fits using different $\tau_2$ values to test the sensitivity of the analysis. The slow exponential component is plotted separately for two representative $\tau_2$ values to evaluate the fit quality and highlight its influence on the measured response.}
  \label{fig:S9.1.1}
  \end{figure}

}}
\newpage

\vspace{0.5em}
{\fontsize{12pt}{16pt}\selectfont \hspace*{0.5em}{

}}

\begin{figure}[H]
  \centering
  \includegraphics[width= 1\linewidth]{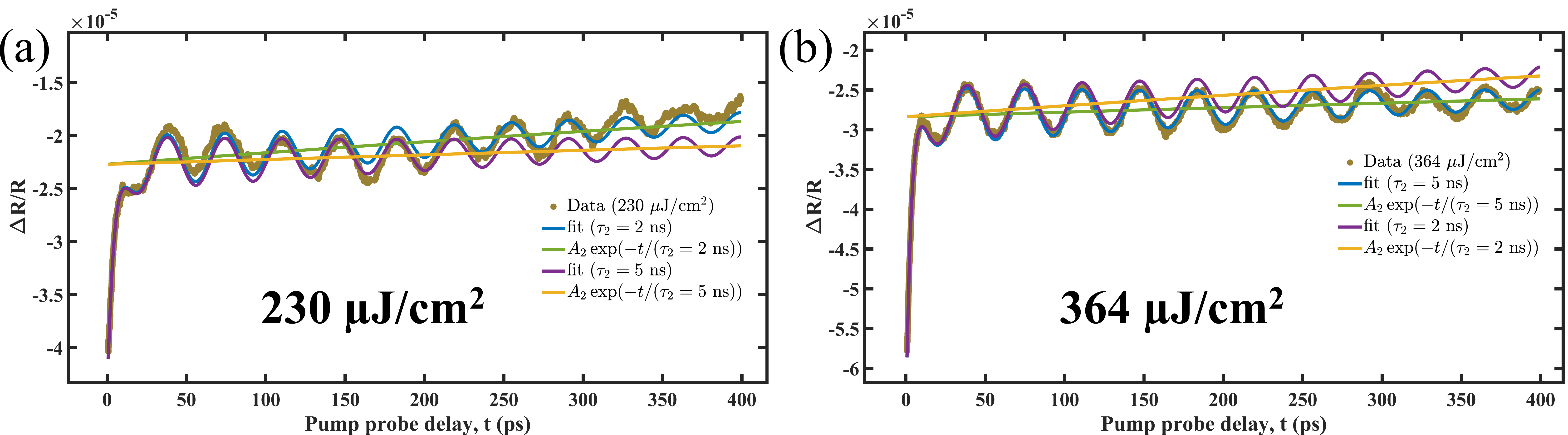}
  \caption*{\justifying FIG.~S9.1.2. Time-resolved reflectivity data of NiPS$_3$ at 135\,K for two fluences: (a) 260\,\(\mu\)J/cm$^2$ and (b) 364\,\(\mu\)J/cm$^2$. Fits using varied $\tau_{2}$ values are shown to assess the sensitivity of the slow exponential component. This component is plotted separately to illustrate its influences on the data and the overall fit quality.}
  \label{fig:S9.1.2}
  \end{figure}

}}

\vspace{0.5em}
{\fontsize{12pt}{16pt}\selectfont \hspace*{1em}\textbf{S9.2: Goodness of the fit}}

\phantomsection
\label{sec:S9.2}

\vspace{0.5em}
{\fontsize{12pt}{16pt}\selectfont \hspace*{0.5em}{ The goodness of fit was evaluated through graphical analysis and the coefficient of determination, $R^2$, defined as:
\begin{equation}
R^2 = 1- \frac{\sum \left( \frac{\Delta R}{R} - (\frac{\Delta R}{R})_{\text{biexp-conv}} \right)^2}{\sum \left( \frac{\Delta R}{R} - \left\langle \frac{\Delta R}{R} \right\rangle \right)^2}
\end{equation}

}}

\putbib[supp-refs]     
\end{bibunit}


\begin{thebibliography}{0}%
\makeatletter
\providecommand \@ifxundefined [1]{%
 \@ifx{#1\undefined}
}%
\providecommand \@ifnum [1]{%
 \ifnum #1\expandafter \@firstoftwo
 \else \expandafter \@secondoftwo
 \fi
}%
\providecommand \@ifx [1]{%
 \ifx #1\expandafter \@firstoftwo
 \else \expandafter \@secondoftwo
 \fi
}%
\providecommand \natexlab [1]{#1}%
\providecommand \enquote  [1]{``#1''}%
\providecommand \bibnamefont  [1]{#1}%
\providecommand \bibfnamefont [1]{#1}%
\providecommand \citenamefont [1]{#1}%
\providecommand \href@noop [0]{\@secondoftwo}%
\providecommand \href [0]{\begingroup \@sanitize@url \@href}%
\providecommand \@href[1]{\@@startlink{#1}\@@href}%
\providecommand \@@href[1]{\endgroup#1\@@endlink}%
\providecommand \@sanitize@url [0]{\catcode `\\12\catcode `\$12\catcode `\&12\catcode `\#12\catcode `\^12\catcode `\_12\catcode `\%12\relax}%
\providecommand \@@startlink[1]{}%
\providecommand \@@endlink[0]{}%
\providecommand \url  [0]{\begingroup\@sanitize@url \@url }%
\providecommand \@url [1]{\endgroup\@href {#1}{\urlprefix }}%
\providecommand \urlprefix  [0]{URL }%
\providecommand \Eprint [0]{\href }%
\providecommand \doibase [0]{https://doi.org/}%
\providecommand \selectlanguage [0]{\@gobble}%
\providecommand \bibinfo  [0]{\@secondoftwo}%
\providecommand \bibfield  [0]{\@secondoftwo}%
\providecommand \translation [1]{[#1]}%
\providecommand \BibitemOpen [0]{}%
\providecommand \bibitemStop [0]{}%
\providecommand \bibitemNoStop [0]{.\EOS\space}%
\providecommand \EOS [0]{\spacefactor3000\relax}%
\providecommand \BibitemShut  [1]{\csname bibitem#1\endcsname}%
\let\auto@bib@innerbib\@empty
\end{thebibliography}%


\begin{thebibliography}{45}%
\makeatletter
\providecommand \@ifxundefined [1]{%
 \@ifx{#1\undefined}
}%
\providecommand \@ifnum [1]{%
 \ifnum #1\expandafter \@firstoftwo
 \else \expandafter \@secondoftwo
 \fi
}%
\providecommand \@ifx [1]{%
 \ifx #1\expandafter \@firstoftwo
 \else \expandafter \@secondoftwo
 \fi
}%
\providecommand \natexlab [1]{#1}%
\providecommand \enquote  [1]{``#1''}%
\providecommand \bibnamefont  [1]{#1}%
\providecommand \bibfnamefont [1]{#1}%
\providecommand \citenamefont [1]{#1}%
\providecommand \href@noop [0]{\@secondoftwo}%
\providecommand \href [0]{\begingroup \@sanitize@url \@href}%
\providecommand \@href[1]{\@@startlink{#1}\@@href}%
\providecommand \@@href[1]{\endgroup#1\@@endlink}%
\providecommand \@sanitize@url [0]{\catcode `\\12\catcode `\$12\catcode `\&12\catcode `\#12\catcode `\^12\catcode `\_12\catcode `\%12\relax}%
\providecommand \@@startlink[1]{}%
\providecommand \@@endlink[0]{}%
\providecommand \url  [0]{\begingroup\@sanitize@url \@url }%
\providecommand \@url [1]{\endgroup\@href {#1}{\urlprefix }}%
\providecommand \urlprefix  [0]{URL }%
\providecommand \Eprint [0]{\href }%
\providecommand \doibase [0]{https://doi.org/}%
\providecommand \selectlanguage [0]{\@gobble}%
\providecommand \bibinfo  [0]{\@secondoftwo}%
\providecommand \bibfield  [0]{\@secondoftwo}%
\providecommand \translation [1]{[#1]}%
\providecommand \BibitemOpen [0]{}%
\providecommand \bibitemStop [0]{}%
\providecommand \bibitemNoStop [0]{.\EOS\space}%
\providecommand \EOS [0]{\spacefactor3000\relax}%
\providecommand \BibitemShut  [1]{\csname bibitem#1\endcsname}%
\let\auto@bib@innerbib\@empty
\bibitem [{\citenamefont {Kang}\ \emph {et~al.}(2020)\citenamefont {Kang}, \citenamefont {Kim}, \citenamefont {Kim}, \citenamefont {Kim}, \citenamefont {Sim}, \citenamefont {Lee}, \citenamefont {Lee}, \citenamefont {Park}, \citenamefont {Yun}, \citenamefont {Kim} \emph {et~al.}}]{11}%
  \BibitemOpen
  \bibfield  {author} {\bibinfo {author} {\bibfnamefont {S.}~\bibnamefont {Kang}}, \bibinfo {author} {\bibfnamefont {K.}~\bibnamefont {Kim}}, \bibinfo {author} {\bibfnamefont {B.~H.}\ \bibnamefont {Kim}}, \bibinfo {author} {\bibfnamefont {J.}~\bibnamefont {Kim}}, \bibinfo {author} {\bibfnamefont {K.~I.}\ \bibnamefont {Sim}}, \bibinfo {author} {\bibfnamefont {J.-U.}\ \bibnamefont {Lee}}, \bibinfo {author} {\bibfnamefont {S.}~\bibnamefont {Lee}}, \bibinfo {author} {\bibfnamefont {K.}~\bibnamefont {Park}}, \bibinfo {author} {\bibfnamefont {S.}~\bibnamefont {Yun}}, \bibinfo {author} {\bibfnamefont {T.}~\bibnamefont {Kim}}, \emph {et~al.},\ }\href {https://doi.org/10.1038/s41586-020-2520-5} {\bibfield  {journal} {\bibinfo  {journal} {Nature (London)}\ }\textbf {\bibinfo {volume} {583}},\ \bibinfo {pages} {785} (\bibinfo {year} {2020})}\BibitemShut {NoStop}%
\bibitem [{\citenamefont {Klaproth}\ \emph {et~al.}(2023)\citenamefont {Klaproth}, \citenamefont {Aswartham}, \citenamefont {Shemerliuk}, \citenamefont {Selter}, \citenamefont {Janson}, \citenamefont {van~den Brink}, \citenamefont {B{\"u}chner}, \citenamefont {Knupfer}, \citenamefont {Pazek}, \citenamefont {Mikhailova},\ and\ \citenamefont {Efimenko}}]{13}%
  \BibitemOpen
  \bibfield  {author} {\bibinfo {author} {\bibfnamefont {T.}~\bibnamefont {Klaproth}}, \bibinfo {author} {\bibfnamefont {S.}~\bibnamefont {Aswartham}}, \bibinfo {author} {\bibfnamefont {Y.}~\bibnamefont {Shemerliuk}}, \bibinfo {author} {\bibfnamefont {S.}~\bibnamefont {Selter}}, \bibinfo {author} {\bibfnamefont {O.}~\bibnamefont {Janson}}, \bibinfo {author} {\bibfnamefont {J.}~\bibnamefont {van~den Brink}}, \bibinfo {author} {\bibfnamefont {B.}~\bibnamefont {B{\"u}chner}}, \bibinfo {author} {\bibfnamefont {M.}~\bibnamefont {Knupfer}}, \bibinfo {author} {\bibfnamefont {S.}~\bibnamefont {Pazek}}, \bibinfo {author} {\bibfnamefont {D.}~\bibnamefont {Mikhailova}},\ and\ \bibinfo {author} {\bibfnamefont {A.}~\bibnamefont {Efimenko}},\ }\href {https://doi.org/10.1103/physrevlett.131.256504} {\bibfield  {journal} {\bibinfo  {journal} {Phys. Rev. Lett.}\ }\textbf {\bibinfo {volume} {131}},\ \bibinfo {pages} {256504} (\bibinfo {year} {2023})}\BibitemShut {NoStop}%
\bibitem [{\citenamefont {He}\ \emph {et~al.}(2024)\citenamefont {He}, \citenamefont {Shen}, \citenamefont {Wohlfeld}, \citenamefont {Sears}, \citenamefont {Li}, \citenamefont {Pelliciari}, \citenamefont {Walicki}, \citenamefont {Johnston}, \citenamefont {Baldini}, \citenamefont {Bisogni},\ and\ \citenamefont {Mitrano}}]{12}%
  \BibitemOpen
  \bibfield  {author} {\bibinfo {author} {\bibfnamefont {W.}~\bibnamefont {He}}, \bibinfo {author} {\bibfnamefont {Y.}~\bibnamefont {Shen}}, \bibinfo {author} {\bibfnamefont {K.}~\bibnamefont {Wohlfeld}}, \bibinfo {author} {\bibfnamefont {J.}~\bibnamefont {Sears}}, \bibinfo {author} {\bibfnamefont {J.}~\bibnamefont {Li}}, \bibinfo {author} {\bibfnamefont {J.}~\bibnamefont {Pelliciari}}, \bibinfo {author} {\bibfnamefont {M.}~\bibnamefont {Walicki}}, \bibinfo {author} {\bibfnamefont {S.}~\bibnamefont {Johnston}}, \bibinfo {author} {\bibfnamefont {E.}~\bibnamefont {Baldini}}, \bibinfo {author} {\bibfnamefont {V.}~\bibnamefont {Bisogni}},\ and\ \bibinfo {author} {\bibfnamefont {M.}~\bibnamefont {Mitrano}},\ }\href {https://doi.org/10.1038/s41467-024-47852-x} {\bibfield  {journal} {\bibinfo  {journal} {Nat. Commun.}\ }\textbf {\bibinfo {volume} {15}},\ \bibinfo {pages} {3496} (\bibinfo {year} {2024})}\BibitemShut {NoStop}%
\bibitem [{\citenamefont {Song}\ \emph {et~al.}(2024)\citenamefont {Song}, \citenamefont {Lv}, \citenamefont {Sun}, \citenamefont {Pang}, \citenamefont {Chang}, \citenamefont {Guan}, \citenamefont {Lai}, \citenamefont {Wang}, \citenamefont {Wu}, \citenamefont {Hu},\ and\ \citenamefont {Yuan}}]{15}%
  \BibitemOpen
  \bibfield  {author} {\bibinfo {author} {\bibfnamefont {F.}~\bibnamefont {Song}}, \bibinfo {author} {\bibfnamefont {Y.}~\bibnamefont {Lv}}, \bibinfo {author} {\bibfnamefont {Y.}~\bibnamefont {Sun}}, \bibinfo {author} {\bibfnamefont {S.}~\bibnamefont {Pang}}, \bibinfo {author} {\bibfnamefont {H.}~\bibnamefont {Chang}}, \bibinfo {author} {\bibfnamefont {S.}~\bibnamefont {Guan}}, \bibinfo {author} {\bibfnamefont {J.}~\bibnamefont {Lai}}, \bibinfo {author} {\bibfnamefont {X.}~\bibnamefont {Wang}}, \bibinfo {author} {\bibfnamefont {B.}~\bibnamefont {Wu}}, \bibinfo {author} {\bibfnamefont {C.}~\bibnamefont {Hu}},\ and\ \bibinfo {author} {\bibfnamefont {Z.}~\bibnamefont {Yuan}},\ }\href {https://doi.org/10.1038/s41467-024-52220-w} {\bibfield  {journal} {\bibinfo  {journal} {Nat. Commun.}\ }\textbf {\bibinfo {volume} {15}},\ \bibinfo {pages} {7841} (\bibinfo {year} {2024})}\BibitemShut {NoStop}%
\bibitem [{\citenamefont {Hwangbo}\ \emph {et~al.}(2021)\citenamefont {Hwangbo}, \citenamefont {Zhang}, \citenamefont {Jiang}, \citenamefont {Wang}, \citenamefont {Fonseca}, \citenamefont {Wang}, \citenamefont {Diederich}, \citenamefont {Gamelin}, \citenamefont {Xiao}, \citenamefont {Chu}, \citenamefont {Yao},\ and\ \citenamefont {Xu}}]{21}%
  \BibitemOpen
  \bibfield  {author} {\bibinfo {author} {\bibfnamefont {K.}~\bibnamefont {Hwangbo}}, \bibinfo {author} {\bibfnamefont {Q.}~\bibnamefont {Zhang}}, \bibinfo {author} {\bibfnamefont {Q.}~\bibnamefont {Jiang}}, \bibinfo {author} {\bibfnamefont {Y.}~\bibnamefont {Wang}}, \bibinfo {author} {\bibfnamefont {J.}~\bibnamefont {Fonseca}}, \bibinfo {author} {\bibfnamefont {C.}~\bibnamefont {Wang}}, \bibinfo {author} {\bibfnamefont {G.~M.}\ \bibnamefont {Diederich}}, \bibinfo {author} {\bibfnamefont {D.~R.}\ \bibnamefont {Gamelin}}, \bibinfo {author} {\bibfnamefont {D.}~\bibnamefont {Xiao}}, \bibinfo {author} {\bibfnamefont {J.-H.}\ \bibnamefont {Chu}}, \bibinfo {author} {\bibfnamefont {W.}~\bibnamefont {Yao}},\ and\ \bibinfo {author} {\bibfnamefont {X.}~\bibnamefont {Xu}},\ }\href {https://doi.org/10.1038/s41565-021-00873-9} {\bibfield  {journal} {\bibinfo  {journal} {Nat. Nanotechnol.}\ }\textbf {\bibinfo {volume} {16}},\ \bibinfo {pages} {655} (\bibinfo {year} {2021})}\BibitemShut {NoStop}%
\bibitem [{\citenamefont {Wang}\ \emph {et~al.}(2021)\citenamefont {Wang}, \citenamefont {Cao}, \citenamefont {Lu}, \citenamefont {Cohen}, \citenamefont {Kitadai}, \citenamefont {Li}, \citenamefont {Tan}, \citenamefont {Wilson}, \citenamefont {Lui}, \citenamefont {Smirnov} \emph {et~al.}}]{31}%
  \BibitemOpen
  \bibfield  {author} {\bibinfo {author} {\bibfnamefont {X.}~\bibnamefont {Wang}}, \bibinfo {author} {\bibfnamefont {J.}~\bibnamefont {Cao}}, \bibinfo {author} {\bibfnamefont {Z.}~\bibnamefont {Lu}}, \bibinfo {author} {\bibfnamefont {A.}~\bibnamefont {Cohen}}, \bibinfo {author} {\bibfnamefont {H.}~\bibnamefont {Kitadai}}, \bibinfo {author} {\bibfnamefont {T.}~\bibnamefont {Li}}, \bibinfo {author} {\bibfnamefont {Q.}~\bibnamefont {Tan}}, \bibinfo {author} {\bibfnamefont {M.}~\bibnamefont {Wilson}}, \bibinfo {author} {\bibfnamefont {C.~H.}\ \bibnamefont {Lui}}, \bibinfo {author} {\bibfnamefont {D.}~\bibnamefont {Smirnov}}, \emph {et~al.},\ }\href {https://doi.org/10.1038/s41563-021-00968-7} {\bibfield  {journal} {\bibinfo  {journal} {Nat. Mater.}\ }\textbf {\bibinfo {volume} {20}},\ \bibinfo {pages} {964} (\bibinfo {year} {2021})}\BibitemShut {NoStop}%
\bibitem [{\citenamefont {Dirnberger}\ \emph {et~al.}(2022)\citenamefont {Dirnberger}, \citenamefont {Bushati}, \citenamefont {Datta}, \citenamefont {Kumar}, \citenamefont {MacDonald}, \citenamefont {Baldini},\ and\ \citenamefont {Menon}}]{SOEE1}%
  \BibitemOpen
  \bibfield  {author} {\bibinfo {author} {\bibfnamefont {F.}~\bibnamefont {Dirnberger}}, \bibinfo {author} {\bibfnamefont {R.}~\bibnamefont {Bushati}}, \bibinfo {author} {\bibfnamefont {B.}~\bibnamefont {Datta}}, \bibinfo {author} {\bibfnamefont {A.}~\bibnamefont {Kumar}}, \bibinfo {author} {\bibfnamefont {A.~H.}\ \bibnamefont {MacDonald}}, \bibinfo {author} {\bibfnamefont {E.}~\bibnamefont {Baldini}},\ and\ \bibinfo {author} {\bibfnamefont {V.~M.}\ \bibnamefont {Menon}},\ }\href {https://doi.org/10.1038/s41565-022-01204-2} {\bibfield  {journal} {\bibinfo  {journal} {Nat. Nanotechnol.}\ }\textbf {\bibinfo {volume} {17}},\ \bibinfo {pages} {1060} (\bibinfo {year} {2022})}\BibitemShut {NoStop}%
\bibitem [{\citenamefont {Kim}\ \emph {et~al.}(2023)\citenamefont {Kim}, \citenamefont {Na}, \citenamefont {Kim}, \citenamefont {Park}, \citenamefont {Zhang}, \citenamefont {Hwang}, \citenamefont {Son}, \citenamefont {Kim}, \citenamefont {Cheong},\ and\ \citenamefont {Park}}]{SOEE2}%
  \BibitemOpen
  \bibfield  {author} {\bibinfo {author} {\bibfnamefont {J.}~\bibnamefont {Kim}}, \bibinfo {author} {\bibfnamefont {W.}~\bibnamefont {Na}}, \bibinfo {author} {\bibfnamefont {J.}~\bibnamefont {Kim}}, \bibinfo {author} {\bibfnamefont {P.}~\bibnamefont {Park}}, \bibinfo {author} {\bibfnamefont {K.}~\bibnamefont {Zhang}}, \bibinfo {author} {\bibfnamefont {I.}~\bibnamefont {Hwang}}, \bibinfo {author} {\bibfnamefont {Y.-W.}\ \bibnamefont {Son}}, \bibinfo {author} {\bibfnamefont {J.~H.}\ \bibnamefont {Kim}}, \bibinfo {author} {\bibfnamefont {H.}~\bibnamefont {Cheong}},\ and\ \bibinfo {author} {\bibfnamefont {J.-G.}\ \bibnamefont {Park}},\ }\href {https://doi.org/10.1021/acs.nanolett.3c02677} {\bibfield  {journal} {\bibinfo  {journal} {Nano Lett.}\ }\textbf {\bibinfo {volume} {23}},\ \bibinfo {pages} {10189} (\bibinfo {year} {2023})}\BibitemShut {NoStop}%
\bibitem [{\citenamefont {Lee}\ \emph {et~al.}(2024)\citenamefont {Lee}, \citenamefont {Lee}, \citenamefont {Choi}, \citenamefont {Gries}, \citenamefont {Klingeler}, \citenamefont {Raju}, \citenamefont {Ulaganathan}, \citenamefont {Sankar}, \citenamefont {Seong},\ and\ \citenamefont {Choi}}]{SOEE3}%
  \BibitemOpen
  \bibfield  {author} {\bibinfo {author} {\bibfnamefont {J.-H.}\ \bibnamefont {Lee}}, \bibinfo {author} {\bibfnamefont {S.}~\bibnamefont {Lee}}, \bibinfo {author} {\bibfnamefont {Y.}~\bibnamefont {Choi}}, \bibinfo {author} {\bibfnamefont {L.}~\bibnamefont {Gries}}, \bibinfo {author} {\bibfnamefont {R.}~\bibnamefont {Klingeler}}, \bibinfo {author} {\bibfnamefont {K.}~\bibnamefont {Raju}}, \bibinfo {author} {\bibfnamefont {R.~K.}\ \bibnamefont {Ulaganathan}}, \bibinfo {author} {\bibfnamefont {R.}~\bibnamefont {Sankar}}, \bibinfo {author} {\bibfnamefont {M.-J.}\ \bibnamefont {Seong}},\ and\ \bibinfo {author} {\bibfnamefont {K.-Y.}\ \bibnamefont {Choi}},\ }\href {https://doi.org/10.1002/adfm.202405153} {\bibfield  {journal} {\bibinfo  {journal} {Adv. Funct. Mater.}\ }\textbf {\bibinfo {volume} {34}},\ \bibinfo {pages} {2405153} (\bibinfo {year} {2024})}\BibitemShut {NoStop}%
\bibitem [{\citenamefont {Chandrasekaran}\ \emph {et~al.}(2025)\citenamefont {Chandrasekaran}, \citenamefont {DeLaney}, \citenamefont {Trinh}, \citenamefont {Parobek}, \citenamefont {Lane}, \citenamefont {Zhu}, \citenamefont {Li}, \citenamefont {Zhao}, \citenamefont {Campbell}, \citenamefont {Martin} \emph {et~al.}}]{SOEE4}%
  \BibitemOpen
  \bibfield  {author} {\bibinfo {author} {\bibfnamefont {V.}~\bibnamefont {Chandrasekaran}}, \bibinfo {author} {\bibfnamefont {C.~R.}\ \bibnamefont {DeLaney}}, \bibinfo {author} {\bibfnamefont {C.~T.}\ \bibnamefont {Trinh}}, \bibinfo {author} {\bibfnamefont {D.}~\bibnamefont {Parobek}}, \bibinfo {author} {\bibfnamefont {C.~A.}\ \bibnamefont {Lane}}, \bibinfo {author} {\bibfnamefont {J.}~\bibnamefont {Zhu}}, \bibinfo {author} {\bibfnamefont {X.}~\bibnamefont {Li}}, \bibinfo {author} {\bibfnamefont {H.}~\bibnamefont {Zhao}}, \bibinfo {author} {\bibfnamefont {M.~A.}\ \bibnamefont {Campbell}}, \bibinfo {author} {\bibfnamefont {L.}~\bibnamefont {Martin}}, \emph {et~al.},\ }\href {https://doi.org/10.1039/D4NH00390J} {\bibfield  {journal} {\bibinfo  {journal} {Nanoscale Horiz.}\ }\textbf {\bibinfo {volume} {10}},\ \bibinfo {pages} {150} (\bibinfo {year} {2025})}\BibitemShut {NoStop}%
\bibitem [{\citenamefont {Wang}\ \emph {et~al.}(2025)\citenamefont {Wang}, \citenamefont {Peng}, \citenamefont {Huang}, \citenamefont {Zhou}, \citenamefont {Sun}, \citenamefont {Tang}, \citenamefont {Li}, \citenamefont {Xu}, \citenamefont {Du}, \citenamefont {Wang},\ and\ \citenamefont {Yang}}]{SOEE5}%
  \BibitemOpen
  \bibfield  {author} {\bibinfo {author} {\bibfnamefont {K.}~\bibnamefont {Wang}}, \bibinfo {author} {\bibfnamefont {Y.}~\bibnamefont {Peng}}, \bibinfo {author} {\bibfnamefont {B.}~\bibnamefont {Huang}}, \bibinfo {author} {\bibfnamefont {C.}~\bibnamefont {Zhou}}, \bibinfo {author} {\bibfnamefont {Q.}~\bibnamefont {Sun}}, \bibinfo {author} {\bibfnamefont {F.}~\bibnamefont {Tang}}, \bibinfo {author} {\bibfnamefont {Z.}~\bibnamefont {Li}}, \bibinfo {author} {\bibfnamefont {W.}~\bibnamefont {Xu}}, \bibinfo {author} {\bibfnamefont {K.}~\bibnamefont {Du}}, \bibinfo {author} {\bibfnamefont {X.}~\bibnamefont {Wang}},\ and\ \bibinfo {author} {\bibfnamefont {Y.}~\bibnamefont {Yang}},\ }\href {https://arxiv.org/abs/2508.06246} {\bibfield  {journal} {\bibinfo  {journal} {arXiv}\ } (\bibinfo {year} {2025})},\ \Eprint {https://arxiv.org/abs/arXiv:2508.06246} {arXiv:2508.06246 [cond-mat.str-el]} \BibitemShut {NoStop}%
\bibitem [{\citenamefont {Kim}\ \emph {et~al.}(2018)\citenamefont {Kim}, \citenamefont {Kim}, \citenamefont {Sandilands}, \citenamefont {Sinn}, \citenamefont {Lee}, \citenamefont {Son}, \citenamefont {Lee}, \citenamefont {Choi}, \citenamefont {Kim}, \citenamefont {Park},\ and\ \citenamefont {Jeon}}]{5}%
  \BibitemOpen
  \bibfield  {author} {\bibinfo {author} {\bibfnamefont {S.}~\bibnamefont {Kim}}, \bibinfo {author} {\bibfnamefont {T.}~\bibnamefont {Kim}}, \bibinfo {author} {\bibfnamefont {L.}~\bibnamefont {Sandilands}}, \bibinfo {author} {\bibfnamefont {S.}~\bibnamefont {Sinn}}, \bibinfo {author} {\bibfnamefont {M.}~\bibnamefont {Lee}}, \bibinfo {author} {\bibfnamefont {J.}~\bibnamefont {Son}}, \bibinfo {author} {\bibfnamefont {S.}~\bibnamefont {Lee}}, \bibinfo {author} {\bibfnamefont {K.}~\bibnamefont {Choi}}, \bibinfo {author} {\bibfnamefont {W.}~\bibnamefont {Kim}}, \bibinfo {author} {\bibfnamefont {B.}~\bibnamefont {Park}},\ and\ \bibinfo {author} {\bibfnamefont {C.}~\bibnamefont {Jeon}},\ }\href {https://doi.org/10.1103/physrevlett.120.136402} {\bibfield  {journal} {\bibinfo  {journal} {Phys. Rev. Lett.}\ }\textbf {\bibinfo {volume} {120}},\ \bibinfo {pages} {136402} (\bibinfo {year} {2018})}\BibitemShut {NoStop}%
\bibitem [{\citenamefont {Lane}\ and\ \citenamefont {Zhu}(2020)}]{33}%
  \BibitemOpen
  \bibfield  {author} {\bibinfo {author} {\bibfnamefont {C.}~\bibnamefont {Lane}}\ and\ \bibinfo {author} {\bibfnamefont {J.}~\bibnamefont {Zhu}},\ }\href {https://doi.org/10.1103/physrevb.102.075124} {\bibfield  {journal} {\bibinfo  {journal} {Phys. Rev. B}\ }\textbf {\bibinfo {volume} {102}},\ \bibinfo {pages} {075124} (\bibinfo {year} {2020})}\BibitemShut {NoStop}%
\bibitem [{\citenamefont {Belvin}\ \emph {et~al.}(2021)\citenamefont {Belvin}, \citenamefont {Baldini}, \citenamefont {Ozel}, \citenamefont {Mao}, \citenamefont {Po}, \citenamefont {Allington}, \citenamefont {Son}, \citenamefont {Kim}, \citenamefont {Kim}, \citenamefont {Hwang} \emph {et~al.}}]{7}%
  \BibitemOpen
  \bibfield  {author} {\bibinfo {author} {\bibfnamefont {C.~A.}\ \bibnamefont {Belvin}}, \bibinfo {author} {\bibfnamefont {E.}~\bibnamefont {Baldini}}, \bibinfo {author} {\bibfnamefont {I.~O.}\ \bibnamefont {Ozel}}, \bibinfo {author} {\bibfnamefont {D.}~\bibnamefont {Mao}}, \bibinfo {author} {\bibfnamefont {H.~C.}\ \bibnamefont {Po}}, \bibinfo {author} {\bibfnamefont {C.~J.}\ \bibnamefont {Allington}}, \bibinfo {author} {\bibfnamefont {S.}~\bibnamefont {Son}}, \bibinfo {author} {\bibfnamefont {B.~H.}\ \bibnamefont {Kim}}, \bibinfo {author} {\bibfnamefont {J.}~\bibnamefont {Kim}}, \bibinfo {author} {\bibfnamefont {I.}~\bibnamefont {Hwang}}, \emph {et~al.},\ }\href {https://doi.org/10.1038/s41467-021-25164-8} {\bibfield  {journal} {\bibinfo  {journal} {Nat. Commun.}\ }\textbf {\bibinfo {volume} {12}},\ \bibinfo {pages} {4837} (\bibinfo {year} {2021})}\BibitemShut {NoStop}%
\bibitem [{\citenamefont {Afanasiev}\ \emph {et~al.}(2021)\citenamefont {Afanasiev}, \citenamefont {Hortensius}, \citenamefont {Matthiesen}, \citenamefont {Ma{\~n}as-Valero}, \citenamefont {{\v{S}}i{\v{s}}kins}, \citenamefont {Lee}, \citenamefont {Lesne}, \citenamefont {van~der Zant}, \citenamefont {Steeneken}, \citenamefont {Ivanov} \emph {et~al.}}]{6}%
  \BibitemOpen
  \bibfield  {author} {\bibinfo {author} {\bibfnamefont {D.}~\bibnamefont {Afanasiev}}, \bibinfo {author} {\bibfnamefont {J.}~\bibnamefont {Hortensius}}, \bibinfo {author} {\bibfnamefont {M.}~\bibnamefont {Matthiesen}}, \bibinfo {author} {\bibfnamefont {S.}~\bibnamefont {Ma{\~n}as-Valero}}, \bibinfo {author} {\bibfnamefont {M.}~\bibnamefont {{\v{S}}i{\v{s}}kins}}, \bibinfo {author} {\bibfnamefont {M.}~\bibnamefont {Lee}}, \bibinfo {author} {\bibfnamefont {E.}~\bibnamefont {Lesne}}, \bibinfo {author} {\bibfnamefont {H.}~\bibnamefont {van~der Zant}}, \bibinfo {author} {\bibfnamefont {P.}~\bibnamefont {Steeneken}}, \bibinfo {author} {\bibfnamefont {B.}~\bibnamefont {Ivanov}}, \emph {et~al.},\ }\href {https://doi.org/10.1126/sciadv.abf3096} {\bibfield  {journal} {\bibinfo  {journal} {Sci. Adv.}\ }\textbf {\bibinfo {volume} {7}},\ \bibinfo {pages} {eabf3096} (\bibinfo {year} {2021})}\BibitemShut {NoStop}%
\bibitem [{\citenamefont {Erge{\c{c}}en}\ \emph {et~al.}(2022)\citenamefont {Erge{\c{c}}en}, \citenamefont {Ilyas}, \citenamefont {Mao}, \citenamefont {Po}, \citenamefont {Yilmaz}, \citenamefont {Kim}, \citenamefont {Park}, \citenamefont {Senthil},\ and\ \citenamefont {Gedik}}]{8}%
  \BibitemOpen
  \bibfield  {author} {\bibinfo {author} {\bibfnamefont {E.}~\bibnamefont {Erge{\c{c}}en}}, \bibinfo {author} {\bibfnamefont {B.}~\bibnamefont {Ilyas}}, \bibinfo {author} {\bibfnamefont {D.}~\bibnamefont {Mao}}, \bibinfo {author} {\bibfnamefont {H.}~\bibnamefont {Po}}, \bibinfo {author} {\bibfnamefont {M.}~\bibnamefont {Yilmaz}}, \bibinfo {author} {\bibfnamefont {J.}~\bibnamefont {Kim}}, \bibinfo {author} {\bibfnamefont {J.}~\bibnamefont {Park}}, \bibinfo {author} {\bibfnamefont {T.}~\bibnamefont {Senthil}},\ and\ \bibinfo {author} {\bibfnamefont {N.}~\bibnamefont {Gedik}},\ }\href {https://doi.org/10.1038/s41467-021-27741-3} {\bibfield  {journal} {\bibinfo  {journal} {Nat. Commun.}\ }\textbf {\bibinfo {volume} {13}},\ \bibinfo {pages} {98} (\bibinfo {year} {2022})}\BibitemShut {NoStop}%
\bibitem [{\citenamefont {Warshauer}\ \emph {et~al.}(2025{\natexlab{a}})\citenamefont {Warshauer}, \citenamefont {Chen}, \citenamefont {Bustamante~Lopez}, \citenamefont {Tan}, \citenamefont {Tang}, \citenamefont {Ling},\ and\ \citenamefont {Hu}}]{9}%
  \BibitemOpen
  \bibfield  {author} {\bibinfo {author} {\bibfnamefont {J.}~\bibnamefont {Warshauer}}, \bibinfo {author} {\bibfnamefont {H.}~\bibnamefont {Chen}}, \bibinfo {author} {\bibfnamefont {D.}~\bibnamefont {Bustamante~Lopez}}, \bibinfo {author} {\bibfnamefont {Q.}~\bibnamefont {Tan}}, \bibinfo {author} {\bibfnamefont {J.}~\bibnamefont {Tang}}, \bibinfo {author} {\bibfnamefont {X.}~\bibnamefont {Ling}},\ and\ \bibinfo {author} {\bibfnamefont {W.}~\bibnamefont {Hu}},\ }\href {https://doi.org/10.1103/physrevlett.134.016901} {\bibfield  {journal} {\bibinfo  {journal} {Phys. Rev. Lett.}\ }\textbf {\bibinfo {volume} {134}},\ \bibinfo {pages} {016901} (\bibinfo {year} {2025}{\natexlab{a}})}\BibitemShut {NoStop}%
\bibitem [{\citenamefont {Li}\ \emph {et~al.}(2024)\citenamefont {Li}, \citenamefont {Liang}, \citenamefont {Kong}, \citenamefont {Sun},\ and\ \citenamefont {Zhang}}]{17}%
  \BibitemOpen
  \bibfield  {author} {\bibinfo {author} {\bibfnamefont {Y.}~\bibnamefont {Li}}, \bibinfo {author} {\bibfnamefont {G.}~\bibnamefont {Liang}}, \bibinfo {author} {\bibfnamefont {C.}~\bibnamefont {Kong}}, \bibinfo {author} {\bibfnamefont {B.}~\bibnamefont {Sun}},\ and\ \bibinfo {author} {\bibfnamefont {X.}~\bibnamefont {Zhang}},\ }\href {https://doi.org/10.1002/adfm.202402161} {\bibfield  {journal} {\bibinfo  {journal} {Adv. Funct. Mater.}\ }\textbf {\bibinfo {volume} {34}},\ \bibinfo {pages} {2402161} (\bibinfo {year} {2024})}\BibitemShut {NoStop}%
\bibitem [{\citenamefont {Allington}\ \emph {et~al.}(2025)\citenamefont {Allington}, \citenamefont {Belvin}, \citenamefont {Seifert}, \citenamefont {Ye}, \citenamefont {Tai}, \citenamefont {Baldini}, \citenamefont {Son}, \citenamefont {Kim}, \citenamefont {Park}, \citenamefont {Park} \emph {et~al.}}]{allington2025distinct}%
  \BibitemOpen
  \bibfield  {author} {\bibinfo {author} {\bibfnamefont {C.~J.}\ \bibnamefont {Allington}}, \bibinfo {author} {\bibfnamefont {C.~A.}\ \bibnamefont {Belvin}}, \bibinfo {author} {\bibfnamefont {U.~F.}\ \bibnamefont {Seifert}}, \bibinfo {author} {\bibfnamefont {M.}~\bibnamefont {Ye}}, \bibinfo {author} {\bibfnamefont {T.}~\bibnamefont {Tai}}, \bibinfo {author} {\bibfnamefont {E.}~\bibnamefont {Baldini}}, \bibinfo {author} {\bibfnamefont {S.}~\bibnamefont {Son}}, \bibinfo {author} {\bibfnamefont {J.}~\bibnamefont {Kim}}, \bibinfo {author} {\bibfnamefont {J.}~\bibnamefont {Park}}, \bibinfo {author} {\bibfnamefont {J.-G.}\ \bibnamefont {Park}}, \emph {et~al.},\ }\href {https://doi.org/10.1103/PhysRevLett.134.066903} {\bibfield  {journal} {\bibinfo  {journal} {Phys. Rev. Lett.}\ }\textbf {\bibinfo {volume} {134}},\ \bibinfo {pages} {066903} (\bibinfo {year} {2025})}\BibitemShut {NoStop}%
\bibitem [{\citenamefont {Warshauer}\ \emph {et~al.}(2025{\natexlab{b}})\citenamefont {Warshauer}, \citenamefont {Chen}, \citenamefont {Tan}, \citenamefont {Tang}, \citenamefont {Ling},\ and\ \citenamefont {Hu}}]{38}%
  \BibitemOpen
  \bibfield  {author} {\bibinfo {author} {\bibfnamefont {J.}~\bibnamefont {Warshauer}}, \bibinfo {author} {\bibfnamefont {H.}~\bibnamefont {Chen}}, \bibinfo {author} {\bibfnamefont {Q.}~\bibnamefont {Tan}}, \bibinfo {author} {\bibfnamefont {J.}~\bibnamefont {Tang}}, \bibinfo {author} {\bibfnamefont {X.}~\bibnamefont {Ling}},\ and\ \bibinfo {author} {\bibfnamefont {W.}~\bibnamefont {Hu}},\ }\href {https://doi.org/10.1103/PhysRevB.111.155111} {\bibfield  {journal} {\bibinfo  {journal} {Phys. Rev. B}\ }\textbf {\bibinfo {volume} {111}},\ \bibinfo {pages} {155111} (\bibinfo {year} {2025}{\natexlab{b}})}\BibitemShut {NoStop}%
\bibitem [{\citenamefont {Jana}\ \emph {et~al.}(2023)\citenamefont {Jana}, \citenamefont {Kapuscinski}, \citenamefont {Mohelsky}, \citenamefont {Vaclavkova}, \citenamefont {Breslavetz}, \citenamefont {Orlita}, \citenamefont {Faugeras},\ and\ \citenamefont {Potemski}}]{26}%
  \BibitemOpen
  \bibfield  {author} {\bibinfo {author} {\bibfnamefont {D.}~\bibnamefont {Jana}}, \bibinfo {author} {\bibfnamefont {P.}~\bibnamefont {Kapuscinski}}, \bibinfo {author} {\bibfnamefont {I.}~\bibnamefont {Mohelsky}}, \bibinfo {author} {\bibfnamefont {D.}~\bibnamefont {Vaclavkova}}, \bibinfo {author} {\bibfnamefont {I.}~\bibnamefont {Breslavetz}}, \bibinfo {author} {\bibfnamefont {M.}~\bibnamefont {Orlita}}, \bibinfo {author} {\bibfnamefont {C.}~\bibnamefont {Faugeras}},\ and\ \bibinfo {author} {\bibfnamefont {M.}~\bibnamefont {Potemski}},\ }\href {https://doi.org/10.1103/physrevb.108.115149} {\bibfield  {journal} {\bibinfo  {journal} {Phys. Rev. B}\ }\textbf {\bibinfo {volume} {108}},\ \bibinfo {pages} {115149} (\bibinfo {year} {2023})}\BibitemShut {NoStop}%
\bibitem [{\citenamefont {Bardeen}\ \emph {et~al.}(1957)\citenamefont {Bardeen}, \citenamefont {Cooper},\ and\ \citenamefont {Schrieffer}}]{43}%
  \BibitemOpen
  \bibfield  {author} {\bibinfo {author} {\bibfnamefont {J.}~\bibnamefont {Bardeen}}, \bibinfo {author} {\bibfnamefont {L.~N.}\ \bibnamefont {Cooper}},\ and\ \bibinfo {author} {\bibfnamefont {J.~R.}\ \bibnamefont {Schrieffer}},\ }\href {https://doi.org/10.1103/PhysRev.108.1175} {\bibfield  {journal} {\bibinfo  {journal} {Phys. Rev.}\ }\textbf {\bibinfo {volume} {108}},\ \bibinfo {pages} {1175} (\bibinfo {year} {1957})}\BibitemShut {NoStop}%
\bibitem [{\citenamefont {Hamad}\ \emph {et~al.}(2024)\citenamefont {Hamad}, \citenamefont {Helman}, \citenamefont {Manuel}, \citenamefont {Feiguin},\ and\ \citenamefont {Aligia}}]{16}%
  \BibitemOpen
  \bibfield  {author} {\bibinfo {author} {\bibfnamefont {I.}~\bibnamefont {Hamad}}, \bibinfo {author} {\bibfnamefont {C.}~\bibnamefont {Helman}}, \bibinfo {author} {\bibfnamefont {L.}~\bibnamefont {Manuel}}, \bibinfo {author} {\bibfnamefont {A.}~\bibnamefont {Feiguin}},\ and\ \bibinfo {author} {\bibfnamefont {A.}~\bibnamefont {Aligia}},\ }\href {https://doi.org/10.1103/physrevlett.133.146502} {\bibfield  {journal} {\bibinfo  {journal} {Phys. Rev. Lett.}\ }\textbf {\bibinfo {volume} {133}},\ \bibinfo {pages} {146502} (\bibinfo {year} {2024})}\BibitemShut {NoStop}%
\bibitem [{\citenamefont {Nishimori}\ and\ \citenamefont {Ortiz}(2011)}]{29}%
  \BibitemOpen
  \bibfield  {author} {\bibinfo {author} {\bibfnamefont {H.}~\bibnamefont {Nishimori}}\ and\ \bibinfo {author} {\bibfnamefont {G.}~\bibnamefont {Ortiz}},\ }\href@noop {} {\emph {\bibinfo {title} {Elements of Phase Transitions and Critical Phenomena}}}\ (\bibinfo  {publisher} {Oxford University Press},\ \bibinfo {address} {New York},\ \bibinfo {year} {2011})\BibitemShut {NoStop}%
\bibitem [{\citenamefont {Khomskii}(2010)}]{30}%
  \BibitemOpen
  \bibfield  {author} {\bibinfo {author} {\bibfnamefont {D.~I.}\ \bibnamefont {Khomskii}},\ }\href {https://doi.org/10.1017/cbo9780511780271} {\emph {\bibinfo {title} {Basic Aspects of the Quantum Theory of Solids: Order and Elementary Excitations}}}\ (\bibinfo  {publisher} {Cambridge University Press},\ \bibinfo {address} {Cambridge},\ \bibinfo {year} {2010})\BibitemShut {NoStop}%
\bibitem [{foo()}]{foot1}%
  \BibitemOpen
  \href@noop {} {}\bibinfo {note} {The pulse width is determined from the cross-correlation of the pump and probe beams. See SM Section S2 for details.}\BibitemShut {Stop}%
\bibitem [{\citenamefont {Trovatello}\ \emph {et~al.}(2020)\citenamefont {Trovatello}, \citenamefont {Katsch}, \citenamefont {Borys}, \citenamefont {Selig}, \citenamefont {Yao}, \citenamefont {Borrego-Varillas}, \citenamefont {Scotognella}, \citenamefont {Kriegel}, \citenamefont {Yan}, \citenamefont {Zettl} \emph {et~al.}}]{18}%
  \BibitemOpen
  \bibfield  {author} {\bibinfo {author} {\bibfnamefont {C.}~\bibnamefont {Trovatello}}, \bibinfo {author} {\bibfnamefont {F.}~\bibnamefont {Katsch}}, \bibinfo {author} {\bibfnamefont {N.~J.}\ \bibnamefont {Borys}}, \bibinfo {author} {\bibfnamefont {M.}~\bibnamefont {Selig}}, \bibinfo {author} {\bibfnamefont {K.}~\bibnamefont {Yao}}, \bibinfo {author} {\bibfnamefont {R.}~\bibnamefont {Borrego-Varillas}}, \bibinfo {author} {\bibfnamefont {F.}~\bibnamefont {Scotognella}}, \bibinfo {author} {\bibfnamefont {I.}~\bibnamefont {Kriegel}}, \bibinfo {author} {\bibfnamefont {A.}~\bibnamefont {Yan}}, \bibinfo {author} {\bibfnamefont {A.}~\bibnamefont {Zettl}}, \emph {et~al.},\ }\href {https://doi.org/10.1038/s41467-020-18835-5} {\bibfield  {journal} {\bibinfo  {journal} {Nat. Commun.}\ }\textbf {\bibinfo {volume} {11}},\ \bibinfo {pages} {5277} (\bibinfo {year} {2020})}\BibitemShut {NoStop}%
\bibitem [{\citenamefont {Wright}\ \emph {et~al.}(2024)\citenamefont {Wright}, \citenamefont {Ebrahim~Nataj}, \citenamefont {Guzman}, \citenamefont {Polster}, \citenamefont {Bouman}, \citenamefont {Kargar},\ and\ \citenamefont {Balandin}}]{19}%
  \BibitemOpen
  \bibfield  {author} {\bibinfo {author} {\bibfnamefont {D.}~\bibnamefont {Wright}}, \bibinfo {author} {\bibfnamefont {Z.}~\bibnamefont {Ebrahim~Nataj}}, \bibinfo {author} {\bibfnamefont {E.}~\bibnamefont {Guzman}}, \bibinfo {author} {\bibfnamefont {J.}~\bibnamefont {Polster}}, \bibinfo {author} {\bibfnamefont {M.}~\bibnamefont {Bouman}}, \bibinfo {author} {\bibfnamefont {F.}~\bibnamefont {Kargar}},\ and\ \bibinfo {author} {\bibfnamefont {A.~A.}\ \bibnamefont {Balandin}},\ }\href {https://doi.org/10.1063/5.0205946} {\bibfield  {journal} {\bibinfo  {journal} {Appl. Phys. Lett.}\ }\textbf {\bibinfo {volume} {124}},\ \bibinfo {pages} {162401} (\bibinfo {year} {2024})}\BibitemShut {NoStop}%
\bibitem [{\citenamefont {Kim}\ \emph {et~al.}(2019)\citenamefont {Kim}, \citenamefont {Lim}, \citenamefont {Lee}, \citenamefont {Lee}, \citenamefont {Kim}, \citenamefont {Park}, \citenamefont {Jeon}, \citenamefont {Park}, \citenamefont {Park},\ and\ \citenamefont {Cheong}}]{37}%
  \BibitemOpen
  \bibfield  {author} {\bibinfo {author} {\bibfnamefont {K.}~\bibnamefont {Kim}}, \bibinfo {author} {\bibfnamefont {S.}~\bibnamefont {Lim}}, \bibinfo {author} {\bibfnamefont {J.}~\bibnamefont {Lee}}, \bibinfo {author} {\bibfnamefont {S.}~\bibnamefont {Lee}}, \bibinfo {author} {\bibfnamefont {T.}~\bibnamefont {Kim}}, \bibinfo {author} {\bibfnamefont {K.}~\bibnamefont {Park}}, \bibinfo {author} {\bibfnamefont {G.}~\bibnamefont {Jeon}}, \bibinfo {author} {\bibfnamefont {C.}~\bibnamefont {Park}}, \bibinfo {author} {\bibfnamefont {J.}~\bibnamefont {Park}},\ and\ \bibinfo {author} {\bibfnamefont {H.}~\bibnamefont {Cheong}},\ }\href {https://doi.org/10.1038/s41467-018-08284-6} {\bibfield  {journal} {\bibinfo  {journal} {Nat. Commun.}\ }\textbf {\bibinfo {volume} {10}},\ \bibinfo {pages} {345} (\bibinfo {year} {2019})}\BibitemShut {NoStop}%
\bibitem [{\citenamefont {Hu}\ \emph {et~al.}(2023)\citenamefont {Hu}, \citenamefont {Wang}, \citenamefont {Chen}, \citenamefont {Xu}, \citenamefont {Li}, \citenamefont {Liu}, \citenamefont {Gu}, \citenamefont {Wang}, \citenamefont {Zhang}, \citenamefont {Yao},\ and\ \citenamefont {Xiong}}]{20}%
  \BibitemOpen
  \bibfield  {author} {\bibinfo {author} {\bibfnamefont {L.}~\bibnamefont {Hu}}, \bibinfo {author} {\bibfnamefont {H.}~\bibnamefont {Wang}}, \bibinfo {author} {\bibfnamefont {Y.}~\bibnamefont {Chen}}, \bibinfo {author} {\bibfnamefont {K.}~\bibnamefont {Xu}}, \bibinfo {author} {\bibfnamefont {M.}~\bibnamefont {Li}}, \bibinfo {author} {\bibfnamefont {H.}~\bibnamefont {Liu}}, \bibinfo {author} {\bibfnamefont {P.}~\bibnamefont {Gu}}, \bibinfo {author} {\bibfnamefont {Y.}~\bibnamefont {Wang}}, \bibinfo {author} {\bibfnamefont {M.}~\bibnamefont {Zhang}}, \bibinfo {author} {\bibfnamefont {H.}~\bibnamefont {Yao}},\ and\ \bibinfo {author} {\bibfnamefont {Q.}~\bibnamefont {Xiong}},\ }\href {https://doi.org/10.1103/physrevb.107.l220407} {\bibfield  {journal} {\bibinfo  {journal} {Phys. Rev. B}\ }\textbf {\bibinfo {volume} {107}},\ \bibinfo {pages} {L220407} (\bibinfo {year} {2023})}\BibitemShut {NoStop}%
\bibitem [{\citenamefont {Chu}\ \emph {et~al.}(2017)\citenamefont {Chu}, \citenamefont {Zhao}, \citenamefont {de~la Torre}, \citenamefont {Hogan}, \citenamefont {Wilson},\ and\ \citenamefont {Hsieh}}]{23}%
  \BibitemOpen
  \bibfield  {author} {\bibinfo {author} {\bibfnamefont {H.}~\bibnamefont {Chu}}, \bibinfo {author} {\bibfnamefont {L.}~\bibnamefont {Zhao}}, \bibinfo {author} {\bibfnamefont {A.}~\bibnamefont {de~la Torre}}, \bibinfo {author} {\bibfnamefont {T.}~\bibnamefont {Hogan}}, \bibinfo {author} {\bibfnamefont {S.}~\bibnamefont {Wilson}},\ and\ \bibinfo {author} {\bibfnamefont {D.}~\bibnamefont {Hsieh}},\ }\href {https://doi.org/10.1038/nmat4836} {\bibfield  {journal} {\bibinfo  {journal} {Nat. Mater.}\ }\textbf {\bibinfo {volume} {16}},\ \bibinfo {pages} {200} (\bibinfo {year} {2017})}\BibitemShut {NoStop}%
\bibitem [{\citenamefont {Tian}\ \emph {et~al.}(2016)\citenamefont {Tian}, \citenamefont {Zhang}, \citenamefont {Li}, \citenamefont {Wu}, \citenamefont {Wu}, \citenamefont {Sun}, \citenamefont {Zhou}, \citenamefont {Wang}, \citenamefont {Ma}, \citenamefont {Xue} \emph {et~al.}}]{tian2016ultrafast}%
  \BibitemOpen
  \bibfield  {author} {\bibinfo {author} {\bibfnamefont {Y.}~\bibnamefont {Tian}}, \bibinfo {author} {\bibfnamefont {W.}~\bibnamefont {Zhang}}, \bibinfo {author} {\bibfnamefont {F.}~\bibnamefont {Li}}, \bibinfo {author} {\bibfnamefont {Y.}~\bibnamefont {Wu}}, \bibinfo {author} {\bibfnamefont {Q.}~\bibnamefont {Wu}}, \bibinfo {author} {\bibfnamefont {F.}~\bibnamefont {Sun}}, \bibinfo {author} {\bibfnamefont {G.}~\bibnamefont {Zhou}}, \bibinfo {author} {\bibfnamefont {L.}~\bibnamefont {Wang}}, \bibinfo {author} {\bibfnamefont {X.}~\bibnamefont {Ma}}, \bibinfo {author} {\bibfnamefont {Q.-K.}\ \bibnamefont {Xue}}, \emph {et~al.},\ }\href {https://doi.org/10.1103/PhysRevLett.116.107001} {\bibfield  {journal} {\bibinfo  {journal} {Phys. Rev. Lett.}\ }\textbf {\bibinfo {volume} {116}},\ \bibinfo {pages} {107001} (\bibinfo {year} {2016})}\BibitemShut {NoStop}%
\bibitem [{\citenamefont {Ho}\ \emph {et~al.}(2021)\citenamefont {Ho}, \citenamefont {Hsu},\ and\ \citenamefont {Muhimmah}}]{22}%
  \BibitemOpen
  \bibfield  {author} {\bibinfo {author} {\bibfnamefont {C.}~\bibnamefont {Ho}}, \bibinfo {author} {\bibfnamefont {T.}~\bibnamefont {Hsu}},\ and\ \bibinfo {author} {\bibfnamefont {L.}~\bibnamefont {Muhimmah}},\ }\href {https://doi.org/10.1038/s41699-020-00188-8} {\bibfield  {journal} {\bibinfo  {journal} {npj 2D Mater. Appl.}\ }\textbf {\bibinfo {volume} {5}},\ \bibinfo {pages} {8} (\bibinfo {year} {2021})}\BibitemShut {NoStop}%
\bibitem [{\citenamefont {Gong}\ \emph {et~al.}(2024)\citenamefont {Gong}, \citenamefont {Yan}, \citenamefont {He}, \citenamefont {Shen}, \citenamefont {Yang}, \citenamefont {Ge}, \citenamefont {Sun}, \citenamefont {Ma},\ and\ \citenamefont {Zhang}}]{e-h}%
  \BibitemOpen
  \bibfield  {author} {\bibinfo {author} {\bibfnamefont {Z.}~\bibnamefont {Gong}}, \bibinfo {author} {\bibfnamefont {L.}~\bibnamefont {Yan}}, \bibinfo {author} {\bibfnamefont {Q.}~\bibnamefont {He}}, \bibinfo {author} {\bibfnamefont {D.}~\bibnamefont {Shen}}, \bibinfo {author} {\bibfnamefont {Y.}~\bibnamefont {Yang}}, \bibinfo {author} {\bibfnamefont {A.}~\bibnamefont {Ge}}, \bibinfo {author} {\bibfnamefont {L.}~\bibnamefont {Sun}}, \bibinfo {author} {\bibfnamefont {G.}~\bibnamefont {Ma}},\ and\ \bibinfo {author} {\bibfnamefont {S.}~\bibnamefont {Zhang}},\ }\href {https://doi.org/10.1364/OME.525626} {\bibfield  {journal} {\bibinfo  {journal} {Opt. Mater. Express}\ }\textbf {\bibinfo {volume} {14}},\ \bibinfo {pages} {1876} (\bibinfo {year} {2024})}\BibitemShut {NoStop}%
\bibitem [{\citenamefont {Negi}\ \emph {et~al.}(2024)\citenamefont {Negi}, \citenamefont {Badola}, \citenamefont {Paul}, \citenamefont {Pistawala}, \citenamefont {Harnagea},\ and\ \citenamefont {Saha}}]{Saha}%
  \BibitemOpen
  \bibfield  {author} {\bibinfo {author} {\bibfnamefont {D.}~\bibnamefont {Negi}}, \bibinfo {author} {\bibfnamefont {S.}~\bibnamefont {Badola}}, \bibinfo {author} {\bibfnamefont {S.}~\bibnamefont {Paul}}, \bibinfo {author} {\bibfnamefont {N.}~\bibnamefont {Pistawala}}, \bibinfo {author} {\bibfnamefont {L.}~\bibnamefont {Harnagea}},\ and\ \bibinfo {author} {\bibfnamefont {S.}~\bibnamefont {Saha}},\ }\href {https://doi.org/10.1103/PhysRevB.110.094434} {\bibfield  {journal} {\bibinfo  {journal} {Phys. Rev. B}\ }\textbf {\bibinfo {volume} {110}},\ \bibinfo {pages} {094434} (\bibinfo {year} {2024})}\BibitemShut {NoStop}%
\bibitem [{\citenamefont {Rothwarf}\ and\ \citenamefont {Taylor}(1967)}]{40}%
  \BibitemOpen
  \bibfield  {author} {\bibinfo {author} {\bibfnamefont {A.}~\bibnamefont {Rothwarf}}\ and\ \bibinfo {author} {\bibfnamefont {B.~N.}\ \bibnamefont {Taylor}},\ }\href {https://doi.org/10.1103/PhysRevLett.19.27} {\bibfield  {journal} {\bibinfo  {journal} {Phys. Rev. Lett.}\ }\textbf {\bibinfo {volume} {19}},\ \bibinfo {pages} {27} (\bibinfo {year} {1967})}\BibitemShut {NoStop}%
\bibitem [{\citenamefont {Demsar}\ \emph {et~al.}(2003)\citenamefont {Demsar}, \citenamefont {Averitt}, \citenamefont {Taylor}, \citenamefont {Kabanov}, \citenamefont {Kang}, \citenamefont {Kim}, \citenamefont {Choi},\ and\ \citenamefont {Lee}}]{41}%
  \BibitemOpen
  \bibfield  {author} {\bibinfo {author} {\bibfnamefont {J.}~\bibnamefont {Demsar}}, \bibinfo {author} {\bibfnamefont {R.~D.}\ \bibnamefont {Averitt}}, \bibinfo {author} {\bibfnamefont {A.~J.}\ \bibnamefont {Taylor}}, \bibinfo {author} {\bibfnamefont {V.~V.}\ \bibnamefont {Kabanov}}, \bibinfo {author} {\bibfnamefont {W.~N.}\ \bibnamefont {Kang}}, \bibinfo {author} {\bibfnamefont {H.~J.}\ \bibnamefont {Kim}}, \bibinfo {author} {\bibfnamefont {E.~M.}\ \bibnamefont {Choi}},\ and\ \bibinfo {author} {\bibfnamefont {S.~I.}\ \bibnamefont {Lee}},\ }\href {https://doi.org/10.1103/PhysRevLett.91.267002} {\bibfield  {journal} {\bibinfo  {journal} {Phys. Rev. Lett.}\ }\textbf {\bibinfo {volume} {91}},\ \bibinfo {pages} {267002} (\bibinfo {year} {2003})}\BibitemShut {NoStop}%
\bibitem [{\citenamefont {Werdehausen}\ \emph {et~al.}(2018)\citenamefont {Werdehausen}, \citenamefont {Takayama}, \citenamefont {Katsufuji}, \citenamefont {Takagi},\ and\ \citenamefont {Kaiser}}]{42}%
  \BibitemOpen
  \bibfield  {author} {\bibinfo {author} {\bibfnamefont {D.}~\bibnamefont {Werdehausen}}, \bibinfo {author} {\bibfnamefont {T.}~\bibnamefont {Takayama}}, \bibinfo {author} {\bibfnamefont {T.}~\bibnamefont {Katsufuji}}, \bibinfo {author} {\bibfnamefont {H.}~\bibnamefont {Takagi}},\ and\ \bibinfo {author} {\bibfnamefont {S.}~\bibnamefont {Kaiser}},\ }\href {https://doi.org/10.1126/sciadv.aap8652} {\bibfield  {journal} {\bibinfo  {journal} {Sci. Adv.}\ }\textbf {\bibinfo {volume} {4}},\ \bibinfo {pages} {eaap8652} (\bibinfo {year} {2018})}\BibitemShut {NoStop}%
\bibitem [{\citenamefont {Plumley}\ \emph {et~al.}(2024)\citenamefont {Plumley}, \citenamefont {Mardanya}, \citenamefont {Peng}, \citenamefont {Nokelainen}, \citenamefont {Assefa}, \citenamefont {Shen}, \citenamefont {Burdet}, \citenamefont {Porter}, \citenamefont {Petsch}, \citenamefont {Israelski},\ and\ \citenamefont {Chen}}]{27}%
  \BibitemOpen
  \bibfield  {author} {\bibinfo {author} {\bibfnamefont {R.}~\bibnamefont {Plumley}}, \bibinfo {author} {\bibfnamefont {S.}~\bibnamefont {Mardanya}}, \bibinfo {author} {\bibfnamefont {C.}~\bibnamefont {Peng}}, \bibinfo {author} {\bibfnamefont {J.}~\bibnamefont {Nokelainen}}, \bibinfo {author} {\bibfnamefont {T.}~\bibnamefont {Assefa}}, \bibinfo {author} {\bibfnamefont {L.}~\bibnamefont {Shen}}, \bibinfo {author} {\bibfnamefont {N.}~\bibnamefont {Burdet}}, \bibinfo {author} {\bibfnamefont {Z.}~\bibnamefont {Porter}}, \bibinfo {author} {\bibfnamefont {A.}~\bibnamefont {Petsch}}, \bibinfo {author} {\bibfnamefont {A.}~\bibnamefont {Israelski}},\ and\ \bibinfo {author} {\bibfnamefont {H.}~\bibnamefont {Chen}},\ }\href {https://doi.org/10.1038/s41535-024-00696-6} {\bibfield  {journal} {\bibinfo  {journal} {npj Quantum Mater.}\ }\textbf {\bibinfo {volume} {9}},\ \bibinfo {pages} {95} (\bibinfo {year} {2024})}\BibitemShut {NoStop}%
\bibitem [{\citenamefont {Zhou}\ \emph {et~al.}(2022)\citenamefont {Zhou}, \citenamefont {Hwangbo}, \citenamefont {Zhang}, \citenamefont {Wang}, \citenamefont {Shen}, \citenamefont {Zhang}, \citenamefont {Jiang}, \citenamefont {Zong}, \citenamefont {Su}, \citenamefont {Zajac},\ and\ \citenamefont {Ahn}}]{28}%
  \BibitemOpen
  \bibfield  {author} {\bibinfo {author} {\bibfnamefont {F.}~\bibnamefont {Zhou}}, \bibinfo {author} {\bibfnamefont {K.}~\bibnamefont {Hwangbo}}, \bibinfo {author} {\bibfnamefont {Q.}~\bibnamefont {Zhang}}, \bibinfo {author} {\bibfnamefont {C.}~\bibnamefont {Wang}}, \bibinfo {author} {\bibfnamefont {L.}~\bibnamefont {Shen}}, \bibinfo {author} {\bibfnamefont {J.}~\bibnamefont {Zhang}}, \bibinfo {author} {\bibfnamefont {Q.}~\bibnamefont {Jiang}}, \bibinfo {author} {\bibfnamefont {A.}~\bibnamefont {Zong}}, \bibinfo {author} {\bibfnamefont {Y.}~\bibnamefont {Su}}, \bibinfo {author} {\bibfnamefont {M.}~\bibnamefont {Zajac}},\ and\ \bibinfo {author} {\bibfnamefont {Y.}~\bibnamefont {Ahn}},\ }\href {https://doi.org/10.1038/s41467-022-34376-5} {\bibfield  {journal} {\bibinfo  {journal} {Nat. Commun.}\ }\textbf {\bibinfo {volume} {13}},\ \bibinfo {pages} {6598} (\bibinfo {year} {2022})}\BibitemShut {NoStop}%
\bibitem [{\citenamefont {Scheie}\ \emph {et~al.}(2023)\citenamefont {Scheie}, \citenamefont {Park}, \citenamefont {Villanova}, \citenamefont {Granroth}, \citenamefont {Sarkis}, \citenamefont {Zhang}, \citenamefont {Stone}, \citenamefont {Park}, \citenamefont {Okamoto}, \citenamefont {Berlijn},\ and\ \citenamefont {Tennant}}]{24}%
  \BibitemOpen
  \bibfield  {author} {\bibinfo {author} {\bibfnamefont {A.}~\bibnamefont {Scheie}}, \bibinfo {author} {\bibfnamefont {P.}~\bibnamefont {Park}}, \bibinfo {author} {\bibfnamefont {J.}~\bibnamefont {Villanova}}, \bibinfo {author} {\bibfnamefont {G.}~\bibnamefont {Granroth}}, \bibinfo {author} {\bibfnamefont {C.}~\bibnamefont {Sarkis}}, \bibinfo {author} {\bibfnamefont {H.}~\bibnamefont {Zhang}}, \bibinfo {author} {\bibfnamefont {M.}~\bibnamefont {Stone}}, \bibinfo {author} {\bibfnamefont {J.}~\bibnamefont {Park}}, \bibinfo {author} {\bibfnamefont {S.}~\bibnamefont {Okamoto}}, \bibinfo {author} {\bibfnamefont {T.}~\bibnamefont {Berlijn}},\ and\ \bibinfo {author} {\bibfnamefont {D.}~\bibnamefont {Tennant}},\ }\href {https://doi.org/10.1103/physrevb.108.104402} {\bibfield  {journal} {\bibinfo  {journal} {Phys. Rev. B}\ }\textbf {\bibinfo {volume} {108}},\ \bibinfo {pages} {104402} (\bibinfo {year} {2023})}\BibitemShut {NoStop}%
\bibitem [{\citenamefont {Mehlawat}\ \emph {et~al.}(2022)\citenamefont {Mehlawat}, \citenamefont {Alfonsov}, \citenamefont {Selter}, \citenamefont {Shemerliuk}, \citenamefont {Aswartham}, \citenamefont {B{\"u}chner},\ and\ \citenamefont {Kataev}}]{25}%
  \BibitemOpen
  \bibfield  {author} {\bibinfo {author} {\bibfnamefont {K.}~\bibnamefont {Mehlawat}}, \bibinfo {author} {\bibfnamefont {A.}~\bibnamefont {Alfonsov}}, \bibinfo {author} {\bibfnamefont {S.}~\bibnamefont {Selter}}, \bibinfo {author} {\bibfnamefont {Y.}~\bibnamefont {Shemerliuk}}, \bibinfo {author} {\bibfnamefont {S.}~\bibnamefont {Aswartham}}, \bibinfo {author} {\bibfnamefont {B.}~\bibnamefont {B{\"u}chner}},\ and\ \bibinfo {author} {\bibfnamefont {V.}~\bibnamefont {Kataev}},\ }\href {https://doi.org/10.1103/physrevb.105.214427} {\bibfield  {journal} {\bibinfo  {journal} {Phys. Rev. B}\ }\textbf {\bibinfo {volume} {105}},\ \bibinfo {pages} {214427} (\bibinfo {year} {2022})}\BibitemShut {NoStop}%
\bibitem [{\citenamefont {Lovinger}\ \emph {et~al.}(2020)\citenamefont {Lovinger}, \citenamefont {Brahlek}, \citenamefont {Kissin}, \citenamefont {Kennes}, \citenamefont {Millis}, \citenamefont {Engel-Herbert},\ and\ \citenamefont {Averitt}}]{Lovinger}%
  \BibitemOpen
  \bibfield  {author} {\bibinfo {author} {\bibfnamefont {D.}~\bibnamefont {Lovinger}}, \bibinfo {author} {\bibfnamefont {M.}~\bibnamefont {Brahlek}}, \bibinfo {author} {\bibfnamefont {P.}~\bibnamefont {Kissin}}, \bibinfo {author} {\bibfnamefont {D.}~\bibnamefont {Kennes}}, \bibinfo {author} {\bibfnamefont {A.}~\bibnamefont {Millis}}, \bibinfo {author} {\bibfnamefont {R.}~\bibnamefont {Engel-Herbert}},\ and\ \bibinfo {author} {\bibfnamefont {R.}~\bibnamefont {Averitt}},\ }\href {https://doi.org/10.1103/physrevb.102.085138} {\bibfield  {journal} {\bibinfo  {journal} {Phys. Rev. B}\ }\textbf {\bibinfo {volume} {102}},\ \bibinfo {pages} {115143} (\bibinfo {year} {2020})}\BibitemShut {NoStop}%
\bibitem [{\citenamefont {Braun}(1968)}]{45}%
  \BibitemOpen
  \bibfield  {author} {\bibinfo {author} {\bibfnamefont {C.}~\bibnamefont {Braun}},\ }\href {https://doi.org/10.1103/PhysRevLett.21.215} {\bibfield  {journal} {\bibinfo  {journal} {Phys. Rev. Lett.}\ }\textbf {\bibinfo {volume} {21}},\ \bibinfo {pages} {215} (\bibinfo {year} {1968})}\BibitemShut {NoStop}%
\bibitem [{\citenamefont {Bergman}\ and\ \citenamefont {Jortner}(1974)}]{46}%
  \BibitemOpen
  \bibfield  {author} {\bibinfo {author} {\bibfnamefont {A.}~\bibnamefont {Bergman}}\ and\ \bibinfo {author} {\bibfnamefont {J.}~\bibnamefont {Jortner}},\ }\href {https://doi.org/10.1103/PhysRevB.9.4560} {\bibfield  {journal} {\bibinfo  {journal} {Phys. Rev. B}\ }\textbf {\bibinfo {volume} {9}},\ \bibinfo {pages} {4560} (\bibinfo {year} {1974})}\BibitemShut {NoStop}%
\end{thebibliography}%


\begin{thebibliography}{22}%
\makeatletter
\providecommand \@ifxundefined [1]{%
 \@ifx{#1\undefined}
}%
\providecommand \@ifnum [1]{%
 \ifnum #1\expandafter \@firstoftwo
 \else \expandafter \@secondoftwo
 \fi
}%
\providecommand \@ifx [1]{%
 \ifx #1\expandafter \@firstoftwo
 \else \expandafter \@secondoftwo
 \fi
}%
\providecommand \natexlab [1]{#1}%
\providecommand \enquote  [1]{``#1''}%
\providecommand \bibnamefont  [1]{#1}%
\providecommand \bibfnamefont [1]{#1}%
\providecommand \citenamefont [1]{#1}%
\providecommand \href@noop [0]{\@secondoftwo}%
\providecommand \href [0]{\begingroup \@sanitize@url \@href}%
\providecommand \@href[1]{\@@startlink{#1}\@@href}%
\providecommand \@@href[1]{\endgroup#1\@@endlink}%
\providecommand \@sanitize@url [0]{\catcode `\\12\catcode `\$12\catcode `\&12\catcode `\#12\catcode `\^12\catcode `\_12\catcode `\%12\relax}%
\providecommand \@@startlink[1]{}%
\providecommand \@@endlink[0]{}%
\providecommand \url  [0]{\begingroup\@sanitize@url \@url }%
\providecommand \@url [1]{\endgroup\@href {#1}{\urlprefix }}%
\providecommand \urlprefix  [0]{URL }%
\providecommand \Eprint [0]{\href }%
\providecommand \doibase [0]{https://doi.org/}%
\providecommand \selectlanguage [0]{\@gobble}%
\providecommand \bibinfo  [0]{\@secondoftwo}%
\providecommand \bibfield  [0]{\@secondoftwo}%
\providecommand \translation [1]{[#1]}%
\providecommand \BibitemOpen [0]{}%
\providecommand \bibitemStop [0]{}%
\providecommand \bibitemNoStop [0]{.\EOS\space}%
\providecommand \EOS [0]{\spacefactor3000\relax}%
\providecommand \BibitemShut  [1]{\csname bibitem#1\endcsname}%
\let\auto@bib@innerbib\@empty
\bibitem [{\citenamefont {Jenjeti}\ \emph {et~al.}(2018)\citenamefont {Jenjeti}, \citenamefont {Kumar}, \citenamefont {Austeria},\ and\ \citenamefont {Sampath}}]{XRD1}%
  \BibitemOpen
  \bibfield  {author} {\bibinfo {author} {\bibfnamefont {R.~N.}\ \bibnamefont {Jenjeti}}, \bibinfo {author} {\bibfnamefont {R.}~\bibnamefont {Kumar}}, \bibinfo {author} {\bibfnamefont {M.~P.}\ \bibnamefont {Austeria}},\ and\ \bibinfo {author} {\bibfnamefont {S.}~\bibnamefont {Sampath}},\ }\href@noop {} {\bibfield  {journal} {\bibinfo  {journal} {Sci. Rep.}\ }\textbf {\bibinfo {volume} {8}},\ \bibinfo {pages} {8586} (\bibinfo {year} {2018})}\BibitemShut {NoStop}%
\bibitem [{\citenamefont {Basnet}\ \emph {et~al.}(2021)\citenamefont {Basnet}, \citenamefont {Wegner}, \citenamefont {Pandey}, \citenamefont {Storment},\ and\ \citenamefont {Hu}}]{XRD2}%
  \BibitemOpen
  \bibfield  {author} {\bibinfo {author} {\bibfnamefont {R.}~\bibnamefont {Basnet}}, \bibinfo {author} {\bibfnamefont {A.}~\bibnamefont {Wegner}}, \bibinfo {author} {\bibfnamefont {K.}~\bibnamefont {Pandey}}, \bibinfo {author} {\bibfnamefont {S.}~\bibnamefont {Storment}},\ and\ \bibinfo {author} {\bibfnamefont {J.}~\bibnamefont {Hu}},\ }\href@noop {} {\bibfield  {journal} {\bibinfo  {journal} {Phys. Rev. Mater.}\ }\textbf {\bibinfo {volume} {5}},\ \bibinfo {pages} {064413} (\bibinfo {year} {2021})}\BibitemShut {NoStop}%
\bibitem [{\citenamefont {Liu}\ \emph {et~al.}(2019)\citenamefont {Liu}, \citenamefont {Li}, \citenamefont {Xu}, \citenamefont {Ge}, \citenamefont {Wang}, \citenamefont {Zhang}, \citenamefont {Wang}, \citenamefont {Fang}, \citenamefont {Yang}, \citenamefont {Wang} \emph {et~al.}}]{XRD3}%
  \BibitemOpen
  \bibfield  {author} {\bibinfo {author} {\bibfnamefont {J.}~\bibnamefont {Liu}}, \bibinfo {author} {\bibfnamefont {X.}~\bibnamefont {Li}}, \bibinfo {author} {\bibfnamefont {Y.}~\bibnamefont {Xu}}, \bibinfo {author} {\bibfnamefont {Y.}~\bibnamefont {Ge}}, \bibinfo {author} {\bibfnamefont {Y.}~\bibnamefont {Wang}}, \bibinfo {author} {\bibfnamefont {F.}~\bibnamefont {Zhang}}, \bibinfo {author} {\bibfnamefont {Y.}~\bibnamefont {Wang}}, \bibinfo {author} {\bibfnamefont {Y.}~\bibnamefont {Fang}}, \bibinfo {author} {\bibfnamefont {F.}~\bibnamefont {Yang}}, \bibinfo {author} {\bibfnamefont {C.}~\bibnamefont {Wang}}, \emph {et~al.},\ }\href@noop {} {\bibfield  {journal} {\bibinfo  {journal} {Nanoscale}\ }\textbf {\bibinfo {volume} {11}},\ \bibinfo {pages} {14383} (\bibinfo {year} {2019})}\BibitemShut {NoStop}%
\bibitem [{\citenamefont {Belvin}\ \emph {et~al.}(2021)\citenamefont {Belvin}, \citenamefont {Baldini}, \citenamefont {Ozel}, \citenamefont {Mao}, \citenamefont {Po}, \citenamefont {Allington}, \citenamefont {Son}, \citenamefont {Kim}, \citenamefont {Kim}, \citenamefont {Hwang} \emph {et~al.}}]{MT}%
  \BibitemOpen
  \bibfield  {author} {\bibinfo {author} {\bibfnamefont {C.~A.}\ \bibnamefont {Belvin}}, \bibinfo {author} {\bibfnamefont {E.}~\bibnamefont {Baldini}}, \bibinfo {author} {\bibfnamefont {I.~O.}\ \bibnamefont {Ozel}}, \bibinfo {author} {\bibfnamefont {D.}~\bibnamefont {Mao}}, \bibinfo {author} {\bibfnamefont {H.~C.}\ \bibnamefont {Po}}, \bibinfo {author} {\bibfnamefont {C.~J.}\ \bibnamefont {Allington}}, \bibinfo {author} {\bibfnamefont {S.}~\bibnamefont {Son}}, \bibinfo {author} {\bibfnamefont {B.~H.}\ \bibnamefont {Kim}}, \bibinfo {author} {\bibfnamefont {J.}~\bibnamefont {Kim}}, \bibinfo {author} {\bibfnamefont {I.}~\bibnamefont {Hwang}}, \emph {et~al.},\ }\href@noop {} {\bibfield  {journal} {\bibinfo  {journal} {Nat. Commun.}\ }\textbf {\bibinfo {volume} {12}},\ \bibinfo {pages} {4837} (\bibinfo {year} {2021})}\BibitemShut {NoStop}%
\bibitem [{\citenamefont {Bernasconi}\ \emph {et~al.}(1988)\citenamefont {Bernasconi}, \citenamefont {Marra}, \citenamefont {Benedek}, \citenamefont {Miglio}, \citenamefont {Jouanne}, \citenamefont {Julien}, \citenamefont {Scagliotti},\ and\ \citenamefont {Balkanski}}]{Raman}%
  \BibitemOpen
  \bibfield  {author} {\bibinfo {author} {\bibfnamefont {M.}~\bibnamefont {Bernasconi}}, \bibinfo {author} {\bibfnamefont {G.}~\bibnamefont {Marra}}, \bibinfo {author} {\bibfnamefont {G.}~\bibnamefont {Benedek}}, \bibinfo {author} {\bibfnamefont {L.}~\bibnamefont {Miglio}}, \bibinfo {author} {\bibfnamefont {M.}~\bibnamefont {Jouanne}}, \bibinfo {author} {\bibfnamefont {C.}~\bibnamefont {Julien}}, \bibinfo {author} {\bibfnamefont {M.}~\bibnamefont {Scagliotti}},\ and\ \bibinfo {author} {\bibfnamefont {M.}~\bibnamefont {Balkanski}},\ }\href@noop {} {\bibfield  {journal} {\bibinfo  {journal} {Phys. Rev. B}\ }\textbf {\bibinfo {volume} {38}},\ \bibinfo {pages} {12089} (\bibinfo {year} {1988})}\BibitemShut {NoStop}%
\bibitem [{\citenamefont {Negi}\ \emph {et~al.}(2024)\citenamefont {Negi}, \citenamefont {Badola}, \citenamefont {Paul}, \citenamefont {Pistawala}, \citenamefont {Harnagea},\ and\ \citenamefont {Saha}}]{Raman2}%
  \BibitemOpen
  \bibfield  {author} {\bibinfo {author} {\bibfnamefont {D.}~\bibnamefont {Negi}}, \bibinfo {author} {\bibfnamefont {S.}~\bibnamefont {Badola}}, \bibinfo {author} {\bibfnamefont {S.}~\bibnamefont {Paul}}, \bibinfo {author} {\bibfnamefont {N.}~\bibnamefont {Pistawala}}, \bibinfo {author} {\bibfnamefont {L.}~\bibnamefont {Harnagea}},\ and\ \bibinfo {author} {\bibfnamefont {S.}~\bibnamefont {Saha}},\ }\href@noop {} {\bibfield  {journal} {\bibinfo  {journal} {Phys. Rev. B}\ }\textbf {\bibinfo {volume} {110}},\ \bibinfo {pages} {094434} (\bibinfo {year} {2024})}\BibitemShut {NoStop}%
\bibitem [{\citenamefont {Kim}\ \emph {et~al.}(2019)\citenamefont {Kim}, \citenamefont {Lim}, \citenamefont {Lee}, \citenamefont {Lee}, \citenamefont {Kim}, \citenamefont {Park}, \citenamefont {Jeon}, \citenamefont {Park}, \citenamefont {Park},\ and\ \citenamefont {Cheong}}]{Raman3}%
  \BibitemOpen
  \bibfield  {author} {\bibinfo {author} {\bibfnamefont {K.}~\bibnamefont {Kim}}, \bibinfo {author} {\bibfnamefont {S.~Y.}\ \bibnamefont {Lim}}, \bibinfo {author} {\bibfnamefont {J.-U.}\ \bibnamefont {Lee}}, \bibinfo {author} {\bibfnamefont {S.}~\bibnamefont {Lee}}, \bibinfo {author} {\bibfnamefont {T.~Y.}\ \bibnamefont {Kim}}, \bibinfo {author} {\bibfnamefont {K.}~\bibnamefont {Park}}, \bibinfo {author} {\bibfnamefont {G.~S.}\ \bibnamefont {Jeon}}, \bibinfo {author} {\bibfnamefont {C.-H.}\ \bibnamefont {Park}}, \bibinfo {author} {\bibfnamefont {J.-G.}\ \bibnamefont {Park}},\ and\ \bibinfo {author} {\bibfnamefont {H.}~\bibnamefont {Cheong}},\ }\href@noop {} {\bibfield  {journal} {\bibinfo  {journal} {Nat. Commun.}\ }\textbf {\bibinfo {volume} {10}},\ \bibinfo {pages} {345} (\bibinfo {year} {2019})}\BibitemShut {NoStop}%
\bibitem [{\citenamefont {Kuo}\ \emph {et~al.}(2016)\citenamefont {Kuo}, \citenamefont {Neumann}, \citenamefont {Balamurugan}, \citenamefont {Park}, \citenamefont {Kang}, \citenamefont {Shiu}, \citenamefont {Kang}, \citenamefont {Hong}, \citenamefont {Han}, \citenamefont {Noh} \emph {et~al.}}]{Raman4}%
  \BibitemOpen
  \bibfield  {author} {\bibinfo {author} {\bibfnamefont {C.-T.}\ \bibnamefont {Kuo}}, \bibinfo {author} {\bibfnamefont {M.}~\bibnamefont {Neumann}}, \bibinfo {author} {\bibfnamefont {K.}~\bibnamefont {Balamurugan}}, \bibinfo {author} {\bibfnamefont {H.~J.}\ \bibnamefont {Park}}, \bibinfo {author} {\bibfnamefont {S.}~\bibnamefont {Kang}}, \bibinfo {author} {\bibfnamefont {H.~W.}\ \bibnamefont {Shiu}}, \bibinfo {author} {\bibfnamefont {J.~H.}\ \bibnamefont {Kang}}, \bibinfo {author} {\bibfnamefont {B.~H.}\ \bibnamefont {Hong}}, \bibinfo {author} {\bibfnamefont {M.}~\bibnamefont {Han}}, \bibinfo {author} {\bibfnamefont {T.~W.}\ \bibnamefont {Noh}}, \emph {et~al.},\ }\href@noop {} {\bibfield  {journal} {\bibinfo  {journal} {Sci. Rep.}\ }\textbf {\bibinfo {volume} {6}},\ \bibinfo {pages} {20904} (\bibinfo {year} {2016})}\BibitemShut {NoStop}%
\bibitem [{\citenamefont {Li}\ \emph {et~al.}(2024)\citenamefont {Li}, \citenamefont {Liang}, \citenamefont {Kong}, \citenamefont {Sun},\ and\ \citenamefont {Zhang}}]{17}%
  \BibitemOpen
  \bibfield  {author} {\bibinfo {author} {\bibfnamefont {Y.}~\bibnamefont {Li}}, \bibinfo {author} {\bibfnamefont {G.}~\bibnamefont {Liang}}, \bibinfo {author} {\bibfnamefont {C.}~\bibnamefont {Kong}}, \bibinfo {author} {\bibfnamefont {B.}~\bibnamefont {Sun}},\ and\ \bibinfo {author} {\bibfnamefont {X.}~\bibnamefont {Zhang}},\ }\href {https://doi.org/10.1002/adfm.202402161} {\bibfield  {journal} {\bibinfo  {journal} {Adv. Funct. Mater.}\ }\textbf {\bibinfo {volume} {34}},\ \bibinfo {pages} {2402161} (\bibinfo {year} {2024})}\BibitemShut {NoStop}%
\bibitem [{\citenamefont {Kamaraju}\ \emph {et~al.}(2010)\citenamefont {Kamaraju}, \citenamefont {Kumar}, \citenamefont {Anija},\ and\ \citenamefont {Sood}}]{Anh1}%
  \BibitemOpen
  \bibfield  {author} {\bibinfo {author} {\bibfnamefont {N.}~\bibnamefont {Kamaraju}}, \bibinfo {author} {\bibfnamefont {S.}~\bibnamefont {Kumar}}, \bibinfo {author} {\bibfnamefont {M.}~\bibnamefont {Anija}},\ and\ \bibinfo {author} {\bibfnamefont {A.}~\bibnamefont {Sood}},\ }\href@noop {} {\bibfield  {journal} {\bibinfo  {journal} {Phys. Rev. B}\ }\textbf {\bibinfo {volume} {82}},\ \bibinfo {pages} {195202} (\bibinfo {year} {2010})}\BibitemShut {NoStop}%
\bibitem [{\citenamefont {Balkanski}\ \emph {et~al.}(1983)\citenamefont {Balkanski}, \citenamefont {Wallis},\ and\ \citenamefont {Haro}}]{Anh2}%
  \BibitemOpen
  \bibfield  {author} {\bibinfo {author} {\bibfnamefont {M.}~\bibnamefont {Balkanski}}, \bibinfo {author} {\bibfnamefont {R.}~\bibnamefont {Wallis}},\ and\ \bibinfo {author} {\bibfnamefont {E.}~\bibnamefont {Haro}},\ }\href@noop {} {\bibfield  {journal} {\bibinfo  {journal} {Phys. Rev. B}\ }\textbf {\bibinfo {volume} {28}},\ \bibinfo {pages} {1928} (\bibinfo {year} {1983})}\BibitemShut {NoStop}%
\bibitem [{\citenamefont {Men{\'e}ndez}\ and\ \citenamefont {Cardona}(1984)}]{Anh3}%
  \BibitemOpen
  \bibfield  {author} {\bibinfo {author} {\bibfnamefont {J.}~\bibnamefont {Men{\'e}ndez}}\ and\ \bibinfo {author} {\bibfnamefont {M.}~\bibnamefont {Cardona}},\ }\href@noop {} {\bibfield  {journal} {\bibinfo  {journal} {Phys. Rev. B}\ }\textbf {\bibinfo {volume} {29}},\ \bibinfo {pages} {2051} (\bibinfo {year} {1984})}\BibitemShut {NoStop}%
\bibitem [{\citenamefont {Saha}\ \emph {et~al.}(2008)\citenamefont {Saha}, \citenamefont {Singh}, \citenamefont {Dkhil}, \citenamefont {Dhar}, \citenamefont {Suryanarayanan}, \citenamefont {Dhalenne}, \citenamefont {Revcolevschi},\ and\ \citenamefont {Sood}}]{Selfenergy}%
  \BibitemOpen
  \bibfield  {author} {\bibinfo {author} {\bibfnamefont {S.}~\bibnamefont {Saha}}, \bibinfo {author} {\bibfnamefont {S.}~\bibnamefont {Singh}}, \bibinfo {author} {\bibfnamefont {B.}~\bibnamefont {Dkhil}}, \bibinfo {author} {\bibfnamefont {S.}~\bibnamefont {Dhar}}, \bibinfo {author} {\bibfnamefont {R.}~\bibnamefont {Suryanarayanan}}, \bibinfo {author} {\bibfnamefont {G.}~\bibnamefont {Dhalenne}}, \bibinfo {author} {\bibfnamefont {A.}~\bibnamefont {Revcolevschi}},\ and\ \bibinfo {author} {\bibfnamefont {A.}~\bibnamefont {Sood}},\ }\href@noop {} {\bibfield  {journal} {\bibinfo  {journal} {Phys. Rev. B}\ }\textbf {\bibinfo {volume} {78}},\ \bibinfo {pages} {214102} (\bibinfo {year} {2008})}\BibitemShut {NoStop}%
\bibitem [{\citenamefont {Liang}\ \emph {et~al.}(2022)\citenamefont {Liang}, \citenamefont {Zhai}, \citenamefont {Ma}, \citenamefont {Wang}, \citenamefont {Zhao}, \citenamefont {Wu},\ and\ \citenamefont {Zhang}}]{Cd3As2}%
  \BibitemOpen
  \bibfield  {author} {\bibinfo {author} {\bibfnamefont {G.}~\bibnamefont {Liang}}, \bibinfo {author} {\bibfnamefont {G.}~\bibnamefont {Zhai}}, \bibinfo {author} {\bibfnamefont {J.}~\bibnamefont {Ma}}, \bibinfo {author} {\bibfnamefont {H.}~\bibnamefont {Wang}}, \bibinfo {author} {\bibfnamefont {J.}~\bibnamefont {Zhao}}, \bibinfo {author} {\bibfnamefont {X.}~\bibnamefont {Wu}},\ and\ \bibinfo {author} {\bibfnamefont {X.}~\bibnamefont {Zhang}},\ }\href@noop {} {\bibfield  {journal} {\bibinfo  {journal} {J. Phys. Chem. Lett.}\ }\textbf {\bibinfo {volume} {13}},\ \bibinfo {pages} {8783} (\bibinfo {year} {2022})}\BibitemShut {NoStop}%
\bibitem [{\citenamefont {Thomsen}\ \emph {et~al.}(1984)\citenamefont {Thomsen}, \citenamefont {Strait}, \citenamefont {Vardeny}, \citenamefont {Maris}, \citenamefont {Tauc},\ and\ \citenamefont {Hauser}}]{Thomsen}%
  \BibitemOpen
  \bibfield  {author} {\bibinfo {author} {\bibfnamefont {C.}~\bibnamefont {Thomsen}}, \bibinfo {author} {\bibfnamefont {J.}~\bibnamefont {Strait}}, \bibinfo {author} {\bibfnamefont {Z.}~\bibnamefont {Vardeny}}, \bibinfo {author} {\bibfnamefont {H.~J.}\ \bibnamefont {Maris}}, \bibinfo {author} {\bibfnamefont {J.}~\bibnamefont {Tauc}},\ and\ \bibinfo {author} {\bibfnamefont {J.}~\bibnamefont {Hauser}},\ }\href@noop {} {\bibfield  {journal} {\bibinfo  {journal} {Phys. Rev. Lett.}\ }\textbf {\bibinfo {volume} {53}},\ \bibinfo {pages} {989} (\bibinfo {year} {1984})}\BibitemShut {NoStop}%
\bibitem [{\citenamefont {Zhou}\ \emph {et~al.}(2022)\citenamefont {Zhou}, \citenamefont {Hwangbo}, \citenamefont {Zhang}, \citenamefont {Wang}, \citenamefont {Shen}, \citenamefont {Zhang}, \citenamefont {Jiang}, \citenamefont {Zong}, \citenamefont {Su}, \citenamefont {Zajac},\ and\ \citenamefont {Ahn}}]{spin-shear}%
  \BibitemOpen
  \bibfield  {author} {\bibinfo {author} {\bibfnamefont {F.}~\bibnamefont {Zhou}}, \bibinfo {author} {\bibfnamefont {K.}~\bibnamefont {Hwangbo}}, \bibinfo {author} {\bibfnamefont {Q.}~\bibnamefont {Zhang}}, \bibinfo {author} {\bibfnamefont {C.}~\bibnamefont {Wang}}, \bibinfo {author} {\bibfnamefont {L.}~\bibnamefont {Shen}}, \bibinfo {author} {\bibfnamefont {J.}~\bibnamefont {Zhang}}, \bibinfo {author} {\bibfnamefont {Q.}~\bibnamefont {Jiang}}, \bibinfo {author} {\bibfnamefont {A.}~\bibnamefont {Zong}}, \bibinfo {author} {\bibfnamefont {Y.}~\bibnamefont {Su}}, \bibinfo {author} {\bibfnamefont {M.}~\bibnamefont {Zajac}},\ and\ \bibinfo {author} {\bibfnamefont {Y.}~\bibnamefont {Ahn}},\ }\href {https://doi.org/10.1038/s41467-022-34376-5} {\bibfield  {journal} {\bibinfo  {journal} {Nat. Commun.}\ }\textbf {\bibinfo {volume} {13}},\ \bibinfo {pages} {6598} (\bibinfo {year} {2022})}\BibitemShut {NoStop}%
\bibitem [{\citenamefont {Plumley}\ \emph {et~al.}(2024)\citenamefont {Plumley}, \citenamefont {Mardanya}, \citenamefont {Peng}, \citenamefont {Nokelainen}, \citenamefont {Assefa}, \citenamefont {Shen}, \citenamefont {Burdet}, \citenamefont {Porter}, \citenamefont {Petsch}, \citenamefont {Israelski},\ and\ \citenamefont {Chen}}]{criticalexponent}%
  \BibitemOpen
  \bibfield  {author} {\bibinfo {author} {\bibfnamefont {R.}~\bibnamefont {Plumley}}, \bibinfo {author} {\bibfnamefont {S.}~\bibnamefont {Mardanya}}, \bibinfo {author} {\bibfnamefont {C.}~\bibnamefont {Peng}}, \bibinfo {author} {\bibfnamefont {J.}~\bibnamefont {Nokelainen}}, \bibinfo {author} {\bibfnamefont {T.}~\bibnamefont {Assefa}}, \bibinfo {author} {\bibfnamefont {L.}~\bibnamefont {Shen}}, \bibinfo {author} {\bibfnamefont {N.}~\bibnamefont {Burdet}}, \bibinfo {author} {\bibfnamefont {Z.}~\bibnamefont {Porter}}, \bibinfo {author} {\bibfnamefont {A.}~\bibnamefont {Petsch}}, \bibinfo {author} {\bibfnamefont {A.}~\bibnamefont {Israelski}},\ and\ \bibinfo {author} {\bibfnamefont {H.}~\bibnamefont {Chen}},\ }\href@noop {} {\bibfield  {journal} {\bibinfo  {journal} {npj Quantum Mater.}\ }\textbf {\bibinfo {volume} {9}},\ \bibinfo {pages} {95} (\bibinfo {year} {2024})}\BibitemShut {NoStop}%
\bibitem [{\citenamefont {Scheie}\ \emph {et~al.}(2023)\citenamefont {Scheie}, \citenamefont {Park}, \citenamefont {Villanova}, \citenamefont {Granroth}, \citenamefont {Sarkis}, \citenamefont {Zhang}, \citenamefont {Stone}, \citenamefont {Park}, \citenamefont {Okamoto}, \citenamefont {Berlijn},\ and\ \citenamefont {Tennant}}]{SWG1}%
  \BibitemOpen
  \bibfield  {author} {\bibinfo {author} {\bibfnamefont {A.}~\bibnamefont {Scheie}}, \bibinfo {author} {\bibfnamefont {P.}~\bibnamefont {Park}}, \bibinfo {author} {\bibfnamefont {J.}~\bibnamefont {Villanova}}, \bibinfo {author} {\bibfnamefont {G.}~\bibnamefont {Granroth}}, \bibinfo {author} {\bibfnamefont {C.}~\bibnamefont {Sarkis}}, \bibinfo {author} {\bibfnamefont {H.}~\bibnamefont {Zhang}}, \bibinfo {author} {\bibfnamefont {M.}~\bibnamefont {Stone}}, \bibinfo {author} {\bibfnamefont {J.}~\bibnamefont {Park}}, \bibinfo {author} {\bibfnamefont {S.}~\bibnamefont {Okamoto}}, \bibinfo {author} {\bibfnamefont {T.}~\bibnamefont {Berlijn}},\ and\ \bibinfo {author} {\bibfnamefont {D.}~\bibnamefont {Tennant}},\ }\href {https://doi.org/10.1103/physrevb.108.104402} {\bibfield  {journal} {\bibinfo  {journal} {Phys. Rev. B}\ }\textbf {\bibinfo {volume} {108}},\ \bibinfo {pages} {104402} (\bibinfo {year} {2023})}\BibitemShut {NoStop}%
\bibitem [{\citenamefont {Mehlawat}\ \emph {et~al.}(2022)\citenamefont {Mehlawat}, \citenamefont {Alfonsov}, \citenamefont {Selter}, \citenamefont {Shemerliuk}, \citenamefont {Aswartham}, \citenamefont {B{\"u}chner},\ and\ \citenamefont {Kataev}}]{SWG2}%
  \BibitemOpen
  \bibfield  {author} {\bibinfo {author} {\bibfnamefont {K.}~\bibnamefont {Mehlawat}}, \bibinfo {author} {\bibfnamefont {A.}~\bibnamefont {Alfonsov}}, \bibinfo {author} {\bibfnamefont {S.}~\bibnamefont {Selter}}, \bibinfo {author} {\bibfnamefont {Y.}~\bibnamefont {Shemerliuk}}, \bibinfo {author} {\bibfnamefont {S.}~\bibnamefont {Aswartham}}, \bibinfo {author} {\bibfnamefont {B.}~\bibnamefont {B{\"u}chner}},\ and\ \bibinfo {author} {\bibfnamefont {V.}~\bibnamefont {Kataev}},\ }\href {https://doi.org/10.1103/physrevb.105.214427} {\bibfield  {journal} {\bibinfo  {journal} {Phys. Rev. B}\ }\textbf {\bibinfo {volume} {105}},\ \bibinfo {pages} {214427} (\bibinfo {year} {2022})}\BibitemShut {NoStop}%
\bibitem [{\citenamefont {Jana}\ \emph {et~al.}(2023)\citenamefont {Jana}, \citenamefont {Kapuscinski}, \citenamefont {Mohelsky}, \citenamefont {Vaclavkova}, \citenamefont {Breslavetz}, \citenamefont {Orlita}, \citenamefont {Faugeras},\ and\ \citenamefont {Potemski}}]{SWG3}%
  \BibitemOpen
  \bibfield  {author} {\bibinfo {author} {\bibfnamefont {D.}~\bibnamefont {Jana}}, \bibinfo {author} {\bibfnamefont {P.}~\bibnamefont {Kapuscinski}}, \bibinfo {author} {\bibfnamefont {I.}~\bibnamefont {Mohelsky}}, \bibinfo {author} {\bibfnamefont {D.}~\bibnamefont {Vaclavkova}}, \bibinfo {author} {\bibfnamefont {I.}~\bibnamefont {Breslavetz}}, \bibinfo {author} {\bibfnamefont {M.}~\bibnamefont {Orlita}}, \bibinfo {author} {\bibfnamefont {C.}~\bibnamefont {Faugeras}},\ and\ \bibinfo {author} {\bibfnamefont {M.}~\bibnamefont {Potemski}},\ }\href {https://doi.org/10.1103/physrevb.108.115149} {\bibfield  {journal} {\bibinfo  {journal} {Phys. Rev. B}\ }\textbf {\bibinfo {volume} {108}},\ \bibinfo {pages} {115149} (\bibinfo {year} {2023})}\BibitemShut {NoStop}%
\bibitem [{\citenamefont {Boggs}\ \emph {et~al.}(1992)\citenamefont {Boggs}, \citenamefont {Boggs}, \citenamefont {Rogers},\ and\ \citenamefont {Schnabel}}]{1}%
  \BibitemOpen
  \bibfield  {author} {\bibinfo {author} {\bibfnamefont {P.~T.}\ \bibnamefont {Boggs}}, \bibinfo {author} {\bibfnamefont {P.~T.}\ \bibnamefont {Boggs}}, \bibinfo {author} {\bibfnamefont {J.~E.}\ \bibnamefont {Rogers}},\ and\ \bibinfo {author} {\bibfnamefont {R.~B.}\ \bibnamefont {Schnabel}},\ }\href@noop {} {\emph {\bibinfo {title} {User's reference guide for odrpack version 2.01: Software for weighted orthogonal distance regression}}},\ \bibinfo {type} {Tech. Rep.}\ (\bibinfo  {institution} {National Institute of Standards and Technology},\ \bibinfo {year} {1992})\BibitemShut {NoStop}%
\bibitem [{\citenamefont {Watson}\ \emph {et~al.}(2005)\citenamefont {Watson}, \citenamefont {Boggs},\ and\ \citenamefont {Zwolak}}]{2}%
  \BibitemOpen
  \bibfield  {author} {\bibinfo {author} {\bibfnamefont {L.~T.}\ \bibnamefont {Watson}}, \bibinfo {author} {\bibfnamefont {P.~T.}\ \bibnamefont {Boggs}},\ and\ \bibinfo {author} {\bibfnamefont {J.~W.}\ \bibnamefont {Zwolak}},\ }\href@noop {} {\emph {\bibinfo {title} {ODRPACK95: a weighted orthogonal distance regression code with bound constraints}}},\ \bibinfo {type} {Tech. Rep.}\ (\bibinfo  {institution} {Sandia National Laboratories (SNL), Albuquerque, NM, and Livermore, CA},\ \bibinfo {year} {2005})\BibitemShut {NoStop}%
\end{thebibliography}%
\end{document}